\pdfoutput=1
\documentclass[12pt, a4paper]{article}
\usepackage[utf8]{inputenc}
\usepackage[T1]{fontenc}
\usepackage{lmodern}

\usepackage{graphicx, xcolor, tikz, booktabs, float, natbib, subcaption}
\usepackage{amsmath, amssymb, amsthm, dsfont}
\usetikzlibrary{arrows.meta, calc, shapes, positioning}

\usepackage{datetime}
\newdateformat{monthyeardate}{\monthname[\THEMONTH] \THEYEAR}

\theoremstyle{plain}
\newtheorem{theorem}{Theorem}
\newtheorem{lemma}{Lemma}
\newtheorem{corollary}[lemma]{Corollary}
\newtheorem{proposition}[lemma]{Proposition}
\theoremstyle{definition}
\newtheorem{definition}{Definition}
\newtheorem{example}{Example}
\theoremstyle{remark}
\newtheorem{remark}{Remark}

\usepackage{etoolbox}
\AtBeginEnvironment{example}{\pushQED{\qed}}
\AtEndEnvironment{example}{\popQED}

\usepackage{hyperref}

\usepackage[margin=1.2in]{geometry}

\def\imagewidth{0.85\linewidth}

\title{{Platform-Induced Coordination}\thanks{Chitiva acknowledges the support of Hi!PARIS and the ANR (``Investissements d'Avenir programme'' reference ANR-18-EURE-0005/EUR DATA EFM and ``France 2030'' reference ANR-23-IACL-0005).
Hernández acknowledges the support of the Spanish Ministry of Economics and Innovation, grant PID2024-162635NB-I00 funded by MICIU/AEI/10.13039/501100011033/FEDER, UE and by Consellería d'Innovació, Universitats, Ciència i Societat Digital, Generalitat Valenciana, grant CIAICO/2023/031.
This work was partially done when Julián Chitiva was visiting University of Valencia.
}}
\author{Julián Chitiva\thanks{HEC Paris -- \href{mailto:julian.chitiva@hec.edu}{julian.chitiva@hec.edu}.} \and Penélope Hernández\thanks{ERI-CES, Department of Economics, University of Valencia, Spain -- \href{mailto:Penelope.Hernandez@uv.es}{Penelope.Hernandez@uv.es}}}
\date{}

\def\keywordname{{\bf Keywords:}}
\def\jelname{{\bf JEL:}}
\providecommand{\keywords}[1]{\def\and{{\textperiodcentered} }%
\par\addvspace\baselineskip
\noindent\keywordname\enspace\ignorespaces#1}%
\providecommand{\jel}[1]{\def\and{{\textperiodcentered} }%
\par
\noindent\jelname\enspace\ignorespaces#1}%

\begin{document}

\maketitle
\begin{abstract}
This paper studies a platform that influences collective action by selecting which posts each user observes. Each user would like to take an action close to those of the others and to an ideal action determined by her private type and an aggregate state. Posts are non-strategic but correlated with the types of their authors, so each post carries information about the state and its author. We characterise the unique Bayes-Nash equilibrium in closed form. Two explicit networks give the minimum and maximum collective action the platform can induce. Moreover, collective action depends on the network only through the numbers of readers. Hence, many networks induce the same collective action, but they can induce different individual actions. These networks may differ for a regulator, and we compare them by welfare, polarisation and cross-type exposure. In particular, the complete network need not maximise welfare, and the segregated network need not maximise polarisation.
\keywords{Social media platforms \and information networks \and coordination games \and network design}
\jel{C72 \and D82 \and D85}
\end{abstract}

\section{Introduction}\label{sec:intro}

Social media has become a primary channel through which people learn about the world and decide how to act in it \citep{kemp2024digital}. What any individual user sees is not the totality of what is posted but a selection, and the platform makes that selection. A recommender system decides which posts are shown to whom, and in doing so it determines what each user knows about the state of the world and about the people around her. The assumption that platforms have this power is not a theoretical convenience. \citet{gauthier2025twitter} provide causal evidence from a 2023 field experiment that the feed algorithm of X promotes conservative content and shifts users' political opinions towards more conservative positions, an effect that persists after the algorithm is switched off.
Effects of this kind matter in election years, such as the US presidential campaign of 2024.

This paper asks what a platform can achieve by choosing who observes whom, and what the consequences are for the users. The environment we study has two features that distinguish it from the canonical models of information design. The first is that users have a dual motive. Each has a private type that is payoff-relevant in its own right, so that a user predisposed towards political activism has a higher preferred level of participation whatever the aggregate state turns out to be, and each also wants to act in accordance with the people she observes. The second is that the platform manipulates neither the content of posts nor the precision of signals. It chooses only the topology of attention, which posts are shown to which users, and it does so knowing the distribution of types but not the content that will be generated.
These two features interact in a way that drives everything that follows. Because posts are correlated with their authors' types, a post carries two kinds of information. It is evidence about the aggregate state, which a user needs to act consistently with her own preferences, and it is evidence about the type of its author, which she needs to anticipate how that author will act. A platform that changes who sees a post therefore changes beliefs about the world and beliefs about the population simultaneously, without ever revealing a type or altering a message.

Our model features a sequential game in which Nature first determines user types and realises a payoff-relevant state, after which each user privately observes both their type and a correlated signal about the state. The platform observes user types but not their individual signals, while the distributions of types and posts are common knowledge. Users do not observe the network. Each user sees the posts in her feed and how many users see each of them, and she does not infer the types of others from the structure of her feed, although the platform chooses the network knowing the types. Her expectations about the platform are coarse, in the spirit of analogy-based expectations \citep{jehiel2005analogy}, while she is fully Bayesian about the other users. This is a natural assumption on large platforms, where users cannot observe the algorithm that selects their feed. Given this informational structure, the platform designs the network by selecting which signals each user can observe beyond their own, thereby shaping the informational environment for coordination. Users then choose actions based on their private information and beliefs formed through network-mediated signal observation. Equilibrium behaviour arises from the interaction among users' intrinsic heterogeneity, their coordination motives, the information that the network makes available, and the platform's strategic objective of steering collective action towards a target level. 

We characterise the unique Bayes-Nash equilibrium of the user coordination game. Each user's equilibrium action is a weighted aggregation of the posts she observes, with a closed-form expression for the weights, and it satisfies a monotonicity property that drives the platform's leverage. The weight a post receives is increasing in the size of its audience, so that widening the circulation of a post shifts every reader's action in the direction of its content and of the inferred type of its author. Revealing more about high-type users raises collective action, and revealing more about low-type users lowers it, yielding two uniquely defined extreme architectures that bound the range of collective action the platform can induce.

Beyond these extremes, we demonstrate that multiple networks may be platform-equivalent, inducing identical ex ante collective actions despite different architectures, and we characterise operators and construct architectures that generate such equivalence classes. For a platform pursuing an aggregate target, what matters is the composition of the information circulating, not the identity of the users who receive it. From the platform's perspective, the choice within a class is a matter of indifference, and that indifference creates room for a regulator to act without disturbing the platform's objective.

Whether the room is worth occupying depends on how members of a class differ, and we evaluate them along three dimensions. The first is utilitarian welfare. The second is heterophilic density, a measure of the type composition of the links a platform creates. The third is polarisation, measured as the dispersion of equilibrium actions. We show first that welfare and polarisation are not independent objectives, since the conformity component of welfare is population polarisation up to a positive constant. A regulator who lowers polarisation without degrading the accuracy of users' actions therefore raises welfare, and the two criteria can move in opposite directions only through the quality of the information users hold about the aggregate state.

Our main substantive finding is that two conjectures about these criteria, both natural and both supported by existing results in adjacent settings, are false in this environment. The first is that full information sharing maximises welfare, so that a welfare-motivated regulator should aim at the complete network. %
The second is that segregation maximises polarisation, so that the architecture in which each type observes only its own type generates the widest behavioural divergence. %
We exhibit type compositions\footnote{Both counterexamples use rational parameter values and are therefore exact rather than numerical, and both arise at interior parameter values rather than at extremes.} at which muting a minority strictly raises utilitarian welfare, and parameter regions in which the architecture where each type observes only the opposite type is strictly more polarising than the echo chamber. To establish the two failures, we work with a class of architectures in which the feed conditions only on the author type and the reader type. On this class, users of a common type are informationally interchangeable, the equilibrium collapses onto two representative users, and all three criteria admit closed-form expressions in the primitives.

The two failures have different sources. The first rests on the coordination motive together with the fact that users do not see how their feeds are selected. Because users load on beliefs about one another's types as well as on the posts themselves, and do not infer types from what the platform hides, a feed that conditions on types moves the average action of its readers, and not only the precision with which they track the state. A designer who knows the types can use this in two ways. She can offset the shading that conformity induces when the population differs from what users expect, as when a minority is muted, or she can bring the two groups closer together, as when each group is shown only the posts of the other. Both effects disappear as the coordination weight vanishes, which is why full disclosure is optimal in settings without such a motive and is not optimal here. The second failure does not depend on coordination. Dispersion counts disagreement within groups as well as between them, and segregation makes each group internally uniform, so the identification of segregation with polarisation fails whenever users read posts mainly for their content, and continues to fail as the coordination weight vanishes.

\subsection{Related Literature}\label{subsec:literature}

Our paper examines how platforms can use information network design to influence collective action, and the welfare implications of this strategic capability. This study contributes to various areas of economic literature. We organise our discussion by distinguishing between the theoretical foundations on which we build, the literature we extend, and the approaches from which we depart.

Our user model is a coordination problem with dispersed information on networks. Users play a linear-quadratic beauty-contest game \citep{morris2002social, angeletos2007efficient}, in which each agent trades off tracking a fundamental against aligning with others. Closest to us are network versions of this game. \citet{calvo2009information} study organisations whose members coordinate on a common decision and gather information through pairwise channels, and characterise how architecture shapes equilibrium and welfare. \citet{hagenbach2010strategic} study agents who choose whom to reveal their signals to when preferences differ by commonly known biases, and \citet{galeotti2013strategic} show that truthful communication to a receiver becomes harder as she draws information from more sources. As in organisations whose divisions must adapt their decisions to local conditions and coordinate with each other \citep{alonso2008does}, each user's payoff adds an adaptation loss and a coordination loss. We depart from these models in two respects, which together drive most of our
results. First, each user's ideal action depends on a private type that is
payoff-relevant in its own right. Second, posts are correlated with types. A post is
therefore evidence about two things, the aggregate state and the action its author
will take. Existence and uniqueness of equilibrium follow from \citet{marschak1972economic}. Our contribution is the closed-form characterisation for every network, and the finding that the network enters actions only through the reach of each post. The component of actions that responds to the state coincides with the solution of \citet{hagenbach2010strategic}, nevertheless the component that responds to types has no counterpart there, because their biases are common knowledge.

The platform's problem is information design in a game with strategic complementarities. Unrestricted design is well understood in the literature. The outcomes a designer can induce are the Bayes correlated equilibria of the game \citep{bergemann2013robust, bergemann2016bayes}, and optimal structures have been characterised for a single receiver \citep{kamenica2011persuasion} and in games \citep{taneva2019information, mathevet2020information}. Our platform is a restricted designer. It cannot garble, randomise or recommend, and it cannot create evidence. It only decides who observes which of the posts that users generate. In this respect, it is closest to designers limited to public signals \citep{candogan2022persuasion}, to designers who seed information that then spills over a network \citep{galperti2025games}, and to the design of the structure through which information is transmitted \citep{taneva2022organized}. What we add is a characterisation of the attainable set for this instrument. Collective action lies in an interval whose endpoints are attained by two explicit networks, the vector of reaches is a sufficient statistic for it, and the platform-equivalence classes this induces are large and constructible. A second difference concerns what receivers know about the designer. In the persuasion literature receivers know the information policy. Our users instead hold coarse expectations about the platform, in the spirit of analogy-based expectations \citep{jehiel2005analogy, jehiel2008revisiting}. Namely, users are Bayesian about the other users but do not invert the rule that maps types into feeds.

Our first negative result concerns the welfare value of information. In beauty contests, more precise public information can lower welfare \citep{morris2002social}, and \citet{angeletos2007efficient} trace its social value to the gap between the equilibrium and the efficient use of information. In organisations with coordination motives, by contrast, complete information is the Pareto-dominant information structure \citep{hagenbach2016full}. We show that the complete network can be welfare-dominated even though welfare fully accounts for the coordination loss, and even though the platform cannot alter the precision of any signal. The source differs from both strands. What is withdrawn is not a noisy signal about a common fundamental but evidence about the type of an author whose ideal action is far from that of her readers. Circulating it improves inference about the state but raises the cost of conformity, and when the type motive is strong and the minority is small the second effect dominates. The effect vanishes as the coordination weight goes to zero, and the region in which it arises is bounded in the absolute size of the minority rather than in its share.

Theoretical work on platforms has studied the spread of misinformation under engagement-maximising curation \citep{acemoglu2024model}, when a platform should break echo chambers \citep{mostagir2023should}, and how personalisation algorithms shape social learning \citep{risco2024feed}; \citet{aridor2024economics} survey the field. In these models, users learn but do not coordinate, and the questions concern beliefs rather than actions. The network literature has studied how homophily arises \citep{currarini2009economic}, how it slows learning and convergence \citep{golub2012homophily}, and how network position governs the weight of opinions \citep{demarzo2003persuasion}. Our heterophilic density serves a different purpose since it describes the mixing a designer imposes rather than the homophily agents choose. Our second negative result speaks to the empirical evidence, which is mixed. Exposure on platforms is ideologically homogeneous \citep{bakshy2015exposure} and digital media have been linked to affective polarisation \citep{tornberg2022digital}, yet polarisation has grown fastest among the groups least likely to be online \citep{boxell2017greater}, and experimental changes to feed algorithms have produced both null \citep{guess2023social} and persistent \citep{gauthier2025twitter} effects on attitudes. In our model, segregation and behavioural dispersion are distinct objects, related non-monotonically. Dispersion counts disagreement within groups as well as between them, and the echo chamber minimises the former. %

The remainder of the paper is organised as follows. Section~\ref{sec:optimal_action} characterises the equilibrium of the coordination game and establishes the monotonicity properties of actions with respect to observed posts and network reach. Section~\ref{sec:manipulation} analyses the platform's problem, introduces the network transformation operators and identifies the conditions under which distinct architectures induce identical collective action. Section~\ref{sec:policy} evaluates the three criteria and establishes the two failures. Section~\ref{sec:conclusions} concludes. All proofs are in the appendix, and a replication script for every reported computation accompanies the paper.
\section{Coordination inside a Platform}\label{sec:optimal_action}

Consider a social media platform operating during an electoral campaign. Let $N=\{1,2,\dots,n\}$ denote the set of users of a Platform, with $n\geq 2$. The platform's users consist of two types: \emph{right-leaning users}, who are intrinsically predisposed towards right-wing political participation, and \emph{left-leaning users}, who are predisposed towards left-wing political participation. The platform infers each user's type from behavioural history, such as the account settings, declared interests, and prior campaign activity, but does not observe the specific content each user will post on a given day. Formally, each user has a private type $t_i\in T=\{\underbar{t},\overline{t}\}$, where $\underbar{t}<\overline{t}$. The types are drawn \textit{i.i.d.} from a Bernoulli distribution such that $\Pr(t_i=\overline{t})=p$. While the distribution of types is common knowledge, each user is privately informed only about their own type.

At the beginning of the game, each user has a post, a private signal $s_i \in \{\bar{s},\underline{s}\}$, that is positively correlated with their type. Right-leaning users are more likely to post right-leaning content, and left-leaning users are more likely to post left-leaning content. Formally, the signals are correlated with the user types, and the marginal distributions conditional on the type follow a Bernoulli distribution such that $\Pr(s_i=\overline{s}|t_i=\overline{t})=\lambda$ and $\Pr(s_i=\underbar{s}|t_i=\underbar{t})=\mu$, with $\lambda,\mu>\frac{1}{2}$. This regularity is well documented: on digital platforms, the content users generate is systematically aligned with their underlying preferences and characteristics. Users with conservative political predispositions disproportionately share conservative-coded content, and users with  progressive ones do the opposite \citep{barbera2015tweeting}. Similarly, users' sharing behaviour during electoral campaigns is strongly predicted by their prior ideological dispositions  \citep{guess2023social}. From a theoretical standpoint, this correlation structure is standard in the global games and social learning literature, where each agent's private signal is an informative but noisy draw correlated with the payoff-relevant state \citep{carlsson1993global, golub2010naive}. Although many setups treat content generation as a strategic component, this paper abstracts from this by assuming that content is generated automatically according to the given distributions (see Section~\ref{sec:assumptions}). 

We assume that the state of nature is the aggregate term $\theta(\mathbf{s})=\sum_{i\in N} s_i$, where $\mathbf{s}$ is the profile of signals. This assumption is standard in the literature, where the sum of players' signals serves as a proxy for the payoff-relevant state. For example, in organisational settings, this could represent a division's resources (time, budget, etc.), which are private information, whereas the variable that matters to the organisation is the aggregate of those resources. In the context of social media, $\theta(\mathbf{s})$ might represent the ``trending'' potential or public sentiment surrounding a specific topic, which individual users contribute to through their private posts. Moreover, this assumption is a standard simplification in common-value environments, such as auctions, lobbying, or organisational theory (see \cite{mares2003private,levin2004competitiveness,wolinsky2002eliciting}; and more recently 
\cite{bergemann2017firstprice,battaglini2018influencing,dessein2016rational}).

Users participate in a social media platform represented by an exogenous directed network $G=(N,E)$, where $N$ is the set of nodes and $E\subseteq N^2$ is the set of edges. This network determines the set of posts that each user sees, for example, the ``Suggested for you'' feature on Instagram or the ``For you'' feed on $\mathbb{X}$. An edge $(i,j)\in E$ means that user $j$ has access to the tuple $(s_i,r_i)$, where $s_i$ is the post made by user $i$ and $r_i$ the number of users other than $i$ who have access to it. 
Once the agents have observed the information on the platform, they take an action $a_i$ to maximise their utility function

\begin{equation}\label{eq:user_utility}
    u_i(\mathbf{a};\mathbf{s},t_i)=-(1-\alpha)\left[a_i-(\beta_1\theta(\mathbf{s})+\beta_2t_i)\right]^2-\frac{\alpha}{n-1}\sum\limits_{j\neq i}\left(a_i-a_j\right)^2
\end{equation}

This function has two main components. First, the consistency part measures the quadratic loss between user $i$'s action $a_i$ and their ideal action $\beta_1\theta(\mathbf{s})+\beta_2t_i$, where $\beta_1,\beta_2>0$. 
The parameters $\beta_1$ and $\beta_2$ represent the importance of the state of nature relative to the user's type in determining their preferred action. Second, the coordination component indicates that user $i$ incurs an increasing loss in the average quadratic distance between their action and others' actions. The type $t_i$ captures a dimension of user heterogeneity that is payoff-relevant independently of the aggregate state $\theta(\mathbf{s})$. A user with a high type $\overline{t}$ has an intrinsically higher preferred action. They are, for instance, predisposed towards higher engagement, stronger political activism, or greater consumption regardless of what the platform's aggregate signal reveals. This feature distinguishes our specification from the classic beauty-contest framework \cite{morris2002social}, in which all agents share a common ideal action, and heterogeneity enters only through the precision of private signals. It also departs from \cite{hagenbach2010strategic}, where preference heterogeneity is modelled as a bias over a common state, rather than as a separate payoff-relevant dimension. Our specification is instead related to models in political economy and organisational economics in which payoffs add up separate motives. In \citet{feddersen2006theory}, voters' payoffs combine a preference over the act of voting with a payoff from the outcome of the election. In \citet{alonso2008does}, the decisions of a multidivisional organisation must be adapted to local conditions and also coordinated with each other, so that each division bears a quadratic adaptation loss and a quadratic coordination loss. These two losses are the counterpart of the two components of Equation~\eqref{eq:user_utility}. In their model, however, the private information concerns local conditions and is communicated through cheap talk, whereas in ours each user's ideal action depends on both a private type and the aggregate state. The parameter $\alpha$ weights these two quadratic losses. The weight $\alpha$ on coordination relative to consistency is not merely a theoretical parameter since it reflects a behaviourally documented feature of digital platforms. Experimental and observational work finds that users' actions on social media are substantially responsive to perceived peer behaviour: likes, shares, and endorsements shift individuals' actions towards those of visible peers, independent of users' prior assessments (\cite{muchnik2013social}, \cite{lorenz2023systematic}). The parameter $\alpha\in (0,1)$ thus captures the documented tendency towards social conformity in action, while $(1-\alpha)$ captures the residual weight assigned to individual preference consistency.

Since the action $a_i$ depends on the information provided by the network, we use $a_i$ to denote $a_i(G)$ by a slight abuse of notation, since users do not directly observe the network but a set of tuples $(s_i, r_i)$ induced by $G$. 

\begin{figure}[H]
    \centering
    \begin{tikzpicture}[auto,node distance=1.8cm,main node/.style={circle,inner sep=0.5pt, font=\sffamily\large, draw, thin}]

    \node[main node] (0_1) {\textcolor{blue}{1}};
    \node[main node] (0_2) [right of=0_1] {\textcolor{red}{2}};
    \node[main node] (0_3) [below of=0_2] {\textcolor{red}{3}};
    \node[main node] (0_4)  [below of=0_1] {\textcolor{blue}{4}};
    
    \node[label={[label distance=-.01cm]90:{{$G$}}}] (G) at ($(0_1)!0.5!(0_2)$) {};
    \path[shorten >=1pt, thick,-{Latex[length=2.5mm]}]
    (0_1) edge [blue] (0_2)
    edge [blue] (0_3)
    edge [blue] (0_4)
    (0_3) edge [red] (0_2)
    edge [red] (0_4)
    (0_4) edge [blue] (0_1);
    \end{tikzpicture}%
    \caption{Network~$G$ for Example~\ref{ex:political_1}.
             Blue (resp.\ red) nodes are right-leaning
             (resp.\ left-leaning) users.
             An arrow $(i \to j)$ indicates that user~$j$ observes
             user~$i$'s post.}
    \label{fig:ex_network_G}
\end{figure}
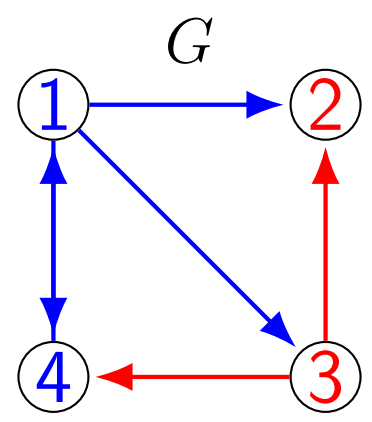

\begin{example}[An electoral campaign network]
\label{ex:political_1}
Let $n = 4$, with users~$1$ and~$4$ being right-leaning
($t_1 = t_4 = \bar{t}$) and users~$2$ and~$3$ being left-leaning
($t_2 = t_3 = \underline{t}$).
The platform configures the feed as follows: user~$1$'s post is visible
to all other users ($r_1 = 3$); user~$3$'s post is visible to
users~$2$ and~$4$ ($r_3 = 2$); user~$4$'s post is visible only to
user~$1$ ($r_4 = 1$); and user~$2$'s post is private, i.e.\ visible
only to user~$2$ themselves ($r_2 = 0$).
Formally, $E = \{(1,2),(1,3),(1,4),(3,2),(3,4),(4,1)\}$.
Figure~\ref{fig:ex_network_G} depicts this network, where blue nodes
denote right-leaning users, and red nodes denote left-leaning users.
\end{example}

The remainder of this section will be devoted to computing the equilibrium actions and analysing their properties.

\subsection{Optimal Action}\label{sec:compute_optimal_action}

To compute the optimal action for each user, we begin by making assumptions and defining additional sets. First, we assume that agents update their beliefs about unknown information in a Bayesian manner, and that this is common knowledge.
Second, let us define the sets of users $j$ whose signals $i$ either observes as  $I_i=\{j\in N: (j,i) \in E \}\cup\{i\}$ or does not as  $\overline{I}_i = N\setminus I_i$. Moreover, let $r_k=|\{j \in N: j\neq k \text{ and }k\in I_j\}|$. Finally, we write
\begin{equation*}
    \mathbb{E}_i[\cdot] = \mathbb{E}\big[\,\cdot \mid t_i,\ (s_k, r_k)_{k\in I_i}\big]
\end{equation*}
for the expectation of user $i$. It conditions on the type of user $i$, on the posts that user $i$ observes, and on the number of users who see each of these posts.

Under these constraints, we can compute the best response from the expected utility maximisation problem as follows:
\begin{equation}\label{eq:best_response_expected}
    a_i=(1-\alpha)\beta_1\mathbb{E}_i[\theta(\mathbf{s})]+ (1-\alpha)\beta_2t_i+\frac{\alpha}{n-1}\sum\limits_{j\neq i} \mathbb{E}_i[a_j]
\end{equation}

From the best response in Equation~\eqref{eq:best_response_expected}, we observe that each user considers the information they hold to update payoff-relevant variables, such as the state of nature and the types of the others. From \citet[Theorem 4]{marschak1972economic}, we know that there exists a unique equilibrium profile of actions. The authors first consider a potential game whose best-response functions correspond to those in the original problem. Then the optimal profile corresponds to the solution of a system of linear equations that depends on the conditional expectations given the information sets. The complete details of this proof are provided in Appendix~\ref{app:equilibrium_existence}. 

The result of \citet{marschak1972economic} gives existence and uniqueness, but the linear
system behind it has one equation for each user and each realisation of her
information, and it does not give the equilibrium in closed form. We therefore
characterise the equilibrium by guessing and verifying. We propose a candidate linear
strategy with the same functional form for every user, in which each user weighs her
own type, the posts she observes and the priors about the posts she does not
observe:\footnote{\citet{hagenbach2010strategic} use the same approach for the part of
the action that depends on the posts. The functional form is a candidate and not a
restriction on strategies. Since the equilibrium is unique, a candidate that is a best
response to itself is the equilibrium.} This functional form is:

\begin{equation}\label{eq:general_action_complete}
    \begin{split}
        a_i =&\quad\; \beta_1\left[\sum\limits_{k\in I_i}\left(\omega_1^k s_k + \omega_2^k\mathbb{E}[s_k]\right)+\sum\limits_{k\in \overline{I}_i} \omega_3^k\mathbb{E}[s_k]\right]\\
        &+ \beta_2\left[\sum\limits_{k\in I_i}\left(\omega_4^k \mathbb{E}[t_k] + \omega_5^k\mathbb{E}[t_k|s_k]\right)+\sum\limits_{k\in \overline{I}_i} \omega_6^k\mathbb{E}[t_k] \right. \\
        &\quad\quad\quad +\left. \vphantom{\sum\limits_{k\in \overline{I}_i} \omega_6^k\mathbb{E}[t_k]} \omega_7^i\mathbb{E}[t_i|s_i]+\omega_8^i\mathbb{E}[t_i] + \omega_9^i t_i\right]
    \end{split}
\end{equation}

The functional form described in Equation~\eqref{eq:general_action_complete} reflects that information plays a dual role in players' optimal actions. 
On the one hand, a direct effect comes from the fact that more signals mean that users have more information about $\theta(\mathbf{s})$ and, therefore, a better choice of their ideal action (consistency motives). On the other hand, since the signals are correlated with the users' types, they transmit information about others' types and, therefore, their preferred actions (coordination motives). 

We then show that this form reproduces itself under best responses. If every other
user plays Equation~\eqref{eq:general_action_complete}, the best response
\eqref{eq:best_response_expected} of user $i$ takes the same form, provided that the
coefficients $\{\omega_l^k\}_{l=1}^{9}$ solve a system of nine linear equations. This
system has the same form for every user, every $n$ and every network, and the network
enters it only through the reach $r_k$ of each author. Its solution gives a profile of
strategies in which each user's strategy is a best response to the others', that is, a
Bayes--Nash equilibrium, and by uniqueness it is the only one. The problem is thereby
reduced from a system with $\mathcal{O}\left(n2^n\right)$ unknowns, which must be solved again for each network, to nine equations solved once in closed form. Theorem~\ref{thm:users_equilibrium} states the result, and Appendix~\ref{app:equilibrium_independent_gamma} gives the proof.

\begin{theorem}[Equilibrium Profile of the Users]\label{thm:users_equilibrium}The Bayes-Nash Equilibrium of the coordination game is given by the profile of actions $\mathbf{a}^*=(a_1^*,a_2^*,\dots,a_n^*)$ such that 

\begin{equation}\label{eq:optimal_action_complete}
\begin{split}
    a_i^* =& \beta_1\Bigg\{\sum\limits_{k \in I_i} \frac{\alpha (n-1-r_k)\mathbb{E}(s_k)+(1-\alpha)(n-1)s_k}{n-1-\alpha r_k} + \sum\limits_{k \in \bar{I}_i}\mathbb{E}(s_k)\Bigg\}\\ 
    +& \beta_2\Bigg\{\frac{\alpha}{n-1+\alpha}\left[\sum\limits_{k \in I_i} \frac{\alpha(n-1-r_k)\mathbb{E}(t_k)+(1-\alpha)(n-1)\mathbb{E}[t_k|s_k]}{(n-1-\alpha r_k)}+\sum\limits_{k \in \bar{I}_i}\mathbb{E}[t_k]
    \right]\\
    +& (1-\alpha)t_i + \frac{-\alpha(1-\alpha)}{n-1+\alpha}\mathbb{E}[t_i|s_i]\Bigg\}
\end{split}
\end{equation}                                                                
\end{theorem}

Theorem~\ref{thm:users_equilibrium} uses two assumptions about what users know. Users do not know the types of others, and they do not infer these types from the structure of their feed. We discuss both assumptions, together with the other assumptions of the model, in Section~\ref{sec:assumptions}.

\subsection{Properties of the Optimal Action}\label{sec:prop_optimal_action}

Let us now analyse the structure of the equilibrium action. First, for the information that users do not observe, they rely on their prior (the expected value). Second, for the information to which they have access, a trade-off emerges between the prior and posterior values, as determined by the number of users with access to that information, $r_k$. Moreover, Equation~\eqref{eq:optimal_action_complete} indicates that these trade-offs between private signals and shared network information share a functional form applicable to both the direct effect (signals regarding the state of nature) and the indirect effect (signals regarding beliefs about others' types).

We start by studying the effect of the number of users with access to a signal. Let us define the function $F$ as in Equation~\eqref{eq:informational_tradeoff}. This function determines how individuals integrate information from different sources when making decisions, accounting for the number of users who receive the same signal. Proposition~\ref{prop:properties_of_F} characterises the monotonicity of the function $F$ with respect to $r_i$. For the direct effect of the information, if the signal is $\overline{s}$, revealing this information to more users will bias their actions upwards. Similarly, for the indirect effect, if the signal revises upward beliefs about the type, it will also bias users' actions in the same direction. Moreover, we observe the opposite effect when the signal is $\underbar{s}$. 

\begin{proposition}[Properties of $F$]\label{prop:properties_of_F} Let $F(x,y,z)$ be defined as in Equation~\eqref{eq:informational_tradeoff}.The function $F$ is increasing in $z$ if and only if $x>y$. 
\begin{equation}\label{eq:informational_tradeoff}
F(x,y,z)=\frac{\alpha (n-1-z)y+(1-\alpha)(n-1)x}{n-1-\alpha z}
\end{equation}
\end{proposition}

On the other hand, we can study the effects of having more tuples $(s_i,r_i)$ which is equivalent to stating that there is more information available in the network. We quantify the direction of this change by moving users between the sets $I_i$ and $\overline{I}_i$. Similarly to the function $F$, the action of a user $i$ will increase when the information of user $j$ is revealed to them, provided the prior $y$ is lower than the posterior $x$. We provide the formal statement and proof of this result in the appendix. 
These properties characterise the role of information in shaping individual actions, and they help us study how a platform can influence collective actions. 

\begin{example}[An electoral campaign network] \label{ex:political_2}
Following the Example~\ref{ex:political_1}, let us apply Theorem~\ref{thm:users_equilibrium} to characterise the equilibrium action of each user, and use the result to highlight the two key mechanisms identified in Section~2. The information sets induced by $G$ are:
$I_1 = \{1,4\}$,\; $I_2 = \{2,1,3\}$,\; $I_3 = \{3,1\}$,\;
$I_4 = \{4,1,3\}$,
with complements $\bar{I}_i = N \setminus I_i$ as defined in
Section~2.1. Let us start by shaping the weight assigned to observed signals.

A central feature of Equation~\eqref{eq:optimal_action_complete} is that the
weight assigned to an observed signal~$s_k$ is not uniform: it depends
on the reach count~$r_k$ through the function~$F$ in
Equation~\eqref{eq:informational_tradeoff}.
Evaluating~$F$ at the four reach values present in this network
illustrates the full range of this effect.

User~$1$'s post, seen by all three other users ($r_1 = 3 = n-1$),
enters every observer's equilibrium action at \emph{full weight}:
\begin{equation*}
    F\!\left(s_1,\,E(s_1),\,3\right)
    = \frac{\alpha(3-3)\,E(s_1)+3(1-\alpha)\,s_1}{3-3\alpha}
    = s_1.
\end{equation*}
When all users respond to the same post, they can fully engage without concern. Any confusion from guessing previous responses will affect everyone equally.

At the opposite extreme, user~$2$'s post is seen by no one else
($r_2 = 0$), so it enters user~$2$'s own action only as a convex
combination of the actual signal and the prior:
\begin{equation*}
    F\!\left(s_2,\,E(s_2),\,0\right)
    = \frac{3\alpha\,E(s_2)+3(1-\alpha)\,s_2}{3}
    = (1-\alpha)\,s_2 + \alpha\,E(s_2).
\end{equation*}
Since no other user can coordinate on $s_2$, user~$2$ must put weight $\alpha$ on the prior in order to avoid a coordination loss with peers who are not observing their post. As a result, user $2$ reduces the impact of their private political signal and moves closer to the common understanding.

The intermediate reach values $r_3 = 2$ and $r_4 = 1$ yield weights
that lie strictly between these extremes:
\begin{align*}
    F\!\left(s_3,\,E(s_3),\,2\right)
    &= \frac{\alpha\,E(s_3)+3(1-\alpha)\,s_3}{3-2\alpha},\\
    F\!\left(s_4,\,E(s_4),\,1\right)
    &= \frac{2\alpha\,E(s_4)+3(1-\alpha)\,s_4}{3-\alpha}.
\end{align*}
Wider reach shifts the weight monotonically from the prior towards the actual signal, as in Proposition~\ref{prop:properties_of_F}.
In the electoral context, this means that a post shared with a large audience carries disproportionate influence on each observer's action, independent of whether that observer is politically aligned with the
sender.

Let us illustrate that signals play a dual informational role. Consider user~$2$, a left-leaning observer whose information set
$I_2 = \{2,1,3\}$ is the richest in this network.
Applying Equation~\eqref{eq:optimal_action_complete}, their equilibrium action is:
\begin{align*}
    a_2^* \;=\;&
    \beta_1 \!\left[
        (1-\alpha)\,s_2 + \alpha\,E(s_2)
        \;+\; s_1
        \;+\; \frac{\alpha\,E(s_3)+3(1-\alpha)\,s_3}{3-2\alpha}
        \;+\; E(s_4)
    \right] \\[6pt]
    +\;& \beta_2 \;\left[
        \frac{\alpha}{3+\alpha}
        \; \Big(
            (1-\alpha)\,E[t_2|s_2]+\alpha\,E(t_2)
            \;+\; E[t_1|s_1]
         \right.\\[2pt]
        &\quad\quad\quad\quad\quad   \;+\; \frac{\alpha\,E(t_3)+3(1-\alpha)\,E[t_3|s_3]}{3-2\alpha}
            \;+\; E(t_4)
        \Big)\\
    &\;\quad \left.
        +\; (1-\alpha)\,\underline{t}
        \;-\; \frac{\alpha(1-\alpha)}{3+\alpha}\,E[t_2|s_2]
    \right].
\end{align*}

This expression exhibits the dual informational role of every observed signal.
Take user~$1$'s post~$s_1$: it enters the \emph{state component} (the $\beta_1$ term) as~$s_1$, providing a direct update on aggregate political momentum $\theta(\mathbf{s})$. Simultaneously, it enters the \emph{type-coordination component} (the $\beta_2$ term) as $E[t_1 \mid s_1]$, user~$2$'s posterior belief about user~$1$'s political predisposition. Because $\lambda > \tfrac{1}{2}$, observing a right-leaning post $s_1 = \bar{s}$ raises $E[t_1 \mid \bar{s}]$ above the prior
$E(t_1)$, signalling that user~$1$ is more likely to be a right-leaning activist.
This updated belief raises user~$2$'s participation intensity, not merely because aggregate momentum is higher, but because their visible peer is now more plausibly a committed right-leaning partisan.
The two effects reinforce each other: a right-leaning post
simultaneously raises the state estimate and the inferred type of the sender, generating a doubly amplified upward adjustment in~$a_2^*$. By contrast, a left-leaning post $s_1 = \underline{s}$ produces a doubly amplified downward adjustment.

The same dual role appears for user~$3$'s post~$s_3$, which
user~$2$ observes at intermediate reach $r_3 = 2$. Here, the actual signal~$s_3$ and the type posterior $E[t_3 \mid s_3]$ are each dampened towards their respective priors by the weight $(3-2\alpha)$ in the denominator, the platform can amplify or suppress this effect by adjusting $r_3$, i.e.\ by deciding how widely to distribute left-leaning user~$3$'s post.

It is precisely this dual mechanism, state informativeness and type informativeness, both scaling with reach, that gives the platform leverage over collective political behaviour through the design of the
feed architecture. By controlling which posts are broadcast and to whom, the platform simultaneously shapes every observer's estimate of aggregate momentum and their beliefs about the political composition of their peer group, steering aggregate participation towards its target~$\mathfrak{a}$ without ever directly revealing user types.
\end{example}

\subsection{Discussion of the Assumptions}\label{sec:assumptions}

In this paper, we consider a scenario in which users do not know others' types. This assumption might appear unrealistic, as on social media platforms, we are often connected with people we know in real life and, therefore, we might be aware of their types. 
Nevertheless, given the size of social media platforms, the set of user types we know is small compared to the total number of users on the platform. For example, Facebook has shown sustained growth in Daily Active Users (DAUs), jumping from 483 million in 2011 to 2.06 billion in 2023 \citep{meta2012,meta2023}. 
Moreover, the information users see on these platforms is highly diverse and comes from both known and unknown users. 
\citet{hagenbach2010strategic} consider a similar model in which preferences differ through biases that are common knowledge, so that these biases play the role of types known to all users. We conjecture that a model incorporating a network of known types would yield a combination of the no-information optimal action and the full-information one. Such a model, in which users also know the types of some of their contacts, is a natural extension, which we leave for future research.

Users do not observe the network either. Each user sees the posts in her feed and the number of users who see each of them. She does not infer the types of others from the structure of her feed. In Section~\ref{sec:manipulation}, however, the platform chooses the network knowing the types, so a user who knew how the platform maps type profiles into networks could learn about types from her feed. We assume that users do not make this inference. Formally, given her own type, the belief of each user about the network does not depend on the types of the other users. Remark~\ref{rem:network_knowledge} in Appendix~\ref{app:equilibrium_independent_gamma} shows that Theorem~\ref{thm:users_equilibrium} holds for any such belief. We can interpret this assumption through analogy-based expectations \citep{jehiel2005analogy, jehiel2008revisiting}. With analogy-based expectations, a player forecasts the average behaviour of an opponent over groups of states, called analogy classes. \citet{jehiel2008revisiting} interpret these forecasts as the result of learning from aggregate feedback about past play. In our setting, a user places all the type profiles of the other users in a single analogy class of the platform, so her forecast of the network does not depend on their types. Theorem~\ref{thm:users_equilibrium} does not require this forecast to be correct on average. This description fits a large platform, whose users observe the feeds they receive and cannot observe the algorithm that produces them. Our results use this assumption. If users knew how the platform maps type profiles into networks, the equilibrium of Theorem~\ref{thm:users_equilibrium} would no longer apply.

In our model, types and posts are binary, types are independent across users, and posts are independent across users given the types. Moreover, the state is the sum of the posts. The proof of Theorem~\ref{thm:users_equilibrium} does not use the binary structure, and it holds for any finite sets of types and posts. It uses the common prior, the independence and the additive state. A user expects an unobserved post at its unconditional mean. The part of the action that depends on the posts is the solution of \citet{hagenbach2010strategic}, who assume the same additive state. Theorem~\ref{thm:aggregate_by_degree} also uses the additive state, since each post enters collective action separately. With binary posts, the expected type given a post is affine in the post. We use this property for the closed forms of welfare and polarisation in Appendix~\ref{app:policy}, and in the proofs of Theorem~\ref{thm:all_high_low_equivalence} and Proposition~\ref{prop:extreme_equivalence}. We do not study how the results change without these assumptions.

Finally, users do not choose their posts strategically. Each post is drawn from a distribution that depends on the type of its author, in line with the evidence cited in Section~\ref{sec:optimal_action}. If users chose their posts, they could distort them to influence the actions of others. A post would then be weaker evidence about the type of its author, and the model would be closer to models of strategic communication in networks \citep{hagenbach2010strategic, galeotti2013strategic}. We leave this extension for future research.

\section{Influencing Collective Actions}\label{sec:manipulation}

In the previous section, we studied how users who obtain information from an exogenous information network play a coordination game. In this section, we take a step back to analyse the processes by which platforms determine the design of such information networks.

Let us consider a platform that wants to design an information network with the objective of influencing the collective actions of its users  $(\sum_{i\in N} a_i^*)$ towards a target value~$\mathfrak{a}$. 
Formally, the platform wants to minimise the distance between these two objects, as shown in Equation~\eqref{eq:platform_objective}. 
We assume that the platform chooses the network $G=(N,E)$ knowing the types of the users but prior to the realisation of the signals. This is a reasonable assumption, as social media platforms can gather various types of information from users but do not know ex ante the content of their posts. As in Section~\ref{sec:optimal_action}, an edge $(i,j)\in E$ means that user $j$ has access to the tuple $(s_i,r_i)$, where $s_i$ is the post made by user $i$ and $r_i$ the number of users other than $i$ who have access to it. 

\begin{equation}\label{eq:platform_objective}
    \min\limits_{G} \left(\mathfrak{a}-\sum\limits_{i\in N} a_i^*\right)^2
\end{equation}

Thanks to the results from Section~\ref{sec:compute_optimal_action}, the platform knows the users' actions for a given network $G$. Moreover, the platform could utilise the monotonicity properties of $a_i^*$, established in Section~\ref{sec:prop_optimal_action}, to influence the collective action of its users by revealing more or less information to approach its target~$\mathfrak{a}$. 
Nevertheless, since the platform does not know the signal realisations, we need to study the effect of information on the action from an ex ante point of view. That is, we need to study the properties of $\mathbb{E}_P[F]$, where $\mathbb{E}_P[\cdot]=\mathbb{E}[\cdot\mid\mathbf{t}]$ conditions on the realised type profile and integrates over the posts. 
We establish this result in Proposition~\ref{prop:properties_of_expected_F} and note that the ex ante results for $F$ follow a similar pattern to their ex post form.  
The interpretation of these results is as follows: if the platform wants to increase the action of a user, it will reveal more information about high-type users because they have a higher probability of obtaining a high signal. Moreover, this increase is monotonic in the number of users who receive it. 

\begin{proposition}[Properties of {$\mathbb{E}_P[F]$}]\label{prop:properties_of_expected_F}
Let $F$ be defined as in Equation~\eqref{eq:informational_tradeoff}. 
The expectation $\mathbb{E}_P[F]$ is increasing in $z$ if and only if $\mathbb{E}_P[x]>\mathbb{E}_P[y]$.
\end{proposition}

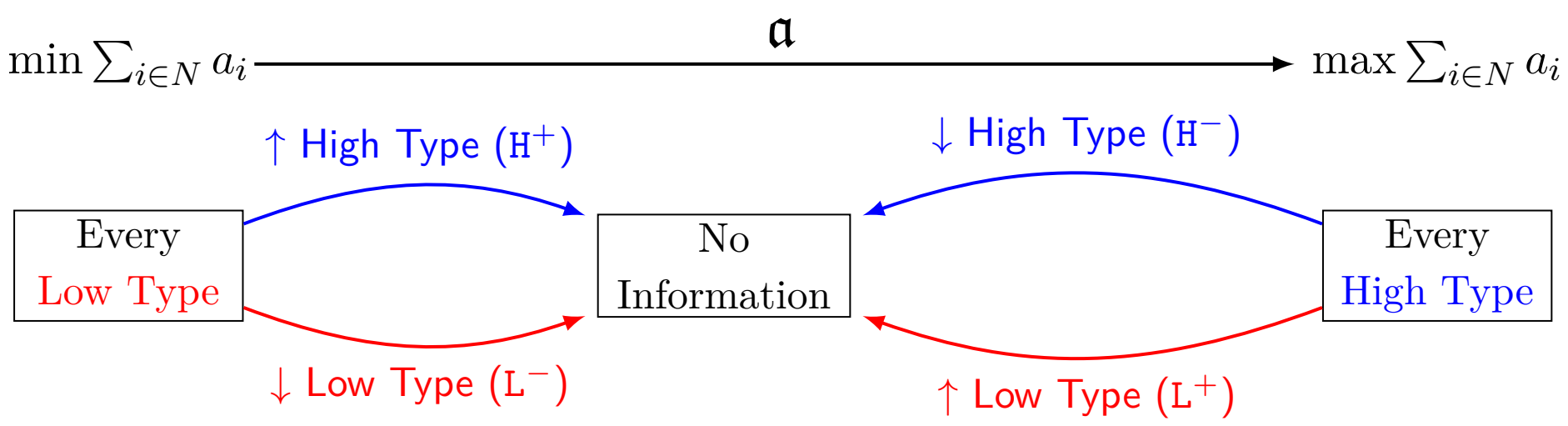
\begin{figure}[H] 
    \centering
\resizebox{0.85\linewidth}{!}{
\begin{tikzpicture}[->,shorten >=1pt]

\node[draw, rectangle, font=\footnotesize,text width= 1.8cm, text centered,inner sep=2pt] (max) {Every \textcolor{blue}{High Type}};
\node[rectangle, font=\small, text centered,inner sep=1pt] (max_val) [above = 1cm of max]{$\max \sum_{i\in N} a_i$};
\node[draw, rectangle, font=\footnotesize, text width= 2cm, text centered,inner sep=2pt] (no) [left = 4cm of  max] {No\\ Information};
\node[draw,rectangle, font=\footnotesize,text width= 1.8cm, text centered, inner sep=2pt] (min) [left = 3cm of no] {Every \textcolor{red}{Low Type}};
\node[rectangle, font=\small, text centered,inner sep=1pt] (min_val) [above = 1cm of min]{$\min \sum_{i\in N} a_i$};

\path[shorten >=3pt,thick, every node/.style={font=\sffamily\footnotesize}]
(max) edge [bend right=20, -latex, blue] node[above,rectangle] {$\downarrow$ High Type ($\texttt{H}^-$)} (no)
edge [bend left=20, -latex,red] node[below,rectangle] {$\uparrow$ Low Type ($\texttt{L}^+$)} (no)
(min) edge [bend left=20, -latex,blue] node[above,rectangle] {$\uparrow$ High Type ($\texttt{H}^+$)} (no)
edge [bend right=20, -latex,red] node[below,rectangle] {$\downarrow$ Low Type ($\texttt{L}^-$)} (no)
(min_val) edge [-latex, thick] node[above,rectangle, font=\large] {$\mathfrak{a}$} (max_val);
\end{tikzpicture}
}
\caption{Effect of Information on Collective Action.}
\label{fig:intuition_alg}
\end{figure}

These ex ante properties provide a valuable framework for the platform to design a network that effectively aligns with its objective $\mathfrak{a}$. The designer knows that if she reveals more information about high-type users, the collective action will increase. On the contrary, if the designer reveals more information about low-type users, the collective action will decrease. Figure~\ref{fig:intuition_alg} depicts the effect of information on collective action. 
An immediate result of these properties is that, if the platform aims to maximise collective action, the optimal network, which we denote $G_{\max}$, connects every high-type user to everyone. Analogously, if the platform has the objective of minimising collective action, then the optimal network, which we denote $G_{\min}$, connects every low-type user to everyone. Moreover, these networks are uniquely defined, as highlighted in Corollary~\ref{cor:uniques_max_min}.
These results outline the characteristics of informational networks that facilitate either the highest or lowest levels of collective action among participants. 

\begin{corollary}\label{cor:uniques_max_min}
    Suppose the set $N$ of users choose their optimal actions in the coordination game according to Equation~\eqref{eq:optimal_action_complete} for a given informational network $G$. Then, 
\begin{align*}
G_{\max} &= (N,E_{\max}) = \arg\max_{G}\sum_{i\in N}\mathbb{E}_P[a_i^*], & E_{\max} &:= \{(i,j): i\neq j,\ t_i=\overline{t}\},\\
G_{\min} &= (N,E_{\min}) = \arg\min_{G}\sum_{i\in N}\mathbb{E}_P[a_i^*], & E_{\min} &:= \{(i,j): i\neq j,\ t_i=\underbar{t}\}.
\end{align*}
\end{corollary}

Returning to the electoral campaign setting of Example~\ref{ex:political_1}, Corollary~\ref{cor:uniques_max_min} identifies two uniquely defined feed architectures that bound the range of collective political participation the platform can induce.

When the platform seeks to maximise aggregate political participation, Corollary~\ref{cor:uniques_max_min} prescribes broadcasting every right-leaning user's post to the entire platform.
In the four-user example, this means that both right-leaning users~$1$ and~$4$ each transmit their post to all three other users, so $r_1 = r_4 = n - 1 = 3$, while the posts of left-leaning users~$2$ and~$3$ remain private ($r_2 = r_3 = 0$).
Formally,
\begin{equation*}
    G_{\max} = \bigl(N,\; E_{\max} = \{(1,2),(1,3),(1,4),(4,1),(4,2),(4,3)\}\bigr).
\end{equation*}
Under $G_{\max}$, every user's feed is saturated with right-leaning content. Because $r_1 = r_4 = 3$, Proposition~\ref{prop:properties_of_F} implies that every observer responds to the right-leaning posts at full weight, with no discounting towards the prior. At the same time, no left-leaning post is distributed, so every user forms beliefs about left-leaning types relying entirely on their prior $E(t_k)$ for $k \in \{2,3\}$. 

The net effect is a systematic upward pull on every user's participation intensity, independently of whether they are themselves right- or left-leaning: even left-leaning users~$2$ and~$3$ observe two right-leaning posts at full weight and update both their estimate of political momentum and their posterior over the type composition of the peer group upwards. This corresponds to the documented phenomenon of algorithmic amplification of politically aligned content, in which a platform's recommendation system systematically privileges posts from one ideological group, shifting aggregate behaviour towards a partisan target \cite{gauthier2025twitter}.

\begin{figure}[H]
    \centering
    \begin{minipage}[t]{0.48\textwidth}
        \centering
        \begin{tikzpicture}[
            auto,
            node distance=1.8cm,
            main node/.style={circle, inner sep=1pt,
                              font=\sffamily\large, draw, thin,
                              minimum size=0.5cm}
        ]
        \node[main node, text=blue]  (1) {$1$};
        \node[main node, text=red]   (2) [right of=1] {$2$};
        \node[main node, text=red]   (3) [below of=2] {$3$};
        \node[main node, text=blue]  (4) [below of=1] {$4$};

        \path[shorten >=1pt, thick, -{Latex[length=2.5mm]}]
            (1) edge [blue] (2)
            (1) edge [blue] (3)
            (1) edge [blue] (4)
            (4) edge [blue] (1)
            (4) edge [blue] (2)
            (4) edge [blue] (3);
        \end{tikzpicture}
        \subcaption{$G_{\max}$: right-leaning users broadcast to all.}
        \label{fig:gmax}
    \end{minipage}
    \hfill
    \begin{minipage}[t]{0.48\textwidth}
        \centering
        \begin{tikzpicture}[
            auto,
            node distance=1.8cm,
            main node/.style={circle, inner sep=1pt,
                              font=\sffamily\large, draw, thin,
                              minimum size=0.5cm}
        ]
        \node[main node, text=blue]  (1) {$1$};
        \node[main node, text=red]   (2) [right of=1] {$2$};
        \node[main node, text=red]   (3) [below of=2] {$3$};
        \node[main node, text=blue]  (4) [below of=1] {$4$};

        \path[shorten >=1pt, thick, -{Latex[length=2.5mm]}]
            (2) edge [red] (1)
            (2) edge [red] (3)
            (2) edge [red] (4)
            (3) edge [red] (1)
            (3) edge [red] (2)
            (3) edge [red] (4);
        \end{tikzpicture}
        \subcaption{$G_{\min}$: left-leaning users broadcast to all.}
        \label{fig:gmin}
    \end{minipage}
    \caption{The two extreme networks for $n = 4$, with users~$1,4$ right-leaning
             (blue) and users~$2,3$ left-leaning (red).
             An arrow $(i \to j)$ indicates that user~$j$ observes user~$i$'s post.
             $G_{\max}$ maximises and $G_{\min}$ minimises expected aggregate
             political participation.}
    \label{fig:gmax_gmin}
\end{figure}
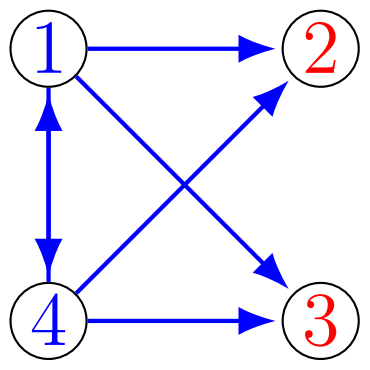
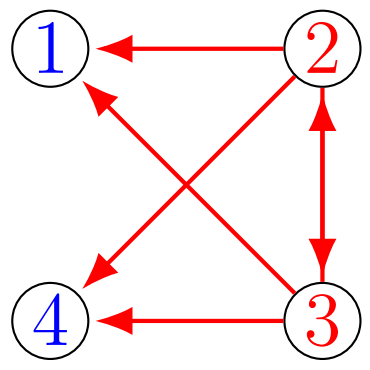

Symmetrically, when the platform seeks to \emph{minimise} aggregate right-leaning participation, the optimal network broadcasts every left-leaning user's post to everyone, while right-leaning posts remain private ($r_2 = r_3 = 3$, $r_1 = r_4 = 0$):
\begin{equation*}
    G_{\min} = \bigl(N,\; E_{\min} = \{(2,1),(2,3),(2,4),(3,1),(3,2),(3,4)\}\bigr).
\end{equation*}
Every user now observes two left-leaning posts at full weight, generating a symmetric downward pull on participation: right-leaning users~$1$ and~$4$ observe left-leaning signals that lower both their state estimate and their posterior that peers are right-leaning partisans.
Figure~\ref{fig:gmax_gmin} depicts both extreme networks.

Neither extreme need correspond to the platform's actual target $\mathfrak{a}$. The network~$G$ introduced in Example~\ref{ex:political_1} (Figure~\ref{fig:ex_network_G}) illustrates an intermediate design: right-leaning user~$1$ broadcasts to all ($r_1 = 3$, as in $G_{\max}$), but right-leaning user~$4$ broadcasts only to user~$1$ ($r_4 = 1$, strictly less than in $G_{\max}$), while left-leaning user~$3$ broadcasts to users~$2$ and~$4$ ($r_3 = 2 > 0$, unlike $G_{\max}$ where no left-leaning post is distributed). Consequently, $G$ lies strictly between the two extremes in the platform's ordering, inducing an expected aggregate participation level in the interior of the feasible interval $\bigl[\mathbb{E}_P[\sum_{i} a_i^*(G_{\min})],\, \mathbb{E}_P[\sum_{i} a_i^*(G_{\max})]\bigr]$.

This illustrates the core message of Corollary~\ref{cor:uniques_max_min}: platforms can approach any target collective action within this interval, and the intermediate networks that do so need not be symmetric, homophilic, or otherwise structurally regular.

Corollary~\ref{cor:uniques_max_min} also answers the question of existence. The platform chooses among the $2^{n(n-1)}$ directed networks on $n$ users without self-loops, for instance $4{,}096$ networks when $n=4$. This set is finite, so for every target $\mathfrak{a}$ there exists at least one network that minimises the distance in Equation~\eqref{eq:platform_objective}. If $\mathfrak{a}$ lies outside the interval, the minimiser is $G_{\max}$ or $G_{\min}$. If $\mathfrak{a}$ lies inside, the minimiser induces the collective action closest to $\mathfrak{a}$, which in general differs from $\mathfrak{a}$, since only finitely many values of collective action can be induced.

The remaining questions concern the structure of the platform's choice. Thanks to our results in Section~\ref{sec:optimal_action}, we can compare networks by comparing the collective actions they induce, and this comparison orders the set of all networks. We say that two networks are {\em platform\nobreakdash-equivalent} if they induce the same ex ante collective action. The platform could find its optimal network by evaluating all $2^{n(n-1)}$ networks, but this is not necessary. In Section~\ref{sec:transformations} we show that collective action depends on the network only through the number of users who see each post, so the platform only needs to choose these numbers. We then study when different architectures are equivalent from the platform's perspective, and in Section~\ref{sec:policy} how they differ in welfare and social mixing.

\subsection{Network Transformation Operators}\label{sec:transformations}

Let us consider a platform with an objective $\mathfrak{a}$ and an informational network $G$. 
If the collective action induced by $G$ does not coincide with the objective $\mathfrak{a}$, the platform may transform the network by adding or removing edges. We begin by focusing on transformations that add or delete a single edge from a high-type or low-type user. We refer to these as atomic transformations, as they constitute the smallest possible change to the network; more complex transformations are compositions of them. 
For a network $G$, each atomic transformation defines a (potentially empty) set, as shown in Definition~\ref{def:atomic_operator}. For example, $\texttt{H}^+(G)$ is the set of networks $G^\prime$ where a new link is created from a high-type user to a generic user. As a result, one user gains access to the signal of a high-type user.
It is important to note that these sets may be empty if applying the transformation to the network is not feasible. 
Example~\ref{ex:transformation_network} depicts, for each atomic transformation, a network generated from $G$. On the other hand, Example~\ref{ex:equivalence_operator} shows all of the networks generated by adding one link from a low-type user to a generic user of the network $G$.  

\begin{definition}[Atomic Transformations]\label{def:atomic_operator}
Let $G=(N,E)$ be a network. For the operators that add or remove edges from a high type or a low-type users, we define the following sets:
\begin{enumerate}
    \item $\texttt{H}^+(G)=\left\{G^\prime=(N,E^\prime): E^\prime=E\cup\{(i,j)\},\ i\neq j,\ E\cap\{(i,j)\}=\emptyset \text{ and } t_i=\overline{t}\right\}$.
    \item $\texttt{H}^-(G)=\left\{G^\prime=(N,E^\prime): E^\prime=E\setminus\{(i,j)\},\ E\cap\{(i,j)\}\neq\emptyset \text{ and } t_i=\overline{t}\right\}$. 
    \item $\texttt{L}^+(G)=\left\{G^\prime=(N,E^\prime): E^\prime=E\cup\{(i,j)\},\ i\neq j,\ E\cap\{(i,j)\}=\emptyset \text{ and } t_i=\underbar{t}\right\}$. 
    \item $\texttt{L}^-(G)=\left\{G^\prime=(N,E^\prime): E^\prime=E\setminus\{(i,j)\},\ E\cap\{(i,j)\}\neq\emptyset \text{ and } t_i=\underbar{t}\right\}$.
\end{enumerate}
\end{definition}

\begin{figure}[H]
    \centering
    \resizebox{\imagewidth}{!}{
    \begin{tikzpicture}[auto,node distance=2cm,main node/.style={circle,inner sep=0.5pt, font=\sffamily\normalsize, draw, thin}]

    \node[main node] (0_1) {\textcolor{blue}{1}};
    \node[main node] (0_2) [right of=0_1] {\textcolor{red}{2}};
    \node[main node] (0_3) [below of=0_2] {\textcolor{red}{3}};
    \node[main node] (0_4)  [below of=0_1] {\textcolor{blue}{4}};
    
    \path[shorten >=1pt, thick,-{Latex[length=2.5mm]}]
    (0_1) edge [blue] (0_2)
    edge [blue] (0_3)
    edge [blue] (0_4)
    (0_3) edge [red] (0_2)
    edge [red] (0_4)
    (0_4) edge [blue] (0_1);

    \node[main node] (1_2) [below = 1.2 cm of 0_4] {\textcolor{red}{2}};
    \node[main node] (1_1) [left of =1_2]{\textcolor{blue}{1}};
    \node[main node] (1_3) [below of=1_2] {\textcolor{red}{3}};
    \node[main node] (1_4)  [below of=1_1] {\textcolor{blue}{4}};
    
    \path[shorten >=1pt, thick,-{Latex[length=2.5mm]}]
    (1_1) edge [blue] (1_2)
    edge [blue] (1_4)
    (1_3) edge [red] (1_2)
    edge [red] (1_4)
    (1_4) edge [blue] (1_1);

    \node[main node] (2) [left of=1_1] {\textcolor{red}{2}};
    \node[main node] (1) [left of =2]{\textcolor{blue}{1}};
    \node[main node] (3) [below of=2] {\textcolor{red}{3}};
    \node[main node] (4)  [below of=1] {\textcolor{blue}{4}};
    
    \path[shorten >=1pt, thick, -{Latex[length=2.5mm]}]
    (1) edge [blue] (2)
    edge [blue] (3)
    edge [blue] (4)
    (3) edge [red] (2)
    edge [red] (4)
    (4) edge [blue] (1)
    edge [blue] (2);

    \node[main node] (2_1) [right of=1_2]{\textcolor{blue}{1}};
    \node[main node] (2_2) [right of=2_1] {\textcolor{red}{2}};
    \node[main node] (2_3) [below of=2_2] {\textcolor{red}{3}};
    \node[main node] (2_4)  [below of=2_1] {\textcolor{blue}{4}};
    
    \path[shorten >=1pt, thick, -{Latex[length=2.5mm]}]
    (2_1) edge [blue] (2_2)
    edge [blue] (2_3)
    edge [blue] (2_4)
    (2_2) edge [red] (2_4)
    (2_3) edge [red] (2_2)
    edge [red] (2_4)
    (2_4) edge [blue] (2_1);

    \node[main node] (3_1) [right of=2_2]{\textcolor{blue}{1}};
    \node[main node] (3_2) [right of=3_1] {\textcolor{red}{2}};
    \node[main node] (3_3) [below of=3_2] {\textcolor{red}{3}};
    \node[main node] (3_4)  [below of=3_1] {\textcolor{blue}{4}};
    
    \path[shorten >=1pt, thick, -{Latex[length=2.5mm]}]
    (3_1) edge [blue] (3_2)
    edge [blue] (3_3)
    edge [blue] (3_4)
    (3_3) edge [red] (3_2)
    (3_4) edge [blue] (3_1);

    \node[label={[label distance=-.01cm]90:{{$G$}}}] (G) at ($(0_1)!0.5!(0_2)$) {};
    \node (S) at ($(0_3)!0.5!(0_4)$) {};
    \node (S_1) at ($(1)!0.5!(2)$) {};
    \node (S_2) at ($(1_1)!0.5!(1_2)$) {};
    \node (S_3) at ($(2_1)!0.5!(2_2)$) {};
    \node (S_4) at ($(3_1)!0.5!(3_2)$) {};

    \path[shorten >=1pt, font=\sffamily\large]
    (S) edge [-latex, out=225, in=90, looseness=0.5] node[above,ellipse,anchor=center,fill=white,inner sep=0.8pt] {$\texttt{H}^+(G)$}(S_1)
        edge [-latex, out=270, in=90, looseness=0.5] node[above,ellipse,anchor=center,fill=white,inner sep=0.5pt] {$\texttt{H}^-(G)$} (S_2)
        edge [-latex, out=270, in=90, looseness=0.5] node[above,ellipse,anchor=center,fill=white,inner sep=0.5pt] {$\texttt{L}^+(G)$} (S_3)
        edge [-latex, out=315, in=90, looseness=0.5] node[above,ellipse,anchor=center,fill=white,inner sep=0.5pt] {$\texttt{L}^-(G)$} (S_4);
    \end{tikzpicture}}
    \caption{Network $G$ and some networks generated from operators in Definition~\ref{def:atomic_operator}.}
    \label{fig:ex_atomic_G}
\end{figure}
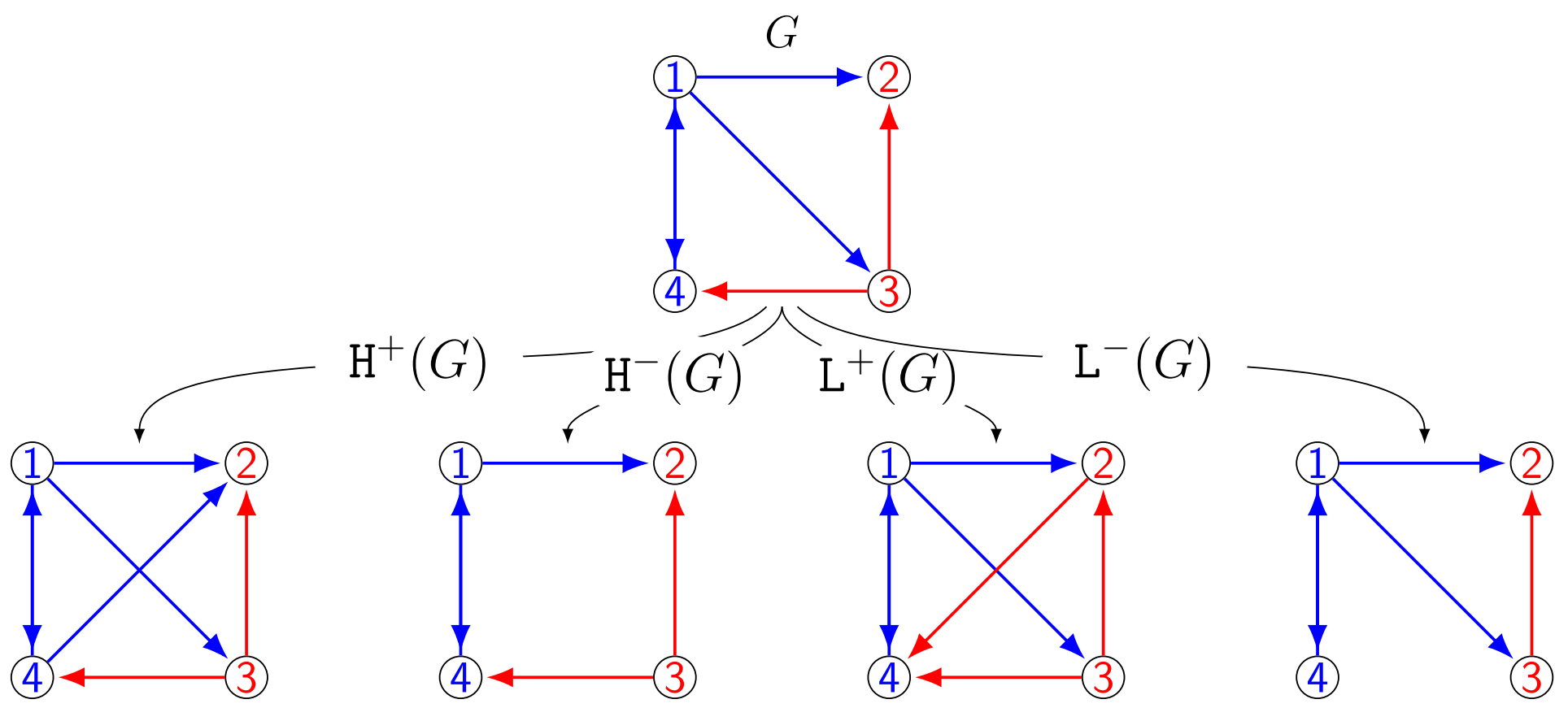

\begin{example}\label{ex:transformation_network} 
Let $n=4$. Figure~\ref{fig:ex_atomic_G} depicts a network $G$ and examples of networks generated by adding or removing edges according to the operators in Definition~\ref{def:atomic_operator}.
\end{example}

This raises an important question: do these networks possess any advantages from the platform's perspective? The answer is affirmative; the transformations discussed here enable us to construct networks that are platform\nobreakdash-equivalent. To understand this better, let us define the user-based transformation $K_i$ (where $K$ represents $\texttt{H}^+, \texttt{H}^-,\texttt{L}^+,$ or $\texttt{L}^-$) as in Definition~\ref{def:atomic_operator} but fixing the user $i$ to which we are adding or removing edges.\footnote{If $t_i=\overline{t}$ then the sets  $\texttt{L}_i^+=\texttt{L}_i^-=\emptyset$. Analogously, if $t_i=\underbar{t}$ then the sets  $\texttt{H}_i^+=\texttt{H}_i^-=\emptyset$.} Moreover, the user-based transformation induces a partition of the atomic transformation given by $K(G)=\bigcup_{i\in N} K_i(G)$. 
All the networks in a set $K_i(G)$ give the post of user $i$ the same number of
readers and leave the readers of every other post unchanged. We show below that they
therefore induce the same collective action for every realisation of the posts, and
hence are platform-equivalent.

The intuition is that the platform cares only about the aggregate, and a post
contributes to the aggregate through how many users read it, not through who they
are. By Theorem~\ref{thm:users_equilibrium}, every reader of the post of user $k$
gives it the same weight, which depends only on $r_k$ and not on the identity or the
type of the reader. A reader's own type enters her action only through her own terms,
which do not depend on the posts she observes. This is not an additional assumption.
It follows from the independence of types, which makes each user's beliefs about the
others independent of her own type. Hence moving a link from one recipient to
another, with its sender fixed, removes the contribution of the post from one action
and adds the same contribution to another. In Example~\ref{ex:equivalence_operator}, we provide a network $G$ and the sets of networks $L_2^+(G)$ and $L_3^+(G)$ included in $L^+(G)$, which are obtained by adding a link from a low-type user. Theorem~\ref{thm:aggregate_by_degree} makes this intuition precise, and Corollary~\ref{cor:equivalence_classes} applies it to the user-based transformations. For any user $i$, any two distinct networks $\widehat{G}$ and $\widetilde{G}$ in $K_i(G)$ induce the same collective action for every realisation of the posts, $\sum_{i\in N} a_i^*(\widehat{G}) = \sum_{i\in N} a_i^*(\widetilde{G})$. Hence they also induce the same expected collective action, and they are platform-equivalent.

\begin{figure}[H]
    \centering
    \resizebox{\imagewidth}{!}{
    \begin{tikzpicture}[auto,node distance=2cm,main node/.style={circle,inner sep=0.5pt, font=\sffamily\large, draw, thin}]

    \node[main node] (1_2) {\textcolor{red}{2}};
    \node[main node] (1_1) [left of =1_2]{\textcolor{blue}{1}};
    \node[main node] (1_3) [below of=1_2] {\textcolor{red}{3}};
    \node[main node] (1_4)  [below of=1_1] {\textcolor{blue}{4}};
    
    \path[shorten >=1pt, thick,-{Latex[length=2.5mm]}]
    (1_1) edge [blue] (1_2)
    edge [blue] (1_3)
    edge [blue] (1_4)
    (1_2) edge [red] (1_1)
    (1_3) edge [red] (1_2)
    edge [red] (1_4)
    (1_4) edge [blue] (1_1);

    \node[main node] (2) [left of=1_1] {\textcolor{red}{2}};
    \node[main node] (1) [left of =2]{\textcolor{blue}{1}};
    \node[main node] (3) [below of=2] {\textcolor{red}{3}};
    \node[main node] (4)  [below of=1] {\textcolor{blue}{4}};
    
    \path[shorten >=1pt, thick, -{Latex[length=2.5mm]}]
    (1) edge [blue] (2)
    edge [blue] (3)
    edge [blue] (4)
    (3) edge [red] (1)
    edge [red] (2)
    edge [red] (4)
    (4) edge [blue] (1);

    \node[main node] (2_1) [right of=1_2]{\textcolor{blue}{1}};
    \node[main node] (2_2) [right of=2_1] {\textcolor{red}{2}};
    \node[main node] (2_3) [below of=2_2] {\textcolor{red}{3}};
    \node[main node] (2_4)  [below of=2_1] {\textcolor{blue}{4}};
    
    \path[shorten >=1pt, thick, -{Latex[length=2.5mm]}]
    (2_1) edge [blue] (2_2)
    edge [blue] (2_3)
    edge [blue] (2_4)
    (2_2) edge [red] (2_4)
    (2_3) edge [red] (2_2)
    edge [red] (2_4)
    (2_4) edge [blue] (2_1);

    \node[main node] (3_1) [right of=2_2]{\textcolor{blue}{1}};
    \node[main node] (3_2) [right of=3_1] {\textcolor{red}{2}};
    \node[main node] (3_3) [below of=3_2] {\textcolor{red}{3}};
    \node[main node] (3_4)  [below of=3_1] {\textcolor{blue}{4}};
    
    \path[shorten >=1pt, thick, -{Latex[length=2.5mm]}]
    (3_1) edge [blue] (3_2)
    edge [blue] (3_3)
    edge [blue] (3_4)
    (3_2) edge [red] (3_3)
    (3_3) edge [red] (3_2)
    edge [red] (3_4)
    (3_4) edge [blue] (3_1);

    \node[font=\sffamily\Large, label={[label distance=0.45cm]90:{{$\texttt{L}^+(G)$}}}] (G) at ($(1)!0.5!(2)$) {};

    \node[font=\sffamily\Large, label={[label distance=0.15cm]270:{{$\texttt{L}_3^+(G)$}}}] (G) at ($(4)!0.5!(3)$) {};
    \node[font=\sffamily\Large, label={[label distance=0.15cm]270:{{$\texttt{L}_2^+(G)$}}}] (G) at ($(1_4)!0.5!(1_3)$) {};
    
    \draw[-, thin ,dashed , line join=round] ($(1.north west)+(-0.8,0.5)$)  [rounded corners] -- ($(3_2.north east)+(0.8,0.5)$)  [rounded corners] -- ($(3_3.south east)+(0.8,-0.9)$)  [rounded corners]-- ($(4.south west)+(-0.8,-0.9)$)[rounded corners] -- ($(1.north west)+(-0.8,0.5)$);
    
    \draw[-, thin ,dashed , line join=round] ($(1.north west)+(-0.2,0.2)$)  [rounded corners] -- ($(2.north east)+(0.2,0.2)$)  [rounded corners] -- ($(3.south east)+(0.2,-0.2)$)  [rounded corners]-- ($(4.south west)+(-0.2,-0.2)$)[rounded corners] -- ($(1.north west)+(-0.2,0.2)$);

    \draw[-, thin ,dashed , line join=round] ($(1_1.north west)+(-0.2,0.2)$)  [rounded corners] -- ($(3_2.north east)+(0.2,0.2)$)  [rounded corners] -- ($(3_3.south east)+(0.2,-0.2)$)  [rounded corners]-- ($(1_4.south west)+(-0.2,-0.2)$)[rounded corners] -- ($(1_1.north west)+(-0.2,0.2)$);
    \end{tikzpicture}}
    \caption{Set of networks $\texttt{L}^+(G)$ generated from Network $G$ by adding a link from a low-type user.}
    \label{fig:ex_atomic_G_L_plus}
\end{figure}
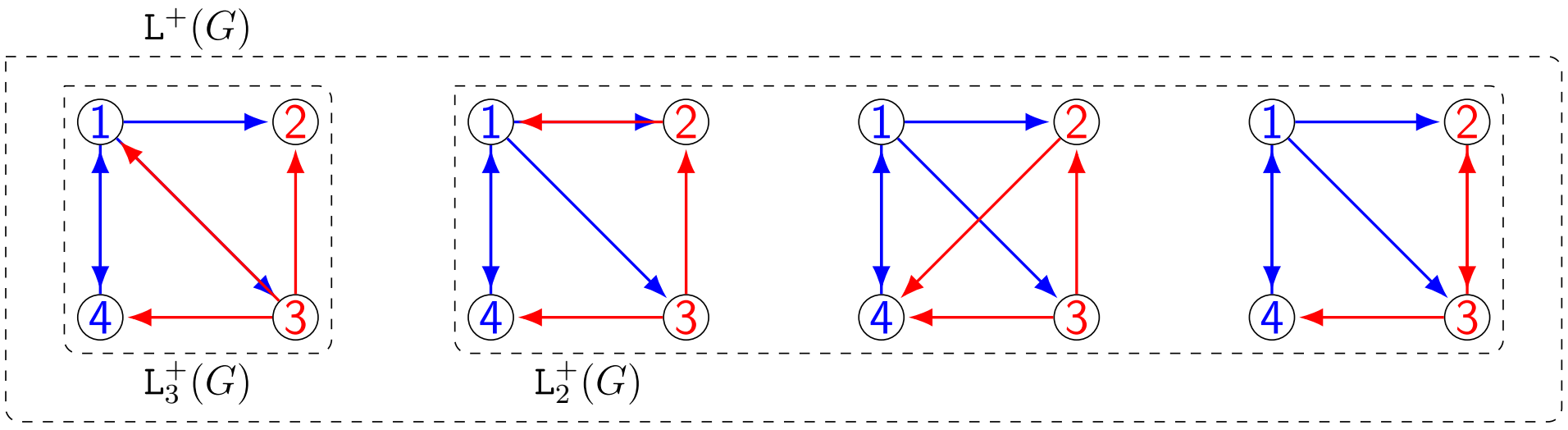

\begin{example}\label{ex:equivalence_operator}
    Let $n=4$ and let $G$ be as in Example~\ref{ex:transformation_network}. The set of networks $\texttt{L}^+(G)$ can be partitioned into $\texttt{L}^+_3(G)$ and $\texttt{L}_2^+(G)$ as shown in Figure~\ref{fig:ex_atomic_G_L_plus}.
\end{example}
\vspace*{-0.2cm}

\begin{theorem}[Collective action and the numbers of readers]\label{thm:aggregate_by_degree}
Let $G$ and $G'$ be two networks such that $r_k(G)=r_k(G')$ for every $k\in N$. Then, for every realisation of the posts $\mathbf{s}$,
\begin{equation*}
    \sum_{i\in N} a_i^*(G) = \sum_{i\in N} a_i^*(G').
\end{equation*}
In particular, $G$ and $G'$ are platform-equivalent.
\end{theorem}

The proof in Appendix~\ref{a_sec:manipulation} writes collective action as a sum over authors, in which the post of author $k$ enters only through $r_k$. The platform therefore chooses how many users see each post, and not which users see it. The user-based transformations are a first application.

\begin{corollary}\label{cor:equivalence_classes} Let $G=(N,E)$ be an informational network. Let $K$ be an operator in $\{\texttt{H}^+, \texttt{H}^-,\texttt{L}^+,\texttt{L}^-\}$. Then, for all $i\in N$, and for all $\widehat{G},\widetilde{G} \in K_i(G)$,  such that $\widehat{G}\neq\widetilde{G}$, we have $$\sum_{j\in N} a_j^*(\widehat{G})=\sum_{j\in N} a_j^*(\widetilde{G}).$$
\end{corollary}

Two senders, by contrast, are not necessarily equivalent. Consider a network in $\texttt{L}_2^+(G)$ and a network in $\texttt{L}_3^+(G)$ from Example~\ref{ex:equivalence_operator}. Theorem~\ref{thm:aggregate_by_degree} does not apply to them, since they give an additional reader to different users, and these two networks are not platform\nobreakdash-equivalent. Specifically, more users have access to user 3's signal ($r_3 > r_2$) in the initial network $G$. In general, from the platform's perspective, two senders are not necessarily equivalent, while Theorem~\ref{thm:aggregate_by_degree} shows that the identity of the receivers does not matter. Platform equivalence is, however, broader than the equality of the numbers of readers. Networks with different numbers of readers can still induce the same expected collective action, as the empty and the complete network do in Proposition~\ref{prop:extreme_equivalence}. Section~\ref{sec:equivalence_operators} studies such equivalences between senders.

\subsection{Equivalence between Transformations}\label{sec:equivalence_operators}

The atomic transformations provide a framework for analysing the platform's incentives when selecting networks to reach its target~$\mathfrak{a}$. Moreover, we know that there is no general equivalence between these transformations, as the change in collective action depends on the type of the sender. In this section, we examine what conditions render a low-type user equivalent to a high-type user among those sending information. This exploration will help clarify the richness of the equivalence classes.

Let us focus on a family of networks in which, whenever the platform sends a signal, it is delivered to all users. Conversely, should the platform decide not to send a user's signal, it remains confidential and known only to that user.
This means we are considering networks where the out\nobreakdash-neighbours of a user are either the empty set or the set of all other users.
Note that the networks $G_{\min}$ and $G_{\max}$ belong to this family. 
By considering these two extremes, full dissemination versus complete confidentiality, we can better understand how information sharing shapes user interactions and decision-making. With full disclosure, all users can align their actions, leading to a more coordinated outcome. 

We can analyse the equivalence between the two signalling operations, with particular focus on the actions of high-type and low-type users. The essence of this equivalence lies in understanding how the aggregate sum of actions can be influenced by adjusting the number of signals sent by each type. We apply the same intuition regarding the platform's ability to reduce the aggregate sum of actions, either by decreasing the number of signals sent by high-type users or by increasing the number of signals sent by low-type users. We want to characterise when these two operations, sending to ``everyone'' or to ``no one'', are equivalent\footnote{The approach of sending a message ``to everyone'' effectively integrates multiple individual operations of $K_i$. Note that while a single operation may not yield equivalence, when these operations are unified under the ``to everyone'' directive, they collectively produce an equivalent effect.}. The balance in probabilities plays a critical role in determining when these operations can be considered equivalent, as stated in Theorem~\ref{thm:all_high_low_equivalence}. If the probabilities of being a high-type or low-type user are equal, the platform can maintain collective actions, making the two operations effectively interchangeable.  

\begin{theorem}\label{thm:all_high_low_equivalence}
    Let $i,j\in N$ such that $t_i\neq t_j$. Consider an informational network $G=(N,E)$ such that $(i,k)\in E$ and $(j,k)\notin E$ for all $k\in N\setminus\{i\}$. Let $\widehat{G}=(N,\widehat{E})$ be the network obtained from $G$ by removing the edge $(i,k)$ for all $k\in N$, and let $\widetilde{G}=(N,\widetilde{E})$ be the network obtained from $G$ by adding the edge $(j,k)$ for all $k\in N\setminus\{j\}$. Then $\widehat{G}$ and $\widetilde{G}$ induce the same ex ante collective action, $\sum_{k\in N} \mathbb{E}_P\left[a_k\right]$, if and only if $p=1/2$.
\end{theorem}

\begin{figure}[h!]
\centering
\resizebox{\imagewidth}{!}{\begin{tikzpicture}[auto,node distance=1.5cm,main node/.style={circle,inner sep=0.5pt, font=\sffamily\large, draw, thin}]

\node[main node] (1)  {\textcolor{blue}{1}};
\node[main node] (2) [right of=1] {\textcolor{red}{2}};
\node[main node] (3) [below of=2] {\textcolor{red}{3}};
\node[main node] (4)  [below of=1] {\textcolor{blue}{4}};

\draw[-, thin ,dashed , line join=round] ($(1.north west)+(-0.2,0.2)$)  [rounded corners] -- ($(2.north east)+(0.2,0.2)$)  [rounded corners] -- ($(3.south east)+(0.2,-0.2)$)  [rounded corners]-- ($(4.south west)+(-0.2,-0.2)$)[rounded corners] -- ($(1.north west)+(-0.2,0.2)$);

\path[shorten >=1pt, thick]
(1) edge [-latex,blue] (2)
edge [-latex,blue] (3)
edge [-latex,blue] (4)
(4) edge [-latex,blue] (2)
edge [-latex,blue] (3)
edge [-latex,blue] (1);

\node[main node] (2_21) [below =0.5cm of 4] {\textcolor{red}{2}};
\node[main node] (2_11) [left of = 2_21]  {\textcolor{blue}{1}};
\node[main node] (2_31) [below of=2_21] {\textcolor{red}{3}};
\node[main node] (2_41)  [below of=2_11] {\textcolor{blue}{4}};
\path[shorten >=1pt, thick]
(2_41) edge [-latex,blue] (2_21)
edge [-latex,blue] (2_31)
edge [-latex,blue] (2_11);

\node[main node] (2_2) [left of=2_11] {\textcolor{red}{2}};
\node[main node] (2_1) [left of=2_2] {\textcolor{blue}{1}};
\node[main node] (2_3) [below of=2_2] {\textcolor{red}{3}};
\node[main node] (2_4)  [below of=2_1] {\textcolor{blue}{4}};
\path[shorten >=1pt, thick]
(2_1) edge [-latex,blue] (2_2)
edge [-latex,blue] (2_3)
edge [-latex,blue] (2_4);

\node[main node] (2_12) [right of=2_21] {\textcolor{blue}{1}};
\node[main node] (2_22) [right of=2_12] {\textcolor{red}{2}};
\node[main node] (2_32) [below of=2_22] {\textcolor{red}{3}};
\node[main node] (2_42)  [below of=2_12] {\textcolor{blue}{4}};

\path[shorten >=1pt, thick]
(2_12) edge [-latex,blue] (2_22)
edge [-latex,blue] (2_32)
edge [-latex,blue] (2_42)
(2_42) edge [-latex,blue] (2_22)
edge [-latex,blue] (2_32)
edge [-latex,blue] (2_12)
(2_22) edge [-latex,red] (2_12)
edge [-latex,red] (2_32)
edge [-latex,red] (2_42);

\node[main node] (2_13) [right of=2_22] {\textcolor{blue}{1}};
\node[main node] (2_23) [right of=2_13] {\textcolor{red}{2}};
\node[main node] (2_33) [below of=2_23] {\textcolor{red}{3}};
\node[main node] (2_43)  [below of=2_13] {\textcolor{blue}{4}};

\path[shorten >=1pt, thick]
(2_13) edge [-latex,blue] (2_23)
edge [-latex,blue] (2_33)
edge [-latex,blue] (2_43)
(2_43) edge [-latex,blue] (2_23)
edge [-latex,blue] (2_33)
edge [-latex,blue] (2_13)
(2_33) edge [-latex,red] (2_13)
edge [-latex,red] (2_23)
edge [-latex,red] (2_43);

\draw[-, thin ,dashed , line join=round] ($(2_1.north west)+(-0.2,0.2)$)  [rounded corners] -- ($(2_23.north east)+(0.2,0.2)$)  [rounded corners] -- ($(2_33.south east)+(0.2,-0.2)$)  [rounded corners]-- ($(2_4.south west)+(-0.2,-0.2)$)[rounded corners] -- ($(2_1.north west)+(-0.2,0.2)$);

\node[main node] (3_14) [below =0.5cm of 2_41] {\textcolor{blue}{1}};
\node[main node] (3_24) [right of=3_14] {\textcolor{red}{2}};
\node[main node] (3_34) [below of=3_24] {\textcolor{red}{3}};
\node[main node] (3_44)  [below of=3_14] {\textcolor{blue}{4}};

\path[shorten >=1pt, thick]
(3_44) edge [-latex,blue] (3_24)
edge [-latex,blue] (3_14)
edge [-latex,blue] (3_34)
(3_24) edge [-latex,red] (3_14)
edge [-latex,red] (3_34)
edge [-latex,red] (3_44);

\node[main node] (3_23) [left of=3_14] {\textcolor{red}{2}};
\node[main node] (3_13) [left of=3_23] {\textcolor{blue}{1}};
\node[main node] (3_33) [below of=3_23] {\textcolor{red}{3}};
\node[main node] (3_43)  [below of=3_13] {\textcolor{blue}{4}};

\path[shorten >=1pt, thick]
(3_13) edge [-latex,blue] (3_23)
edge [-latex,blue] (3_33)
edge [-latex,blue] (3_43)
(3_23) edge [-latex,red] (3_13)
edge [-latex,red] (3_33)
edge [-latex,red] (3_43);

\node[main node] (3_2) [left of=3_13] {\textcolor{red}{2}};
\node[main node] (3_1) [left of=3_2] {\textcolor{blue}{1}};
\node[main node] (3_3) [below of=3_2] {\textcolor{red}{3}};
\node[main node] (3_4)  [below of=3_1] {\textcolor{blue}{4}};

\node[main node] (3_15) [right of=3_24] {\textcolor{blue}{1}};
\node[main node] (3_25) [right of=3_15] {\textcolor{red}{2}};
\node[main node] (3_35) [below of=3_25] {\textcolor{red}{3}};
\node[main node] (3_45)  [below of=3_15] {\textcolor{blue}{4}};

\path[shorten >=1pt, thick]
(3_15) edge [-latex,blue] (3_25)
edge [-latex,blue] (3_35)
edge [-latex,blue] (3_45)
(3_35) edge [-latex,red] (3_15)
edge [-latex,red] (3_25)
edge [-latex,red] (3_45);

\node[main node] (3_16) [right of=3_25] {\textcolor{blue}{1}};
\node[main node] (3_26) [right of=3_16] {\textcolor{red}{2}};
\node[main node] (3_36) [below of=3_26] {\textcolor{red}{3}};
\node[main node] (3_46)  [below of=3_16] {\textcolor{blue}{4}};

\path[shorten >=1pt, thick]
(3_46) edge [-latex,blue] (3_26)
edge [-latex,blue] (3_36)
edge [-latex,blue] (3_16)
(3_36) edge [-latex,red] (3_16)
edge [-latex,red] (3_26)
edge [-latex,red] (3_46);

\node[main node] (3_17) [right of=3_26] {\textcolor{blue}{1}};
\node[main node] (3_27) [right of=3_17] {\textcolor{red}{2}};
\node[main node] (3_37) [below of=3_27] {\textcolor{red}{3}};
\node[main node] (3_47)  [below of=3_17] {\textcolor{blue}{4}};

\path[shorten >=1pt, thick]
(3_47) edge [-latex,blue] (3_27)
edge [-latex,blue] (3_37)
edge [-latex,blue] (3_17)
(3_17) edge [-latex,blue] (3_27)
edge [-latex,blue] (3_37)
edge [-latex,blue] (3_47)
(3_37) edge [-latex,red] (3_17)
edge [-latex,red] (3_27)
edge [-latex,red] (3_47)
(3_27) edge [-latex,red] (3_17)
edge [-latex,red] (3_37)
edge [-latex,red] (3_47);

\draw[-, thin ,dashed , line join=round] ($(3_1.north west)+(-0.2,0.2)$)  [rounded corners] -- ($(3_27.north east)+(0.2,0.2)$)  [rounded corners] -- ($(3_37.south east)+(0.2,-0.2)$)  [rounded corners]-- ($(3_4.south west)+(-0.2,-0.2)$)[rounded corners] -- ($(3_1.north west)+(-0.2,0.2)$);

\node[main node] (4_11) [below =0.5cm of 3_44] {\textcolor{blue}{1}};
\node[main node] (4_21) [right of=4_11] {\textcolor{red}{2}};
\node[main node] (4_31) [below of=4_21] {\textcolor{red}{3}};
\node[main node] (4_41)  [below of=4_11] {\textcolor{blue}{4}};

\path[shorten >=1pt, thick]
(4_31) edge [-latex,red] (4_11)
edge [-latex,red] (4_21)
edge [-latex,red] (4_41);

\node[main node] (4_2) [left of=4_11] {\textcolor{red}{2}};
\node[main node] (4_1) [left of=4_2] {\textcolor{blue}{1}};
\node[main node] (4_3) [below of=4_2] {\textcolor{red}{3}};
\node[main node] (4_4)  [below of=4_1] {\textcolor{blue}{4}};

\path[shorten >=1pt, thick]
(4_2) edge [-latex,red] (4_1)
edge [-latex,red] (4_3)
edge [-latex,red] (4_4);

\node[main node] (4_12) [right of=4_21] {\textcolor{blue}{1}};
\node[main node] (4_22) [right of=4_12] {\textcolor{red}{2}};
\node[main node] (4_32) [below of=4_22] {\textcolor{red}{3}};
\node[main node] (4_42)  [below of=4_12] {\textcolor{blue}{4}};

\path[shorten >=1pt, thick]
(4_12) edge [-latex,blue] (4_22)
edge [-latex,blue] (4_32)
edge [-latex,blue] (4_42)
(4_22) edge [-latex,red] (4_12)
edge [-latex,red] (4_32)
edge [-latex,red] (4_42)
(4_32) edge [-latex,red] (4_12)
edge [-latex,red] (4_22)
edge [-latex,red] (4_42);

\node[main node] (4_13) [right of=4_22] {\textcolor{blue}{1}};
\node[main node] (4_23) [right of=4_13] {\textcolor{red}{2}};
\node[main node] (4_33) [below of=4_23] {\textcolor{red}{3}};
\node[main node] (4_43)  [below of=4_13] {\textcolor{blue}{4}};

\path[shorten >=1pt, thick]
(4_43) edge [-latex,blue] (4_23)
edge [-latex,blue] (4_33)
edge [-latex,blue] (4_13)
(4_23) edge [-latex,red] (4_13)
edge [-latex,red] (4_33)
edge [-latex,red] (4_43)
(4_33) edge [-latex,red] (4_13)
edge [-latex,red] (4_23)
edge [-latex,red] (4_43);

\draw[-, thin ,dashed , line join=round] ($(4_1.north west)+(-0.2,0.2)$)  [rounded corners] -- ($(4_23.north east)+(0.2,0.2)$)  [rounded corners] -- ($(4_33.south east)+(0.2,-0.2)$)  [rounded corners]-- ($(4_4.south west)+(-0.2,-0.2)$)[rounded corners] -- ($(4_1.north west)+(-0.2,0.2)$);

\node[main node] (5_1) [below =0.5cm of 4_31] {\textcolor{blue}{1}};
\node[main node] (5_2) [right of=5_1] {\textcolor{red}{2}};
\node[main node] (5_3) [below of=5_2] {\textcolor{red}{3}};
\node[main node] (5_4)  [below of=5_1] {\textcolor{blue}{4}};

\path[shorten >=1pt, thick]
(5_2) edge [-latex,red] (5_1)
edge [-latex,red] (5_3)
edge [-latex,red] (5_4)
(5_3) edge [-latex,red] (5_2)
edge [-latex,red] (5_4)
edge [-latex,red] (5_1);

\draw[-, thin ,dashed , line join=round] ($(5_1.north west)+(-0.2,0.2)$)  [rounded corners] -- ($(5_2.north east)+(0.2,0.2)$)  [rounded corners] -- ($(5_3.south east)+(0.2,-0.2)$)  [rounded corners]-- ($(5_4.south west)+(-0.2,-0.2)$)[rounded corners] -- ($(5_1.north west)+(-0.2,0.2)$);

\end{tikzpicture}}
\caption{Family of Equivalence Classes for $n=4$, and $p=\frac12$.}
\label{fig:equivalence_classes_n4}
\end{figure}
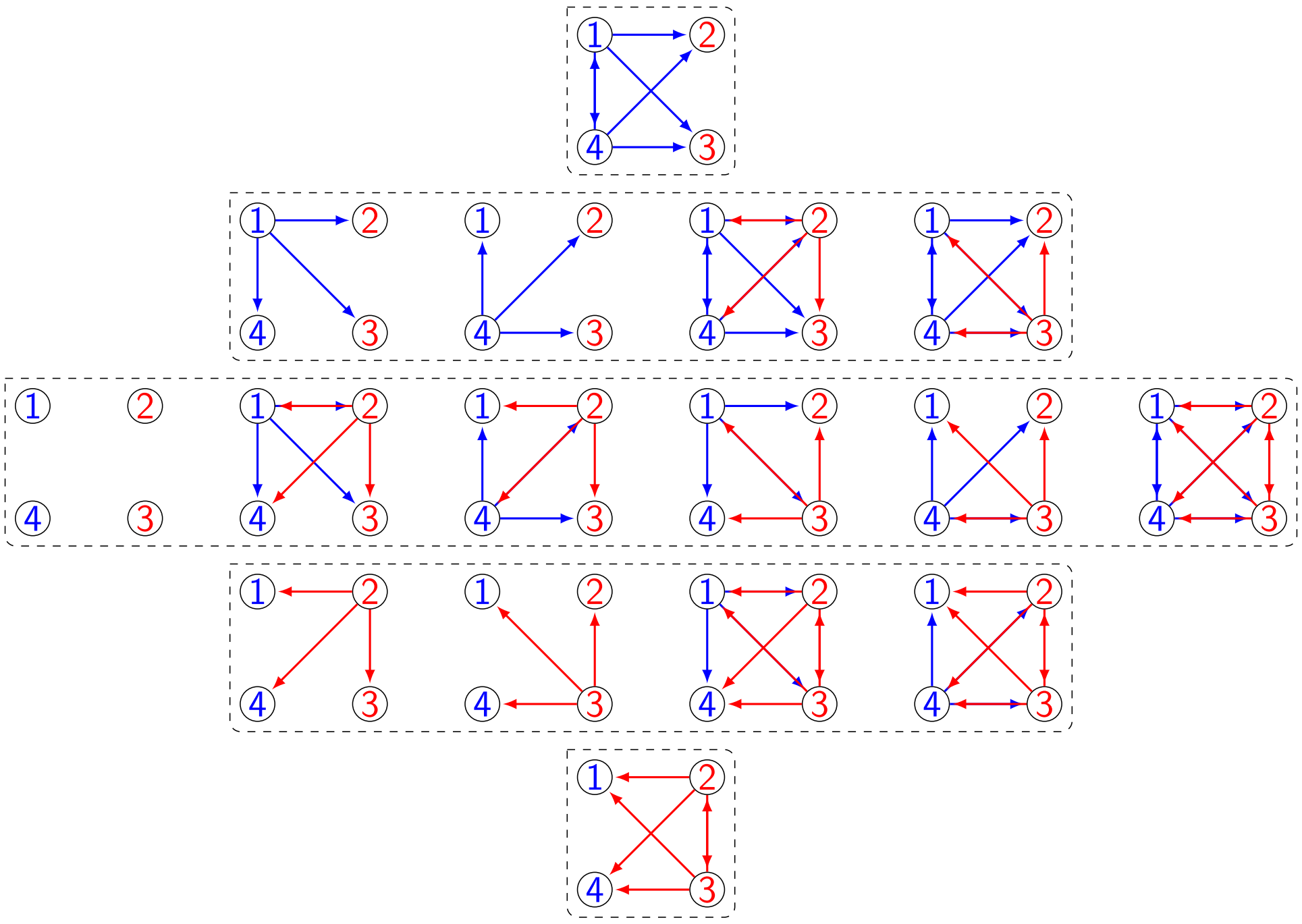

Figure~\ref{fig:equivalence_classes_n4} illustrates how to construct equivalence classes starting at $G_{\max}$ by using the operator to send to ``no one'', and from $G_{\min}$ through the operator to send to ``everyone''. Without loss of generality, let us begin with $G_{\max}$, which constitutes the equivalence class with the highest collective action. In $G_{\max}$, both users 1 and 4 send signals to everyone. Consider now the four possibilities for transformations: user 4, who is high type, switches to sending to ``no one''; user 1, who is also high type, switches to ``no one''; user 2, who is low type, sends ``to everyone''; and finally, user 3, who is low type, sends ``to everyone''. These four networks constitute a second equivalence class. To find the third equivalence class, we begin by eliminating all connections from user 1, leading us to the empty network. Alternatively, we obtain the complete network when starting from the last network in the second equivalence class by sending the signal of user 2 ``to everyone''. Both networks, the empty and the complete, belong to the same platform-equivalence class, inducing the same ex ante collective action. We continue this process to build the remaining equivalence classes until we reach the equivalence class with the lowest induced collective action, constituted by $G_{\min}$. 

Figure~\ref{fig:equivalence_classes_n4} depicts also a striking feature of our model. Let us focus on the extreme informational networks, namely, the complete and the empty networks. Although these two networks represent opposite informational settings, from platform's perspective, they induce the same collective action when the type distributions match prior expectations. This result stems from the fact that, in the empty network, users rely on their private signals and prior beliefs to form expectations. The expected collective action equals $n \cdot \mathbb{E}[a_i^*]$, where the expectation is taken over both signal and type realisations using the prior $p$. In the complete network, users observe all signals and can compute a better posterior over others' types. However, when the realised type distribution matches the prior, the collective action aggregates to the same value as in the empty network, despite users having complete information. We formalise this result in Proposition~\ref{prop:extreme_equivalence}.

\begin{proposition}[Equivalence of Extreme Informational Networks]\label{prop:extreme_equivalence}
Let $G_{\text{complete}}$ and $G_{\emptyset}$ be the complete and empty networks, respectively. Consider a type profile $t = (t_1, \ldots, t_n)$ with $n_h$ high-type users and $n_l$ low-type users, where $n = n_h + n_l$. The two networks induce the same collective action if and only if $\frac{n_h}{n} = p$. Moreover, for every type profile,
\begin{equation}\label{eq:extreme_gap}
\begin{split}
    \mathbb{E}_P\Big[\textstyle\sum_{i} a_i^*(G_{\text{complete}})\Big]
    &- \mathbb{E}_P\Big[\textstyle\sum_{i} a_i^*(G_{\emptyset})\Big]\\
    &= (n_h - np)(\lambda+\mu-1)\Big[\beta_1(n-1+\alpha)(\overline{s}-\underline{s})\\
    &\qquad + \beta_2\,\alpha\big(\mathbb{E}[t\mid\overline{s}]-\mathbb{E}[t\mid\underline{s}]\big)\Big].
\end{split}
\end{equation}
\end{proposition}

\section{Welfare, Information Quality and the Platform}\label{sec:policy}

Social media platforms are not merely neutral information infrastructures since they are dynamic environments that actively shape user behaviour and beliefs. As the literature shows, a growing body of theoretical and experimental research indicates that algorithmic content curation can strongly influence political attitudes, foster ideologically aligned communities, and enhance both political knowledge and the quality of public discourse (see Section~\ref{subsec:literature}). These findings underscore the importance of understanding how design choices on these platforms can drive collective outcomes, ultimately impacting individual welfare, the quality of information aggregation, and the degree of polarisation in our society.

Our result of Theorem~\ref{thm:aggregate_by_degree} shows that many architectures induce the same collective action. Networks in the same equivalence class are indistinguishable to the platform, but they are not necessarily indistinguishable to anyone else, and that gap is where regulation can operate. We measure it along three dimensions: a utilitarian criterion, an informational criterion and a polarisation criterion. We reach two conclusions. First, welfare and polarisation are not independent objectives but two components of a single quadratic loss. Second, the two conjectures set out in the introduction, that full information sharing maximises welfare and that segregation maximises polarisation, both fail in this environment.

\subsection{Three Criteria}\label{subsec:criteria}
This subsection introduces the three criteria and shows that they are neither independent of one another nor determined by the platform's own objective. Welfare and polarisation are two components of a single quadratic loss, so a regulator cannot trade them off freely. Architectures that the platform ranks equally, by contrast, differ substantially on all three criteria, and that gap is where regulatory intervention has purchase.

\noindent\textbf{Social welfare.} The utilitarian criterion is the sum of individual utilities at equilibrium,
\begin{multline}\label{eq:SW}
    SW(G; \mathbf{s}, \mathbf{t}) = \sum_{i=1}^n u_i\bigl(\mathbf{a}^*; \mathbf{s}, t_i\bigr)\\
    = -\sum_{i=1}^{n}\left[(1-\alpha)\bigl(a_i^* - \beta_1\theta(\mathbf{s}) - \beta_2 t_i\bigr)^2
      + \frac{\alpha}{n-1}\sum_{j \neq i}\bigl(a_i^* - a_j^*\bigr)^2\right],
\end{multline}
which measures both the accuracy of users' actions relative to their type-dependent ideal points and the coordination achieved across the network.

\noindent\textbf{Heterophilic density.} Users infer the state $\theta(\mathbf{s})$ from their own post and those of their neighbours, so the quality of information aggregation depends on the pattern of information flows. Density alone is not the right measure, since networks dominated by same-type links may create echo chambers that reinforce homogeneous beliefs \citep{golub2012homophily}. We therefore weight cross-type links fully and same-type links by $\tau \in [0,1]$:
\begin{equation}\label{eq:HD}
    HD(G, \tau;\mathbf{t}) = \frac{\sum_{(i,j)\in E}\bigl(\mathbf{1}_{t_i \neq t_j} + \tau\, \mathbf{1}_{t_i = t_j}\bigr)}
    {2n_h n_l + \tau\bigl(n_h(n_h-1) + n_l(n_l-1)\bigr)} \;\in\; [0,1],
\end{equation}
where $n_h$ and $n_l$ are the numbers of high- and low-type users.\footnote{When all users share one type, every link joins two users of the same type. At $\tau=0$ both the numerator and the denominator of Equation~\eqref{eq:HD} are then zero, and $HD$ is not defined. The measure has no content in this case, since there are no cross-type links to create and every link is one that $\tau=0$ ignores. We therefore use $HD$ at $\tau=0$ only when both types are present.} At $\tau=1$ this is normalised edge density; as $\tau$ falls the metric becomes increasingly sensitive to cross-type exposure. $HD$ also orders networks by the exposure they create. When $\tau>0$, adding any link strictly increases $HD$, which equals $0$ at the empty network and $1$ at the complete network. When $\tau=0$, the same holds for links between users of different types, while links between users of the same type leave $HD$ unchanged. Crucially, $HD$ depends on the architecture and the type profile alone and not on the realised posts, therefore this makes it an instrument to compare networks configurations.

\noindent\textbf{Polarisation.} We measure polarisation as the dispersion of equilibrium actions,
\begin{equation}\label{eq:Var}
    \mathrm{Var}(\mathbf{a}^*) = \frac{1}{n}\sum_{i=1}^n \bigl(a_i^* - \bar{a}^*\bigr)^2,
    \qquad \bar{a}^{*} = \frac{1}{n}\sum_{i} a^{*}_{i},
\end{equation}
so that higher values indicate greater behavioural divergence across the population.
The first and third criteria are linked by an identity. Since $\sum_{i}\sum_{j\neq i}(a_i-a_j)^2 = 2n^{2}\mathrm{Var}(\mathbf{a})$ for any profile of actions, substituting into \eqref{eq:SW} gives the following.

\begin{lemma}\label{lem:sw_var_identity}
For every network $G$ and every realisation $(\mathbf{s},\mathbf{t})$,
\begin{equation}\label{eq:sw_var_identity}
  SW(G;\mathbf{s},\mathbf{t}) \;=\; -(1-\alpha)\sum_{i=1}^{n}\bigl(a_i^* - \beta_1\theta(\mathbf{s}) - \beta_2 t_i\bigr)^{2}
  \;-\; \frac{2\alpha n^{2}}{n-1}\,\mathrm{Var}(\mathbf{a}^*),
\end{equation}
and the same identity holds after taking $\mathbb{E}_P[\cdot]$ on both sides.
\end{lemma}

The conformity component of welfare is population polarisation, up to the
positive constant $2\alpha n^{2}/(n-1)$. Polarisation is therefore not a second objective to be traded off against welfare but a component of it, entering with a weight the model pins down rather than leaving to the regulator's discretion, and one that grows both with the strength of the conformity motive and with the size of the population. A policy maker who maximises utilitarian welfare is already penalising dispersion at that rate, and one who lowers dispersion without degrading the accuracy with which users track the aggregate state necessarily raises welfare.

The two rankings can nonetheless differ, and the identity says exactly how.
Welfare also counts the first term of \eqref{eq:sw_var_identity}, the loss from actions that fail to track the state. An architecture that substantially improves tracking can be welfare-preferred even though it leaves users further apart, so a comparison in which the more dispersed architecture wins on welfare records a gain in accuracy rather than any benefit from disagreement itself.

The three criteria have bite because they separate architectures that the platform's own objective does not. Moreover, all three criteria are evaluated with the operator $\mathbb{E}_P[\cdot]$. The regulator's question is not which network is best in the abstract but which member of a platform-equivalence class should be selected, and those classes are defined by the platform's own evaluation, made after types are observed and before any post is written. For a comparison to be meaningful, the regulator should evaluate candidate networks at the same stage and with the same information as the platform. We follow this rule throughout, and every example below compares architectures under $\mathbb{E}_P$ at a fixed type profile. Heterophilic density does not depend on posts, so the operator is immaterial for it.

\begin{figure}[h!]
\centering
\includegraphics[width=0.72\textwidth]{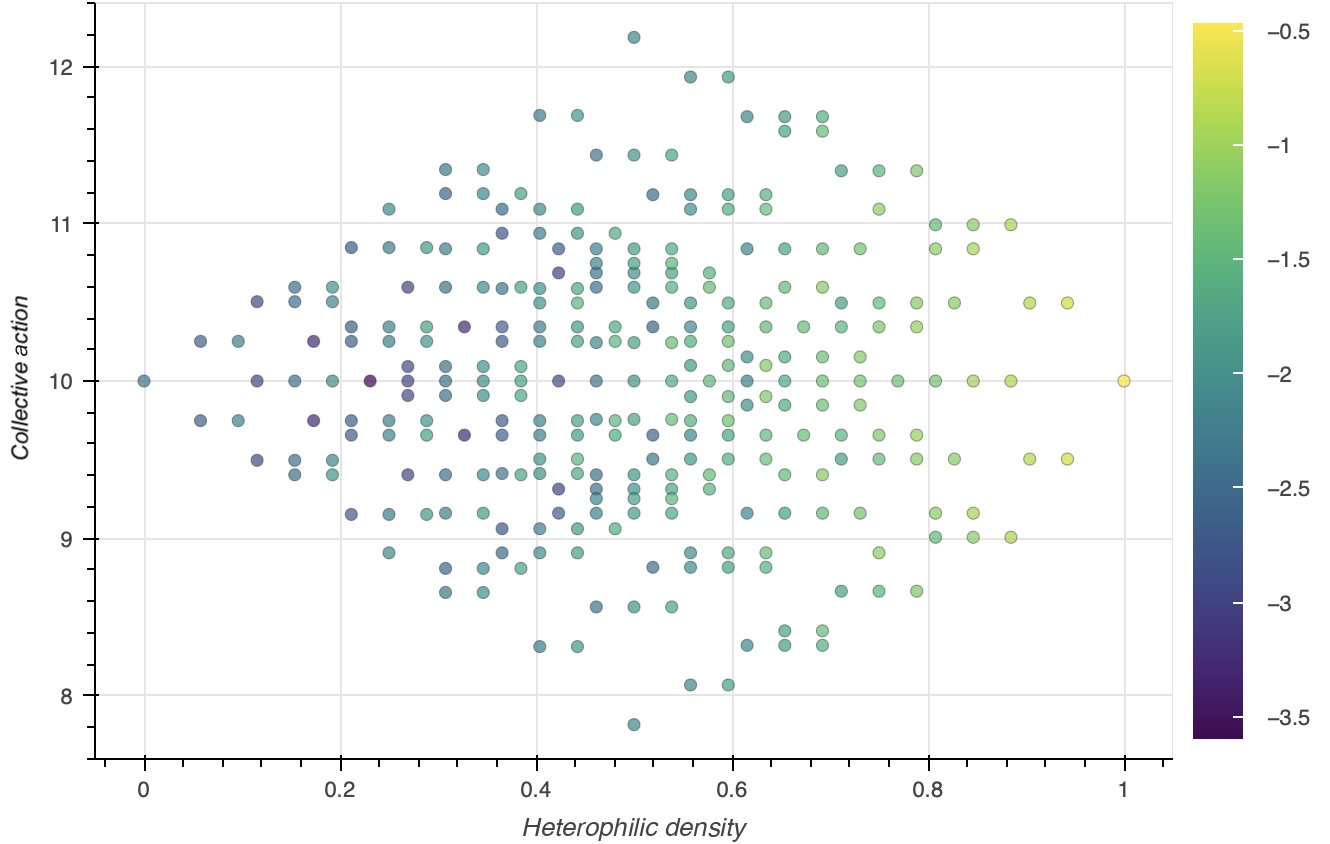}
\caption{Collective action against heterophilic density for all networks on $n=4$ users, at $\alpha=0.4$, $\beta_1=\beta_2=1$, $\lambda=\mu=0.8$, $p=1/2$, $\tau=0.6$ and $n_h = n_l = 2$, with $\overline{t}=\overline{s}=1$ and $\underline{t}=\underline{s}=0$. Colour indicates social welfare $\mathbb{E}_P[SW]$. Horizontal bands are platform-equivalence classes.}
\label{fig:ca_hd}
\end{figure}

Figure~\ref{fig:ca_hd} plots every network on four users against its heterophilic density, with colour indicating social welfare. The points fall into discrete horizontal bands. Each band is a platform-equivalence class: all networks in it induce the same ex ante collective action, so the platform is indifferent along a horizontal line. Colour varies markedly within a single band. Among the $53$ bands with more than one network, the median band spans $30\%$ of the range of welfare across all networks, and the central band spans all of it. Those choices are therefore not equivalent for users and policymakers. The figure also shows that $HD$ is not related to the objective of the platform in a systematic way. Networks in the same band can have different values of $HD$, and networks with the same value of $HD$ can belong to different bands. The $4{,}096$ networks form $55$ bands and take $45$ values of $HD$, and a value of $HD$ is shared by six bands at the median. We therefore use $HD$ as a descriptive measure of cross-type exposure that a regulator can compute, and not as a sufficient statistic for welfare or polarisation. The band at the centre of the figure contains both the empty and the complete network, as stated in Proposition~\ref{prop:extreme_equivalence}: at $p = 1/2$, maximal information sharing and complete isolation induce the same collective action, while differing in everything else.

The atomic operators of Definition~\ref{def:atomic_operator} make this concrete. They move a single link between recipients while keeping the sender fixed, and, by Corollary~\ref{cor:equivalence_classes}, they leave collective action unchanged. They change everything else. Next Example \ref{ex:equivalence_class} provides the computations. 

\begin{example}\label{ex:equivalence_class}
Take the four-user network $G$ of Example~\ref{ex:political_1} and the class $L^{+}_{2}(G)$ of Example~\ref{ex:equivalence_operator}, obtained by giving the low-type user $2$ a single out-link. Its three members send that link to user $1$ (high type), user $3$ (low type) or user $4$ (high type). At $\alpha = 0.7$, $\beta_1 = 1$, $\beta_2 = 5$, $p = 1/2$, $\lambda = 0.9$, $\mu = 0.7$ and $\tau = 0.6$:
\begin{center}
\begin{tabular}{llccc}
\toprule
new link & \multicolumn{1}{c}{$\sum_i \mathbb{E}_P[a_i^*]$} & $\mathbb{E}_P[SW]$ & $\mathbb{E}_P[\mathrm{Var}(\mathbf{a}^*)]$ & $HD(\cdot,0.6;\mathbf{t})$ \\
\midrule
($G$) &  $20.9343$ & $-9.9712$ & $0.8487$ & $6/13$ \\
$(2,1)$ & $20.7039$ & $-9.5714$ & $0.7919$ & $29/52$ \\
$(2,4)$ & $20.7039$ & $-9.7888$ & $0.8170$ & $29/52$ \\
$(2,3)$ & $20.7039$ & $-10.5076$ & $0.9646$ & $27/52$ \\
\bottomrule
\end{tabular}
\end{center}

The three architectures are exactly platform-equivalent. Each is obtained from $G$ by adding one link from user $2$, and each lowers collective action by the same amount. For users, however, the effect of this link depends on its recipient. If the post of user $2$ reaches a high-type user, welfare rises and polarisation falls. Welfare rises by $0.40$ and polarisation falls by $6.7\%$ when the recipient is user $1$, and by $0.18$ and $3.7\%$ when it is user $4$. If the post reaches the other low-type user, welfare falls by $0.54$ and polarisation rises by $13.7\%$.\footnote{Percentages of polarisation are relative to its value under $G$. Welfare is a quadratic loss with no natural zero, so we report its changes as differences.} The first two architectures, which route the same post to two different high-type readers, have \emph{identical} heterophilic density while differing in both other criteria.
\end{example}

\subsection{Block Architectures and Two Failed Conjectures}\label{subsec:block}

This subsection establishes that both conjectures fail. We proceed in two steps. We first identify a class of networks on which the equilibrium collapses onto two representative users, one of each type, and we characterise the class by a single object, a map recording for each pair of types whether posts written by the first are shown to the second.\footnote{Comparing all architectures is not feasible in general. Theorem~\ref{thm:users_equilibrium} gives the equilibrium actions in closed form, and from them we obtain closed forms for the welfare and the polarisation of each network (Appendix~\ref{app:general}). However, the design space has $2^{n(n-1)}$ networks, and, unlike collective action (Theorem~\ref{thm:aggregate_by_degree}), these criteria depend on the whole profile of information sets and not only on the numbers of readers.} We then show that on this class all three criteria admit closed-form expressions, so that architectures can be compared by evaluating explicit functions of the primitives rather than by solving the game. Because the parameters in the comparisons below are rational, the resulting inequalities are exact, and the two conjectures are refuted by computation rather than by approximation.

\begin{definition}\label{def:block}
A network $G$ is a \emph{block architecture} if there is a map $B : T \times T \to \{0,1\}$ such that, for all $i \neq j$, $(i,j) \in E$ if and only if $B(t_i,t_j) = 1$. We write $G[B]$, and identify $B$ with the quadruple $\bigl(B(\bar t,\bar t),\,B(\bar t,\underline t),\,B(\underline t,\bar t),\,B(\underline t,\underline t)\bigr)$.
\end{definition}

Here $B(\sigma,\sigma')=1$ means that posts written by type-$\sigma$ users are shown to type-$\sigma'$ users, so a block architecture is a feed rule conditioning only on the type of the author and the type of the reader, which is exactly the information the platform holds at the design stage. There are sixteen of them, and they include every extreme structure of interest. For our analysis\footnote{The general case, in which the feed conditions on anything finer than these two types, does not admit the reduction below and lies outside the scope of this paper.}, we consider the following structures: 
\begin{center}
\begin{tabular}{lcl}
\toprule
$G_{\mathrm{complete}} = G[1,1,1,1]$ & \quad & every post is shown to everyone \\
$G_{\emptyset} = G[0,0,0,0]$ & & no post is shared \\
$G_{\max} = G[1,1,0,0]$ & & high-type posts broadcast, low-type posts private \\
$G_{\min} = G[0,0,1,1]$ & & low-type posts broadcast, high-type posts private \\
$G_{\mathrm{echo}} = G[1,0,0,1]$ & & each type sees only its own type \\
$G_{\mathrm{het}} = G[0,1,1,0]$ & & each type sees only the opposite type \\
\bottomrule
\end{tabular}
\end{center}

The class is tractable because reach depends only on the author's type. Every type-$\sigma$ user is read by the same number of others, and every type-$\sigma$ user reads the same collection of authors up to relabelling, so each user loads on a post with a weight that depends only on her own type and on the type of its author (Lemma~\ref{lem:exchangeability} in the appendix). Two users of a common type generically receive different posts and obtain different utilities ex post, but their expected utilities given the type profile coincide, and the same holds for the expected squared distance between the actions of two users of given types. Two families of quantities are therefore well defined, the interim utility $U_\sigma = \mathbb{E}_P[u_i]$ of a user $i$ of type $\sigma$, and the pair moment $P_{\sigma\sigma'} = \mathbb{E}_P\bigl[(a_i^* - a_j^*)^2\bigr]$ for distinct users $i$ and $j$ of types $\sigma$ and $\sigma'$.

\begin{proposition}\label{prop:closed_forms}
Let $G[B]$ be a block architecture and let $\mathbf{t}$ be a type profile with $n_h, n_l \geq 1$. The interim utilities $U_{\bar t}$ and $U_{\underline t}$ and the pair moments $P_{\bar t\bar t}$, $P_{\underline t\,\underline t}$ and $P_{\bar t\,\underline t}$ are rational functions of the parameters of the model and of the composition $(n_h, n_l)$. Social welfare and polarisation are expressed through them as
\[
  \mathbb{E}_P\bigl[SW(G[B])\bigr] = n_h\, U_{\bar t} + n_l\, U_{\underline t}
\]
and
\[
  \mathbb{E}_P\bigl[\mathrm{Var}(\mathbf{a}^*)\bigr]
  = \frac{n_h(n_h-1)\, P_{\bar t\bar t} + n_l(n_l-1)\, P_{\underline t\,\underline t} + 2 n_h n_l\, P_{\bar t\,\underline t}}{2n^{2}} .
\]
\end{proposition}

Proposition~\ref{prop:closed_forms} reduces welfare and polarisation to five numbers, each of which can be evaluated exactly, and this is what allows the comparisons below to be stated as exact inequalities, given explicitly in Appendix~\ref{app:policy}. It also makes the distribution of welfare across types visible. The utilitarian criterion aggregates $U_{\bar t}$ and $U_{\underline t}$ with population weights, so an architecture can raise the sum while lowering one of its two components, and this is precisely what happens in Example~\ref{ex:welfare_fails}.

Heterophilic density requires no equilibrium computation. The edge set of $G[B]$ consists of the ordered pairs $(i,j)$ with $i \neq j$ and $B(t_i,t_j) = 1$, so counting the edges in \eqref{eq:HD} by the types of their endpoints gives
\begin{equation}\label{eq:HD_block}
  HD(G[B],\tau;\mathbf{t}) = \frac{\bigl[B(\bar t,\underline t) + B(\underline t,\bar t)\bigr] n_h n_l + \tau\bigl[B(\bar t,\bar t)\, n_h(n_h-1) + B(\underline t,\underline t)\, n_l(n_l-1)\bigr]}{2 n_h n_l + \tau\bigl[n_h(n_h-1) + n_l(n_l-1)\bigr]} .
\end{equation}

\medskip
\noindent\textbf{The complete network does not maximise welfare.}

The first conjecture is natural in our model. When every user observes every post, no user forms beliefs from a proper subset of the evidence, and the strategic shading by which users discount their own post to stay aligned with less-informed peers disappears. The complete network also exposes users to one another as much as possible, since setting $B \equiv 1$ in \eqref{eq:HD_block} gives $HD(G_{\mathrm{complete}},\tau;\mathbf{t}) = 1$, the maximum over all networks, for every $\tau$. For welfare the conjecture is nonetheless false, and the two examples below show that it fails in two opposite ways.

\begin{example}\label{ex:welfare_fails}
Let $n=6$ with five high types and one low type, and let
\[
  \alpha = 0.7,\quad \beta_1 = 1,\quad \beta_2 = 5,\quad p = \tfrac12,\quad \lambda = 0.9,\quad \mu = 0.7,
\]
with $\underline t = \underline s = 0$ and $\bar t = \bar s = 1$. Muting the single low-type user raises welfare,
\[
  \mathbb{E}_P\bigl[SW(G_{\max})\bigr] = -7.0450 \;>\; -7.1805 = \mathbb{E}_P\bigl[SW(G_{\mathrm{complete}})\bigr].
\]
The interim utilities are $U_{\bar t} = -0.8416$ and $U_{\underline t} = -2.9726$ under the complete network, against $U_{\bar t} = -0.6715$ and $U_{\underline t} = -3.6873$ under $G_{\max}$. Each of the five high types gains $0.1700$, for an aggregate gain of $0.8502$, while the muted user loses $0.7147$.\footnote{Appendix~\ref{app:examples} reports the intermediate quantities. Since all parameters are rational, the comparison is exact rather than approximate, the welfare difference being $31296969/231040000$.}
\end{example}

\begin{example}\label{ex:welfare_fails_het}
Let $n=6$ with three users of each type, and let
\[
  \alpha=\tfrac12,\quad \beta_1=\tfrac1{10},\quad \beta_2=1,\quad
  p=\tfrac12,\quad \lambda=\mu=\tfrac7{10},
\]
with $\underline t=\underline s=0$ and $\bar t=\bar s=1$, so that the realised composition matches the prior. Then $G_{\mathrm{het}}$, in which each type observes only the other, maximises welfare among the sixteen block architectures, and
\[
  \mathbb{E}_P\bigl[SW(G_{\mathrm{het}})\bigr] = -0.5972
  \;>\; -0.6348 = \mathbb{E}_P\bigl[SW(G_{\mathrm{complete}})\bigr].
\]
The example is symmetric in the two types, so the interim utility of every user rises, by $0.0063$. Polarisation falls from $0.0608$ to $0.0455$.\footnote{The welfare difference is exactly $44577/1185800$ (Appendix~\ref{app:examples}).}
\end{example}

Table~\ref{tab:welfare_decomposition} splits each comparison into three components. The consistency loss of Lemma~\ref{lem:sw_var_identity} divides into a level component, the average distance between actions and ideal points, and a precision component, the noise around that average. The third component is the coordination loss, which is polarisation up to a positive constant.

\begin{table}[h!]
\centering
\small
\begin{tabular}{lcccccc}
\toprule
 & \multicolumn{3}{c}{Example~\ref{ex:welfare_fails}} & \multicolumn{3}{c}{Example~\ref{ex:welfare_fails_het}}\\
\cmidrule(lr){2-4}\cmidrule(lr){5-7}
 & $G_{\mathrm{complete}}$ & $G_{\max}$ & change & $G_{\mathrm{complete}}$ & $G_{\mathrm{het}}$ & change\\
\midrule
Consistency, level     & $-4.1416$ & $-3.1680$ & $+0.9736$ & $-0.1930$ & $-0.2568$ & $-0.0638$\\
Consistency, precision & $-0.1601$ & $-0.4550$ & $-0.2949$ & $-0.0044$ & $-0.0129$ & $-0.0085$\\
Coordination           & $-2.8788$ & $-3.4221$ & $-0.5433$ & $-0.4374$ & $-0.3275$ & $+0.1099$\\
\midrule
$\mathbb{E}_P[SW]$     & $-7.1805$ & $-7.0450$ & $+0.1355$ & $-0.6348$ & $-0.5972$ & $+0.0376$\\
\bottomrule
\end{tabular}
\caption{The three components of expected welfare in Examples~\ref{ex:welfare_fails} and \ref{ex:welfare_fails_het}. Corollary~\ref{cor:three_components} gives the expressions, and Appendix~\ref{app:examples} the exact values. Changes are computed from exact values and may not add up because of rounding.}
\label{tab:welfare_decomposition}
\end{table}

The two failures are mirror images of each other. In Example~\ref{ex:welfare_fails} the gain comes entirely from the level of actions, while precision and coordination both deteriorate. In Example~\ref{ex:welfare_fails_het} the gain comes entirely from coordination, while accuracy deteriorates in level and in precision. Restricting information can therefore raise welfare by making users act more accurately or by making them agree more, but in neither case by making them better informed.

Both failures have a common source. A post is evidence about the type of its author, and the regulator, like the platform, evaluates architectures knowing the type profile. Users do not, and they do not infer types from the structure of their feeds (Section~\ref{sec:assumptions}). A user who is not shown a post therefore expects it at its population average, and a user who is shown some posts treats them as a representative sample. A feed rule that conditions on types thus moves the average action of each reader, and not only the precision with which she tracks the state (Lemma~\ref{lem:mean_action}). In Example~\ref{ex:welfare_fails}, conformity makes each high-type user shade her action towards the types she expects the others to have, and since five of the six users are of the high type while the prior is $p=\tfrac12$, she expects them to be lower than they are. Muting the low-type user removes from the feed of the majority the post most likely to be low, so each high-type user overestimates the state and raises her action, which partly offsets the shading. This gain relies on the realised composition differing from the prior, and it disappears when welfare is evaluated before types are drawn.\footnote{At the parameters of Example~\ref{ex:welfare_fails} the complete network is then preferred to $G_{\max}$, and on the grid of Appendix~\ref{app:grid} no architecture that hides the posts of a type from every user maximises welfare evaluated before types are drawn.} In Example~\ref{ex:welfare_fails_het}, where the composition matches the prior, each user of $G_{\mathrm{het}}$ reads only posts written by the other type and does not correct for this selection, so the two groups move towards one another. Accuracy falls, but conformity becomes cheaper, and the trade is favourable when users care little about the state relative to their type, that is, when $\beta_1/\beta_2$ is small.\footnote{In \eqref{eq:SW} a disagreement between users $i$ and $j$ enters the utility of both, whereas the best response of each user accounts only for her own coordination loss. Holding the information of users fixed, the utilitarian criterion therefore values a reduction in dispersion at twice the rate at which users do. This wedge between the private and the social value of coordination is the one that \citet{angeletos2007efficient} identify as governing the social value of information in economies with a beauty-contest motive.} Both failures rest on the coordination motive. Without it, each user acts on her own estimate of the state, and the complete network gives every user her ideal action.\footnote{Setting $\alpha = 0$ in Lemma~\ref{lem:innovation} gives $a_i^* = \beta_1\theta(\mathbf{s}) + \beta_2 t_i$ under the complete network, so that $SW = 0$, its largest possible value. On the grid of Appendix~\ref{app:grid}, at $\alpha = 0.001$ the complete network maximises welfare among the block architectures in every configuration.}

The two examples also differ in who gains, which Proposition~\ref{prop:closed_forms} makes visible. In Example~\ref{ex:welfare_fails_het} every user is better off. In Example~\ref{ex:welfare_fails} every high-type user gains, while the muted user loses more than four times as much as each of them, so a regulator applying the utilitarian criterion will endorse architectures that suppress the informational influence of small minorities. Which architecture is optimal depends on the composition of the population and not only on the parameters, and what matters is the absolute size of the minority rather than its share. At the parameters of Example~\ref{ex:welfare_fails}, the complete network is dominated by $G_{\max}$ once the majority is large enough when the minority has at most four members, and it is preferred at every majority size we examine when the minority has five or six. Appendix~\ref{app:composition} reports the welfare-maximising architecture across compositions, together with the corresponding thresholds.

\medskip
\noindent\textbf{The echo chamber does not maximise polarisation.}

The second conjecture is equally plausible. When users see only their own type, intra-group coordination is maximal, the two groups never moderate one another, and the gap between their mean actions appears as wide as the information structure permits. The architecture that overturns it is the structural opposite of the echo chamber.

\begin{example}\label{ex:echo_fails}
Let $n=6$ with three users of each type, and let
\[
  \alpha = 0.9,\quad \beta_1 = 5,\quad \beta_2 = 0.3,\quad p = \tfrac12,\quad \lambda = \mu = 0.7 .
\]
Then the maximally heterophilic architecture is strictly more polarising than the echo chamber,\footnote{Again, the parameters are rational and the comparison exact, the difference being $4839816904489/58926368000000$.}
\[
  \mathbb{E}_P\bigl[\mathrm{Var}(\mathbf{a}^*; G_{\mathrm{het}})\bigr] = 0.5108
  \;>\; 0.4287 = \mathbb{E}_P\bigl[\mathrm{Var}(\mathbf{a}^*; G_{\mathrm{echo}})\bigr]. \qedhere
\]
\end{example}

The echo chamber loses when posts are read mainly for their content rather than for what they reveal about their authors, and strong conformity makes that outcome more frequent without being necessary (Appendix~\ref{app:grid}). This is because the variance criterion does not measure segregation. It measures dispersion, and by Proposition~\ref{prop:closed_forms} it counts the $n_h(n_h-1) + n_l(n_l-1)$ same-type pairs alongside the $2n_hn_l$ cross-type ones. The echo chamber minimises the same-type contribution, because two users of a common type observe exactly the same posts when each observes the whole of her own group, so their actions differ only through the correction each applies to her own post. Under the heterophilic architecture, two users of a common type never observe one another, so each responds to a post the other cannot see. At the parameters of Example~\ref{ex:echo_fails} the three pair moments are
\[
\begin{array}{lccc}
\toprule
 & P_{\bar t \bar t} & P_{\underline t\,\underline t} & P_{\bar t\,\underline t} \\
\midrule
G_{\mathrm{echo}} & 0.000001 & 0.000001 & 1.7148 \\
G_{\mathrm{het}}  & 0.4982   & 0.4982   & 1.7112 \\
\bottomrule
\end{array}
\]
The between-group moment is essentially the same under the two architectures, and marginally larger under the echo chamber, so the entire difference in polarisation is generated within groups. Segregating the types does maximise the separation of their average actions, but it does so by making each group internally uniform, and uniformity is itself a reduction in dispersion. When the type motive is strong the first effect dominates and the echo chamber is the most polarising architecture, and when the signal motive dominates the second prevails, so that isolating users from their own type generates more divergence than isolating them from the other type.

The comparison with Example~\ref{ex:welfare_fails_het}, where $G_{\mathrm{het}}$ lowers polarisation, is instructive. There it narrows the gap between the average actions of the two groups, from $0.4927$ under the complete network to $0.4148$ (Appendix~\ref{app:examples}), but it also makes two users of a common type respond to different posts. The first effect prevails when users care mainly about their type, as in Example~\ref{ex:welfare_fails_het}, and the second when they care mainly about the state, as in Example~\ref{ex:echo_fails}.

The two conjectures fail for different reasons, and the difference is informative. The welfare conjecture fails because a feed that conditions on types moves the average actions of its readers, and not only their precision. Users load on beliefs about one another's types as well as on the posts themselves, and they do not infer types from what the platform hides, so a regulator who knows the types can use the feed to correct the level of actions, as when a minority is muted, or to bring the two groups closer together, as when users see only the other type. Both gains come from the coordination motive and disappear with it. The polarisation conjecture fails for a reason that does not depend on coordination. Dispersion counts disagreement within groups as well as between them, and a segregated architecture makes each group internally uniform, so it can be less polarising than one that exposes each group only to the other. Strong conformity makes this outcome more frequent, but it persists as $\alpha$ tends to zero, because it rests on what the variance criterion measures rather than on how users coordinate.

This has implications for how the empirical literature on echo chambers should be read. \citet{bakshy2015exposure} and \citet{tornberg2022digital} document ideologically homogeneous communities and associate them with polarisation, while \citet{boxell2017greater} find no association between internet use and the growth of political polarisation across demographic groups. These findings need not be in tension. In our model, segregation and behavioural divergence are distinct objects, related non-monotonically, and which of the two an architecture produces depends on whether users read posts mainly for their content or mainly for what they reveal about their authors. Where the type motive dominates, the two coincide and a segregated feed is the most polarising one. Where the signal motive dominates, they come apart, and a feed that exposes each group only to the other can generate more behavioural divergence than one that exposes each group only to itself. Measuring cross-type exposure therefore does not by itself identify the polarisation consequences of a feed design, and the sign of the relationship between the two is a parametric question rather than a structural one.

\section{Conclusions}\label{sec:conclusions}

The analytical core of this paper is the characterisation of equilibrium. In a coordination game where users hold private types that are payoff-relevant in their own right, and where the information each user receives is mediated by a network chosen by a third party, we establish existence and uniqueness and derive the equilibrium in closed form. Three properties of that characterisation carry the rest of the paper. Each user's action is a weighted aggregation of the posts she observes, so that the effect of an architecture can be read off a single profile of weights. The weight attached to a post is increasing in the size of its audience, which converts the platform's control of visibility into control of behaviour. And a post enters a reader's action through two channels at once, one carrying evidence about the aggregate state and one carrying evidence about the author's type, so that the same weight governs what a user believes about the world and what she believes about the people she must coordinate with.

The consequences we draw rest on different parts of that characterisation. The platform's ability to steer collective action, and the equivalence of architectures that induce the same target, follow from the first two properties, since the aggregate action depends on the network only through how widely each author is read. The failure of the welfare conjecture rests on the third. Because a post partly reveals the type of its author, showing or hiding it moves the average action of every reader, and not only her information about the state. A regulator who knows the types can use this either to correct the level of actions or to narrow the gap between groups, and both gains exist only because users coordinate and do not perfectly infer types from their feeds. In Appendix~\ref{app:numerical}, we find that the first gain disappears when welfare is evaluated before types are drawn. The second gain survives, but only in a small region of the parameter space and by small margins. The failure of the polarisation conjecture rests on none of them in particular, but on what dispersion measures, since it counts disagreement within groups and segregation suppresses exactly that. We leave four questions open. For each of them, our closed forms give a concrete starting point.

The first concerns how the welfare-maximising architecture depends on the composition of the population and on the primitives. Within the class of block architectures, the maximiser varies with both. In the configurations of Table~\ref{tab:optimal_by_composition}, it withdraws the informational influence of the minority progressively as the minority shrinks.
We have not found a structural property of the maximiser that holds across the parameter space. A natural candidate is that the maximiser never shows the posts of a type to its own members while hiding them from the other type. This property would tie welfare maximisation to the absence of echo chambers. It holds on every configuration of our grid and in all but one of $25{,}000$ further configurations drawn over wider ranges. However, this one configuration is an exact counterexample (Appendix~\ref{app:numerical}), so the property is not general. We also find that $G_{\max}$ dominates the complete network only when the minority is small in absolute size. Nevertheless, we do not know in general how large a minority must be before muting it ceases to pay. Proposition~\ref{prop:closed_forms} reduces the characterisation of the maximiser to the comparison of explicit polynomials. Therefore, the partition of the space of compositions and primitives into at most sixteen regions is a well-posed problem, which we leave for future research.

The second concerns architectures outside the class. The reduction to two representative users requires that users of a common type be informationally interchangeable, which fails as soon as a feed conditions on anything finer than the type of the author and the type of the reader. Over the unrestricted space of networks we have no method other than enumeration, which is feasible only for small $n$. Whether our conclusions hold there is an open question. Example~\ref{ex:equivalence_class} suggests that block architectures may understate the room for regulation. In that example, the platform ranks three architectures equally, but they differ in welfare and in polarisation. Moreover, two of them send the same post to two users of the same type and still differ in both criteria. Therefore, differences in information within a group may matter as much as differences between groups.

The third concerns measurement. Heterophilic density is computable from the architecture and the type distribution alone, which is what makes it usable by a regulator, but it records the type composition of links without recording where in the network a post lands. In Example~\ref{ex:equivalence_class}, the two architectures that send the same post to two users of the same type also share a value of the metric. However, they differ in welfare and in polarisation, because the two readers already observe different posts. The variation left unexplained within a level of the metric is therefore systematic rather than random, which points to a refinement weighting each link by the position of its recipient as well as by the types at its two ends. Identifying the weighting that recovers this variation, while preserving computability from observable structure, is a natural next step for work on regulatory metrics.

The fourth concerns the choice of welfare criterion. Throughout the paper, we use the utilitarian criterion. In Example~\ref{ex:welfare_fails}, the design that raises the sum makes every majority user better off and the minority user substantially worse off. In Example~\ref{ex:welfare_fails_het}, by contrast, every user gains. Proposition~\ref{prop:closed_forms} makes other criteria easy to apply, since it gives welfare as the pair of interim utilities and not only as their population-weighted sum. Therefore, a natural question is whether the failures documented here persist under a Rawlsian or an equality-weighted objective.
\bibliographystyle{apalike} 
\bibliography{biblio}

@article{hagenbach2010strategic,
  title     = {Strategic communication networks},
  author    = {Hagenbach, Jeanne and Koessler, Fr{\'e}d{\'e}ric},
  journal   = {The Review of Economic Studies},
  volume    = {77},
  number    = {3},
  pages     = {1072--1099},
  year      = {2010},
  publisher = {Wiley-Blackwell}
}

@article{calvo2009information,
  title     = {Information gathering in organizations: Equilibrium, welfare, and optimal network structure},
  author    = {Calv{\'o}-Armengol, Antoni and de Mart{\'\i}, Joan},
  journal   = {Journal of the European Economic Association},
  volume    = {7},
  number    = {1},
  pages     = {116--161},
  year      = {2009},
  publisher = {Oxford University Press}
}

@book{marschak1972economic,
  title     = {Economic Theory of Teams},
  author    = {Marschak, J. and Radner, R.},
  isbn      = {9780835781077},
  lccn      = {78099832},
  series    = {Cowles Foundation for Research in Economics at Yale University. Monograph Series},
  url       = {https://books.google.es/books?id=FIAongEACAAJ},
  year      = {1972},
  publisher = {Yale University Press}
}

@techreport{taneva2022organized,
  title       = {Organized Information Transmission},
  author      = {Taneva, Ina and Mathevet, Laurent},
  year        = {2022},
  institution = {CEPR Discussion Papers}
}

@article{acemoglu2024model,
  title     = {A model of online misinformation},
  author    = {Acemoglu, Daron and Ozdaglar, Asuman and Siderius, James},
  journal   = {Review of Economic Studies},
  volume    = {91},
  number    = {6},
  pages     = {3117--3150},
  year      = {2024},
  publisher = {Oxford University Press UK}
}

@article{mostagir2023should,
  title={When should platforms break echo chambers?},
  author={Mostagir, Mohamed and Siderius, James},
  journal={University of Michigan and Massachusetts Institute of Technology Working Paper, Massachusetts Institute of Technology, Cambridge, MA},
  year={2023}
}

@article{aridor2024economics,
  author  = {Aridor, Guy and Jiménez-Durán, Rafael and Levy, Ro'ee and Song, Lena},
  title   = {The Economics of Social Media},
  journal = {Journal of Economic Literature},
  volume  = {62},
  number  = {4},
  year    = {2024},
  month   = {December},
  pages   = {1422–74},
  doi     = {10.1257/jel.20241743},
  url     = {https://www.aeaweb.org/articles?id=10.1257/jel.20241743}
}

@misc{kemp2024digital,
  title  = {Digital 2024: Global Overview Report. {A}ccessed {D}ecember 31, 2023},
  author = {Kemp, Simon},
  year   = {2024}
}

@article{guess2023social,
  author   = {Andrew M. Guess  and Neil Malhotra  and Jennifer Pan  and Pablo Barberá  and Hunt Allcott  and Taylor Brown  and Adriana Crespo-Tenorio  and Drew Dimmery  and Deen Freelon  and Matthew Gentzkow  and Sandra González-Bailón  and Edward Kennedy  and Young Mie Kim  and David Lazer  and Devra Moehler  and Brendan Nyhan  and Carlos Velasco Rivera  and Jaime Settle  and Daniel Robert Thomas  and Emily Thorson  and Rebekah Tromble  and Arjun Wilkins  and Magdalena Wojcieszak  and Beixian Xiong  and Chad Kiewiet de Jonge  and Annie Franco  and Winter Mason  and Natalie Jomini Stroud  and Joshua A. Tucker },
  title    = {How do social media feed algorithms affect attitudes and behavior in an election campaign?},
  journal  = {Science},
  volume   = {381},
  number   = {6656},
  pages    = {398-404},
  year     = {2023},
  doi      = {10.1126/science.abp9364},
  url      = {https://www.science.org/doi/abs/10.1126/science.abp9364},
  eprint   = {https://www.science.org/doi/pdf/10.1126/science.abp9364}
}

@techreport{meta2012,
  author      = {{Facebook Inc.}},
  title       = {Facebook Annual Report 2012},
  institution = {Facebook Inc.},
  year        = 2012
}

@techreport{meta2023,
  author      = {{Meta Platforms, Inc.}},
  title       = {Meta Reports Second Quarter 2023 Results},
  institution = {Meta Platforms, Inc.},
  year        = 2023
}

@article{galeotti2013strategic,
  title={Strategic information transmission networks},
  author={Galeotti, Andrea and Ghiglino, Christian and Squintani, Francesco},
  journal={Journal of Economic Theory},
  volume={148},
  number={5},
  pages={1751--1769},
  year={2013},
  publisher={Elsevier}
}

@techreport{risco2024feed,
  title={Feed for good? {O}n the effects of personalization algorithms in social platforms},
  author={Risco, Miguel and Lleonart-Anguix, Manuel},
  year={2024},
  institution={University of Bonn and University of Mannheim, Germany}
}

@article{morris2002social,
  title={Social value of public information},
  author={Morris, Stephen and Shin, Hyun Song},
  journal={American Economic Review},
  volume={92},
  number={5},
  pages={1521--1534},
  year={2002},
  publisher={American Economic Association}
}

@article{tornberg2022digital,
  title={How digital media drive affective polarization through partisan sorting},
  author={T{\"o}rnberg, Petter},
  journal={Proceedings of the National Academy of Sciences},
  volume={119},
  number={42},
  pages={e2207159119},
  year={2022},
  publisher={National Academy of Sciences}
}

@article{boxell2017greater,
  title={Greater {I}nternet use is not associated with faster growth in political polarization among {US} demographic groups},
  author={Boxell, Levi and Gentzkow, Matthew and Shapiro, Jesse M},
  journal={Proceedings of the National Academy of Sciences},
  volume={114},
  number={40},
  pages={10612--10617},
  year={2017},
  publisher={National Academy of Sciences}
}

@article{hagenbach2016full,
title = {Full disclosure in decentralized organizations},
journal = {Economics Letters},
volume = {139},
pages = {5-7},
year = {2016},
issn = {0165-1765},
doi = {https://doi.org/10.1016/j.econlet.2015.12.002},
url = {https://www.sciencedirect.com/science/article/pii/S0165176515005054},
author = {Jeanne Hagenbach and Frédéric Koessler}
}

@article{currarini2009economic,
  title={An economic model of friendship: Homophily, minorities, and segregation},
  author={Currarini, Sergio and Jackson, Matthew O and Pin, Paolo},
  journal={Econometrica},
  volume={77},
  number={4},
  pages={1003--1045},
  year={2009},
  publisher={Wiley Online Library}
}

@article{demarzo2003persuasion,
  title={Persuasion bias, social influence, and unidimensional opinions},
  author={DeMarzo, Peter M and Vayanos, Dimitri and Zwiebel, Jeffrey},
  journal={The Quarterly Journal of Economics},
  volume={118},
  number={3},
  pages={909--968},
  year={2003},
  publisher={MIT Press}
}

@article{bakshy2015exposure,
  title={Exposure to ideologically diverse news and opinion on {F}acebook},
  author={Bakshy, Eytan and Messing, Solomon and Adamic, Lada A.},
  journal={Science},
  volume={348},
  number={6239},
  pages={1130--1132},
  year={2015},
  publisher={American Association for the Advancement of Science},
  doi={10.1126/science.aaa1160},
  url={https://www.science.org/doi/abs/10.1126/science.aaa1160},
  eprint={https://www.science.org/doi/pdf/10.1126/science.aaa1160}
}

@article{golub2012homophily,
  title={How homophily affects the speed of learning and best-response dynamics},
  author={Golub, Benjamin and Jackson, Matthew O},
  journal={The Quarterly Journal of Economics},
  volume={127},
  number={3},
  pages={1287--1338},
  year={2012},
  publisher={MIT Press}
}

@article{gauthier2025twitter,
  title={The Political Effect of {X}'s Recommender Algorithm},
  author={Gauthier, Germain and Hodler, Roland and Widmer, Philine and Zhuravskaya, Ekaterina},
  journal={Nature},
  year={2026},
  pubstate={forthcoming}
}

@article{kamenica2011persuasion,
Author = {Kamenica, Emir and Gentzkow, Matthew},
Title = {Bayesian Persuasion},
Journal = {American Economic Review},
Volume = {101},
Number = {6},
Year = {2011},
Month = {October},
Pages = {2590–2615},
DOI = {10.1257/aer.101.6.2590},
URL = {https://www.aeaweb.org/articles?id=10.1257/aer.101.6.2590}}

@article{angeletos2007efficient,
  title={Efficient use of information and social value of information},
  author={Angeletos, George-Marios and Pavan, Alessandro},
  journal={Econometrica},
  volume={75},
  number={4},
  pages={1103--1142},
  year={2007},
  publisher={Wiley Online Library}
}

@article{alonso2008does,
  title={When does coordination require centralization?},
  author={Alonso, Ricardo and Dessein, Wouter and Matouschek, Niko},
  journal={American Economic Review},
  volume={98},
  number={1},
  pages={145--179},
  year={2008},
  publisher={American Economic Association}
}

@article{feddersen2006theory,
  title={A theory of participation in elections},
  author={Feddersen, Timothy and Sandroni, Alvaro},
  journal={American Economic Review},
  volume={96},
  number={4},
  pages={1271--1282},
  year={2006},
  publisher={American Economic Association}
}

@article{barbera2015tweeting,
  title={Tweeting from left to right: Is online political communication more than an echo chamber?},
  author={Barber{\'a}, Pablo and Jost, John T and Nagler, Jonathan and Tucker, Joshua A and Bonneau, Richard},
  journal={Psychological Science},
  volume={26},
  number={10},
  pages={1531--1542},
  year={2015},
  publisher={Sage Publications Sage CA: Los Angeles, CA}
}

@article{carlsson1993global,
  title={Global games and equilibrium selection},
  author={Carlsson, Hans and Van Damme, Eric},
  journal={Econometrica: Journal of the Econometric Society},
  pages={989--1018},
  year={1993},
  publisher={JSTOR}
}

@article{lorenz2023systematic,
  title={A systematic review of worldwide causal and correlational evidence on digital media and democracy},
  author={Lorenz-Spreen, Philipp and Oswald, Lisa and Lewandowsky, Stephan and Hertwig, Ralph},
  journal={Nature Human Behaviour},
  volume={7},
  number={1},
  pages={74--101},
  year={2023},
  publisher={Nature Publishing Group UK London}
}

@article{muchnik2013social,
  title={Social influence bias: A randomized experiment},
  author={Muchnik, Lev and Aral, Sinan and Taylor, Sean J},
  journal={Science},
  volume={341},
  number={6146},
  pages={647--651},
  year={2013},
  publisher={American Association for the Advancement of Science}
}

@article{golub2010naive,
  title={Naive learning in social networks and the wisdom of crowds},
  author={Golub, Benjamin and Jackson, Matthew O},
  journal={American Economic Journal: Microeconomics},
  volume={2},
  number={1},
  pages={112--149},
  year={2010},
  publisher={American Economic Association}
}

@article{levin2004competitiveness,
  title={The competitiveness of joint bidding in multi-unit uniform-price auctions},
  author={Levin, Dan},
  journal={RAND Journal of Economics},
  pages={373--385},
  year={2004},
  publisher={JSTOR}
}

@article{mares2003private,
  title={Private information revelation in common-value auctions},
  author={Mares, Vlad and Harstad, Ronald M},
  journal={Journal of Economic Theory},
  volume={109},
  number={2},
  pages={264--282},
  year={2003},
  publisher={Elsevier}
}

@article{wolinsky2002eliciting,
  title={Eliciting information from multiple experts},
  author={Wolinsky, Asher},
  journal={Games and Economic Behavior},
  volume={41},
  number={1},
  pages={141--160},
  year={2002},
  publisher={Elsevier}
}

@article{bergemann2017firstprice,
  author  = {Bergemann, Dirk and Brooks, Benjamin and Morris, Stephen},
  title   = {First-Price Auctions with General Information Structures:
             Implications for Bidding and Revenue},
  journal = {Econometrica},
  year    = {2017},
  volume  = {85},
  number  = {1},
  pages   = {107--143}
}

@article{dessein2016rational,
  author  = {Dessein, Wouter and Galeotti, Andrea and Santos, Tano},
  title   = {Rational Inattention and Organizational Focus},
  journal = {American Economic Review},
  year    = {2016},
  volume  = {106},
  number  = {6},
  pages   = {1522--1536}
}

@article{battaglini2018influencing,
  author  = {Battaglini, Marco and Patacchini, Eleonora},
  title   = {Influencing Connected Legislators},
  journal = {Journal of Political Economy},
  year    = {2018},
  volume  = {126},
  number  = {6},
  pages   = {2677--2729}
}

@article{bergemann2013robust,
  author  = {Bergemann, Dirk and Morris, Stephen},
  title   = {Robust Predictions in Games With Incomplete Information},
  journal = {Econometrica},
  year    = {2013},
  volume  = {81},
  number  = {4},
  pages   = {1251--1308}
}

@article{taneva2019information,
  author  = {Taneva, Ina},
  title   = {Information Design},
  journal = {American Economic Journal: Microeconomics},
  year    = {2019},
  volume  = {11},
  number  = {4},
  pages   = {151--185}
}

@article{mathevet2020information,
  author  = {Mathevet, Laurent and Perego, Jacopo and Taneva, Ina},
  title   = {On Information Design in Games},
  journal = {Journal of Political Economy},
  year    = {2020},
  volume  = {128},
  number  = {4},
  pages   = {1370--1404}
}

@article{bergemann2016bayes,
  author  = {Bergemann, Dirk and Morris, Stephen},
  title   = {Bayes Correlated Equilibrium and the Comparison of Information Structures in Games},
  journal = {Theoretical Economics},
  year    = {2016},
  volume  = {11},
  number  = {2},
  pages   = {487--522}
}

@article{galperti2025games,
  author  = {Galperti, Simone and Perego, Jacopo},
  title   = {Games with Information Constraints: Seeds and Spillovers},
  journal = {Theoretical Economics},
  year    = {2025},
  volume  = {20},
  number  = {2},
  pages   = {667--711}
}

@article{candogan2022persuasion,
  author  = {Candogan, Ozan},
  title   = {Persuasion in Networks: Public Signals and Cores},
  journal = {Operations Research},
  year    = {2022},
  volume  = {70},
  number  = {4},
  pages   = {2264--2298}
}

@article{jehiel2005analogy,
  title   = {Analogy-based expectation equilibrium},
  author  = {Jehiel, Philippe},
  journal = {Journal of Economic Theory},
  volume  = {123},
  number  = {2},
  pages   = {81--104},
  year    = {2005}
}

@article{jehiel2008revisiting,
  title   = {Revisiting games of incomplete information with analogy-based expectations},
  author  = {Jehiel, Philippe and Koessler, Fr{\'e}d{\'e}ric},
  journal = {Games and Economic Behavior},
  volume  = {62},
  number  = {2},
  pages   = {533--557},
  year    = {2008}
}
\appendix

\section{Supplementary Materials for Section~\ref{sec:optimal_action}}\label{app:optimal_action}
\subsection{Existence and Uniqueness of the Equilibrium}\label{app:equilibrium_existence}

Consider the n-player potential game with potential function given by 
\begin{equation}\label{eq:utility_potential}
    u(\mathbf{a};\mathbf{s},\mathbf{t}) = -(1-\alpha) \sum\limits_{i\in N} \left[a_i-(\beta_1\theta(\mathbf{s})+\beta_2 t_i)\right]^2-\frac{\alpha}{2(n-1)}\sum\limits_{i\in N}\sum\limits_{j\neq i}\left(a_i-a_j\right)^2
\end{equation}

This game has the same best response functions for every player $i$ as the game described in Section~\ref{sec:optimal_action}. Therefore, the set of equilibrium strategies is the same. We know that this potential game has an equilibrium, and it is unique and linear thanks to \cite[Theorem 4]{marschak1972economic}.
To formally prove it, we need to rewrite Equation~\eqref{eq:utility_potential} as Equation~\eqref{eq:utility_potential_MR} to be able to apply the result from \cite{marschak1972economic}. 

\begin{equation}\label{eq:utility_potential_MR}
    u(\mathbf{a};\mathbf{s},\mathbf{t})={\displaystyle \tilde\lambda(\mathbf{s}) + 2 \sum\limits_{i\in N} \tilde\mu_i(\mathbf{s}) a_i - \sum\limits_{i,j \in N} v_{ij}(\mathbf{s}) a_ia_j} 
\end{equation}

where we write $\tilde\lambda$ and $\tilde\mu_i$ for the functions that \cite{marschak1972economic} denote $\lambda$ and $\mu_i$, to distinguish them from the signal parameters $\lambda$ and $\mu$,
\begin{align*}
    \tilde\lambda(\mathbf{s}) &= -(1-\alpha)\sum\limits_{i\in N}(\beta_1\theta(\mathbf{s})+\beta_2 t_i)^2\\
    \tilde\mu_i(\mathbf{s})&= (1-\alpha)(\beta_1\theta(\mathbf{s})+\beta_2 t_i)\\ 
    v_{ij}(\mathbf{s})&=\begin{cases} 1 &\text{ if } i=j \\ -\frac{\alpha}{n-1} &\text{ if } i\neq j\end{cases}
\end{align*}

This rewriting process follows from the following algebraic manipulation.

\begingroup
\allowdisplaybreaks
\begin{align*}
    u(\mathbf{a};\mathbf{s},\mathbf{t}) &= {\displaystyle-(1-\alpha) \sum\limits_{i\in N} \left[a_i-(\beta_1\theta(\mathbf{s})+\beta_2 t_i)\right]^2-\frac{\alpha}{2(n-1)}\sum\limits_{i\in N}\sum\limits_{j\neq i}\left(a_i-a_j\right)^2 }\\
    &= {\displaystyle -(1-\alpha) \sum\limits_{i\in N} \left[a_i^2-2a_i(\beta_1\theta(\mathbf{s})+\beta_2 t_i)+(\beta_1\theta(\mathbf{s})+\beta_2 t_i)^2\right]\nonumber}\\
    &{\displaystyle\hspace*{1cm} -\frac{\alpha}{2(n-1)}\sum\limits_{i\in N}\sum\limits_{j\neq i}\left[a_i^2-2a_ia_j+a_j^2\right]}\\
    &= {\displaystyle-(1-\alpha)\sum\limits_{i\in N} a_i^2 - \frac{\alpha}{2}\sum\limits_{i\in N} a_i^2 - \frac{\alpha}{2}\sum\limits_{j\in N} a_j^2 + \frac{\alpha}{(n-1)}\sum\limits_{i\in N}\sum\limits_{j\neq i} a_ia_j}\\
    &{\displaystyle\hspace*{1cm} + (1-\alpha) \sum\limits_{i\in N}2a_i(\beta_1\theta(\mathbf{s})+\beta_2 t_i)- (1-\alpha)\sum\limits_{i\in N}(\beta_1\theta(\mathbf{s})+\beta_2 t_i)^2}\\
    &= {\displaystyle-\sum\limits_{i\in N} a_i^2 + \frac{\alpha}{(n-1)}\sum\limits_{i\in N}\sum\limits_{j\neq i} a_ia_j}\\
    &{\displaystyle\hspace*{1cm} + (1-\alpha) \sum\limits_{i\in N}2a_i(\beta_1\theta(\mathbf{s})+\beta_2 t_i)- (1-\alpha)\sum\limits_{i\in N}(\beta_1\theta(\mathbf{s})+\beta_2 t_i)^2}\\
    &={\displaystyle \tilde\lambda(\mathbf{s}) + 2 \sum\limits_{i\in N} \tilde\mu_i(\mathbf{s}) a_i - \sum\limits_{i,j \in N} v_{ij}(\mathbf{s}) a_ia_j}
\end{align*}
\endgroup

Let us recall that $I_i$ is the information that user $i$ holds regarding the state of nature $\theta(\mathbf{s})$. Then, Theorem~4 in \cite{marschak1972economic} states that if the utility function is given by Equation~\eqref{eq:utility_potential}, the optimal decision functions $\{a_i(I_i)\}_{i\in N}$ are uniquely determined by the following system of linear equations: 
\begin{equation}\label{eq:utility_potential_MR_BR}
    a_i(I_i)+ \sum\limits_{j\neq i} \mathbb{E}[a_j v_{ij}(\mathbf{s})|I_i] = \mathbb{E}[\tilde\mu_i(\mathbf{s})|I_i], \;\text{ for all } i\in N 
\end{equation}

It is worth mentioning that although this proves the existence and uniqueness of the equilibrium strategies, it does not provide their characterisation. The system in Equation~\eqref{eq:utility_potential_MR_BR} has one equation for each user $i$ and each realisation of her information $(t_i,(s_k)_{k\in I_i})$. It therefore has $\sum_{i\in N}2^{|I_i|+1}\le n2^{n+1}$ equations and the same number of unknowns, that is, $\mathcal{O}(n2^n)$. A direct solution by Gaussian elimination requires $\mathcal{O}(n^3 2^{3n})$ operations, and it must be repeated for each network.

\subsection{Proof of Theorem~\ref{thm:users_equilibrium}}\label{app:equilibrium_independent_gamma}

By Appendix~\ref{app:equilibrium_existence}, the game has a unique equilibrium, and it is linear. It is therefore enough to exhibit one profile of strategies that is a fixed point of the best responses. We proceed in four steps. First, we split the problem into a part for the posts and a part for the types. Second, we obtain the part for the posts from \cite{hagenbach2010strategic}. Third, we propose a functional form with generic coefficients for the part for the types, and we show that it reproduces itself under best responses for every $n$ and every network. Fourth, we solve for the coefficients. Throughout, the network $G$ is taken as given. Remark~\ref{rem:network_knowledge} at the end of this subsection shows that users do not need to know it.

\medskip
\noindent\textbf{Step 1: Two separate problems.} User $j$ chooses $a_j$ to maximise her expected utility
\[
    \mathbb{E}_j[u_j(\mathbf{a};\mathbf{s},t_j)]=-(1-\alpha)\mathbb{E}_j\Bigl[\bigl(a_j-\beta_1\theta(\mathbf{s})-\beta_2 t_j\bigr)^2\Bigr]-\frac{\alpha}{n-1}\sum_{i\neq j}\mathbb{E}_j\bigl[(a_j-a_i)^2\bigr].
\]
This objective is strictly concave in $a_j$, and its first-order condition is
\[
    -2(1-\alpha)\bigl(a_j-\beta_1\mathbb{E}_j[\theta(\mathbf{s})]-\beta_2 t_j\bigr)-\frac{2\alpha}{n-1}\sum_{i\neq j}\bigl(a_j-\mathbb{E}_j[a_i]\bigr)=0 .
\]
The coefficient of $a_j$ is $-2\bigl[(1-\alpha)+\alpha\bigr]=-2$, so the best response of user $j$ is Equation~\eqref{eq:best_response_expected},
\[
    a_j=(1-\alpha)\beta_1\mathbb{E}_j[\theta(\mathbf{s})]+(1-\alpha)\beta_2 t_j+\frac{\alpha}{n-1}\sum_{i\neq j}\mathbb{E}_j[a_i].
\]
Consider two auxiliary profiles. The profile $(a_i^{S})_{i\in N}$ solves the same system with $(1-\alpha)\mathbb{E}_j[\theta(\mathbf{s})]$ in place of the first two terms, and the profile $(a_i^{T})_{i\in N}$ solves it with $(1-\alpha)t_j$ in place of them. The operator $\mathbb{E}_j[\cdot]$ is linear, so $a_i=\beta_1 a_i^{S}+\beta_2 a_i^{T}$ solves the original system. We characterise $\mathbf{a}^{S}$ and $\mathbf{a}^{T}$ separately.

\medskip
\noindent\textbf{Step 2: The part for the posts.} The profile $\mathbf{a}^{S}$ is the equilibrium of the second stage of \cite{hagenbach2010strategic} without biases. Their state is the sum of the players' signals, and player $i$ observes the signals of the players in $I_i$, as here. Their derivation uses that a player who does not observe a signal expects it at its unconditional mean. This holds in our model. Types are independent across users and posts are independent conditionally on types, so $s_k$ is independent of $(t_j,(s_l)_{l\neq k})$ for every $j\neq k$, and $\mathbb{E}_j[s_k]=\mathbb{E}[s_k]$ whenever $k\in\overline{I}_j$. Their verification \citep[Appendix A.1]{hagenbach2010strategic} then applies without change, and
\[
    a_i^{S}=\sum_{k\in I_i}\frac{\alpha(n-1-r_k)\mathbb{E}[s_k]+(1-\alpha)(n-1)s_k}{n-1-\alpha r_k}+\sum_{k\in\overline{I}_i}\mathbb{E}[s_k].
\]

\medskip
\noindent\textbf{Step 3: The part for the types.} The profile $\mathbf{a}^{T}$ is new. In \cite{hagenbach2010strategic}, the private information of each player is a signal about the state, and preferences differ through biases that are common knowledge. Our types play the role of these biases, but they are private information, so their model has no counterpart of $\mathbf{a}^{T}$. We start from a linear form with generic coefficients, which distinguishes every category of information that user $i$ holds about others' types:\footnote{Equation~\eqref{eq:general_eq} separates the terms with $k=i$ from the sums, so that a user's own post can receive a different weight from the posts of others. The coefficients of Equations~\eqref{eq:general_action_complete} and~\eqref{eq:general_eq} are related by $\gamma_1^i=\omega_5^i+\omega_7^i$, $\gamma_2^i=\omega_4^i+\omega_8^i$, $\gamma_3=\omega_9^i$, $\gamma_4^k=\omega_5^k$, $\gamma_5^k=\omega_6^k$ and $\gamma_6^k=\omega_4^k$. The weight on $t_i$ carries no superscript. No other user observes $t_i$, so $t_i$ enters $a_i$ only through the term $(1-\alpha)\beta_2 t_i$ of Equation~\eqref{eq:best_response_expected}, which is the same for every $i$.}
\begin{equation}\label{eq:general_eq}
    a_i^{T} = \gamma_1^i\mathbb{E}[t_i|s_i]+\gamma_2^i\mathbb{E}[t_i] + \gamma_3 t_i + \sum\limits_{\substack{k\in I_i\\ k\neq i}}\gamma_4^k\mathbb{E}[t_k|s_k]+ \sum\limits_{k\in \overline{I}_i}\gamma_5^k\mathbb{E}[t_k]+ \sum\limits_{\substack{k\in I_i\\ k\neq i}}\gamma_6^k\mathbb{E}[t_k].
\end{equation}
Each term reflects one way in which user $i$ processes information about types, and each post plays the dual role described in Section~\ref{sec:optimal_action}. Consider first the post of another user $k$. If user $i$ does not observe it, she relies on the prior $\mathbb{E}[t_k]$, with weight $\gamma_5^k$. If she observes it, she updates her belief about the type of $k$ to the posterior $\mathbb{E}[t_k|s_k]$. She also keeps a weight on the prior, since some of the users she coordinates with do not observe the post of $k$ and rely on the prior. The weights $\gamma_4^k$ and $\gamma_6^k$ capture this trade-off between the posterior and the prior. Consider next the type of user $i$ herself. She knows $t_i$, and $\gamma_3$ reflects the effect of her own type on her action. However, her action also depends on the posterior $\mathbb{E}[t_i|s_i]$ and on the prior $\mathbb{E}[t_i]$. The users who observe her post $s_i$ use it to update their beliefs about her type, the other users rely on the prior, and user $i$ coordinates with both groups. The weights $\gamma_1^i$ and $\gamma_2^i$ capture how she anticipates these beliefs about herself. The coefficients of the terms about user $i$ carry the index $i$, and those of the terms about another user $k$ carry the index $k$.

We now derive the conditions on the coefficients under which this form reproduces itself under best responses, for every $n$ and every network. Fix a user $j$ and suppose that every $i\neq j$ plays Equation~\eqref{eq:general_eq}. By Step~1, the best response of $j$ for the part for the types is
\begin{equation}\label{eq:best_response_expected_simplified}
    a_j^{T}=(1-\alpha)t_j+\frac{\alpha}{n-1}\sum\limits_{i\neq j} \mathbb{E}_j[a_i^{T}].
\end{equation}
To equate this best response with the functional form of Equation~\eqref{eq:general_eq}, we need to compute $\mathbb{E}_j[a_i^{T}]$ for every $i\neq j$, and then their sum. User $j$ knows her type $t_j$ and, for every $k\in I_j$, the post $s_k$ and its reach $r_k$. Types are independent across users, and posts are independent conditionally on types. Therefore, the only information that $j$ holds about the type of another user $k$ is the post $s_k$, when she observes it, and for every $k\neq j$,
\begin{equation}\label{eq:Ej_types}
    \mathbb{E}_j\bigl[\mathbb{E}[t_k|s_k]\bigr]=\mathbb{E}_j[t_k]=\begin{cases} \mathbb{E}[t_k|s_k] & \text{if } k\in I_j,\\ \mathbb{E}[t_k] & \text{if } k\in \overline{I}_j.\end{cases}
\end{equation}
We apply Equation~\eqref{eq:Ej_types} to each term of Equation~\eqref{eq:general_eq}, including the terms about user $i$ herself, and we obtain
\begingroup
\allowdisplaybreaks
\begin{equation}\label{eq:Ej_ai}
\begin{split}
    \mathbb{E}_j[a_i^{T}] =\ &\gamma_1^i\left[\mathbb{E}[t_i|s_i]\mathds{1}_{i \in I_j}+ \mathbb{E}[t_i]\mathds{1}_{i \in \overline{I}_j}\right] + \gamma_2^i\mathbb{E}[t_i] + \gamma_3\left[\mathbb{E}[t_i|s_i]\mathds{1}_{i \in I_j}+ \mathbb{E}[t_i]\mathds{1}_{i \in \overline{I}_j}\right] \\
       &   +\sum\limits_{\substack{k\in I_i\cap I_j\\ k\neq i}}\gamma_4^k\mathbb{E}[t_k|s_k] + \sum\limits_{\substack{k\in I_i\cap \overline{I}_j\\ k\neq i}}\gamma_4^k\mathbb{E}[t_k] + \sum\limits_{k\in \overline{I}_i}\gamma_5^k\mathbb{E}[t_k]+ \sum\limits_{\substack{k\in I_i\\ k\neq i}}\gamma_6^k\mathbb{E}[t_k].
\end{split}
\end{equation}
\endgroup
Next, we sum Equation~\eqref{eq:Ej_ai} over all users $i\neq j$. Summing the indicator functions $\mathds{1}_{i\in I_j}$ over $i\neq j$ gives the set $I_j$ without $j$, so that
\begingroup
\allowdisplaybreaks
\begin{equation}\label{eq:expected_action_sum_pairs}
\begin{split}
    \sum\limits_{i\neq j}\mathbb{E}_j[a_i^{T}] &=
    \sum\limits_{\substack{i\in I_j\\ i\neq j}}\gamma_1^i\mathbb{E}[t_i|s_i]+
    \sum\limits_{i\in \overline{I}_j}\gamma_1^i\mathbb{E}[t_i]+
    \sum\limits_{i\neq j}\gamma_2^i\mathbb{E}[t_i] +
    \gamma_3\left[\sum\limits_{\substack{i\in I_j\\ i\neq j}}\mathbb{E}[t_i|s_i]+
    \sum\limits_{i\in \overline{I}_j}\mathbb{E}[t_i]\right] \\
    & \hspace*{0.5cm} +\sum\limits_{i\neq j}\sum\limits_{\substack{k\in I_i\cap I_j\\ k\neq i}}\gamma_4^k\mathbb{E}[t_k|s_k]
    + \sum\limits_{i\neq j}\sum\limits_{\substack{k\in I_i\cap \overline{I}_j\\ k\neq i}}\gamma_4^k\mathbb{E}[t_k] \\
    & \hspace*{0.5cm} + \sum\limits_{i\neq j}\sum\limits_{k\in \overline{I}_i}\gamma_5^k\mathbb{E}[t_k] +\sum\limits_{i\neq j}\sum\limits_{\substack{k\in I_i\\ k\neq i}}\gamma_6^k\mathbb{E}[t_k].
\end{split}
\end{equation}
\endgroup
The first four terms of Equation~\eqref{eq:expected_action_sum_pairs} come from the terms of each user $i$ about herself. The double sums come from the terms of each user $i$ about the other users $k$, and they run over pairs $(i,k)$. To compare Equation~\eqref{eq:expected_action_sum_pairs} with Equation~\eqref{eq:general_eq}, we regroup all its terms by the author $k$ of each post. For a given author $k$, the double sums count how many users $i\notin\{j,k\}$ observe $k$, which contribute through $\gamma_4^k$ and $\gamma_6^k$, and how many do not, which contribute through $\gamma_5^k$. This requires only the following counts:
\begin{itemize}
    \item User $j$ is observed by $r_j$ users other than $j$, and is not observed by $n-1-r_j$ of them.
    \item Let $k\in I_j$ with $k\neq j$. Among the users other than $j$ and $k$, exactly $r_k-1$ observe $k$, since $j$ is one of the $r_k$ observers of $k$, and $n-1-r_k$ do not.
    \item Let $k\in \overline{I}_j$. Among the users other than $j$ and $k$, exactly $r_k$ observe $k$, and $n-2-r_k$ do not, since $j$ is one of the users who do not observe $k$.
\end{itemize}
Together with the terms of user $k$ about herself, through $\gamma_1^k$, $\gamma_2^k$ and $\gamma_3$, these counts give
\begin{equation}\label{eq:expected_action_sum}
\begin{split}
    \sum_{i\neq j}\mathbb{E}_j[a_i^{T}] ={}& r_j\gamma_4^j\,\mathbb{E}[t_j|s_j]+\bigl[(n-1-r_j)\gamma_5^j+r_j\gamma_6^j\bigr]\mathbb{E}[t_j]\\
    &+\sum_{\substack{k\in I_j\\ k\neq j}}\bigl[\gamma_1^k+\gamma_3+(r_k-1)\gamma_4^k\bigr]\mathbb{E}[t_k|s_k]\\
    &+\sum_{\substack{k\in I_j\\ k\neq j}}\bigl[\gamma_2^k+(n-1-r_k)\gamma_5^k+(r_k-1)\gamma_6^k\bigr]\mathbb{E}[t_k]\\
    &+\sum_{k\in \overline{I}_j}\bigl[\gamma_1^k+\gamma_2^k+\gamma_3+r_k(\gamma_4^k+\gamma_6^k)+(n-2-r_k)\gamma_5^k\bigr]\mathbb{E}[t_k].
\end{split}
\end{equation}
Each line of Equation~\eqref{eq:expected_action_sum} collects the terms of one author.
\begin{itemize}
    \item The first line collects the terms about user $j$. The posterior $\mathbb{E}[t_j|s_j]$ comes from the $r_j$ users who observe $j$, each with weight $\gamma_4^j$. The prior $\mathbb{E}[t_j]$ comes from the $n-1-r_j$ users who do not observe $j$, each with weight $\gamma_5^j$, and from the $r_j$ users who observe $j$, each with weight $\gamma_6^j$.
    \item The second and third lines collect the terms about a user $k\in I_j$ with $k\neq j$. The posterior $\mathbb{E}[t_k|s_k]$ comes from user $k$ herself, with weight $\gamma_1^k+\gamma_3$ since $j$ observes $k$, and from the $r_k-1$ users $i\notin\{j,k\}$ who observe $k$, each with weight $\gamma_4^k$. The prior $\mathbb{E}[t_k]$ comes from user $k$ herself, with weight $\gamma_2^k$, from the $n-1-r_k$ users who do not observe $k$, each with weight $\gamma_5^k$, and from the $r_k-1$ users $i\notin\{j,k\}$ who observe $k$, each with weight $\gamma_6^k$.
    \item The last line collects the terms about a user $k\in\overline{I}_j$. User $j$ does not observe $k$, so by Equation~\eqref{eq:Ej_types} every term about $k$ becomes the prior $\mathbb{E}[t_k]$. It comes from user $k$ herself, with weight $\gamma_1^k+\gamma_2^k+\gamma_3$, from the $r_k$ users who observe $k$, each with weight $\gamma_4^k+\gamma_6^k$, and from the $n-2-r_k$ users $i\neq j$ who do not observe $k$, each with weight $\gamma_5^k$.
\end{itemize}

We substitute Equation~\eqref{eq:expected_action_sum} into Equation~\eqref{eq:best_response_expected_simplified}. The best response of $j$ is then a linear combination of $t_j$, $\mathbb{E}[t_j|s_j]$ and $\mathbb{E}[t_j]$, of $\mathbb{E}[t_k|s_k]$ for $k\in I_j\setminus\{j\}$, and of $\mathbb{E}[t_k]$ for every $k\neq j$. Equation~\eqref{eq:general_eq} for user $j$ is a linear combination of the same terms. We require the two expressions to coincide term by term.

Consider first the terms about user $j$ herself. In Equation~\eqref{eq:general_eq}, their coefficients are $\gamma_1^j$, $\gamma_2^j$ and $\gamma_3$. In the best response, they are the coefficients of $\mathbb{E}[t_j|s_j]$ and $\mathbb{E}[t_j]$ in the first line of Equation~\eqref{eq:expected_action_sum}, multiplied by $\alpha/(n-1)$, and the coefficient $1-\alpha$ of $t_j$. In Equation~\eqref{eq:general_eq}, the weight that a user places on the post of another user depends only on the author of the post. The conditions below are therefore conditions on the coefficients of each author. The user $j$ is arbitrary, so we write the resulting conditions for a generic author $k$:
\begin{align}
    \gamma_1^k =& \frac{\alpha}{n-1} r_k \gamma_4^k\label{eq:gamma1}\\
    \gamma_2^k =& \frac{\alpha}{n-1} \left[(n-1-r_k)\gamma_5^k+r_k\gamma_6^k\right] \label{eq:gamma2}\\
    \gamma_3 =& 1-\alpha\label{eq:gamma3}
\end{align}
Equations~\eqref{eq:gamma1} and~\eqref{eq:gamma2} make the role of the own post explicit. User $j$ weighs the posterior about her own type because $r_j$ other users observe her post, and each of them weighs it by $\gamma_4^j$. She weighs the prior about her own type because $n-1-r_j$ users do not observe her post and rely on the prior, and because the $r_j$ users who observe it also keep a weight $\gamma_6^j$ on the prior. Equation~\eqref{eq:gamma3} gives the weight on her own type. No other user observes $t_j$, so $t_j$ enters her action only through the consistency term $(1-\alpha)t_j$ of her best response, and its weight $1-\alpha$ is the same for every user.

Consider next the terms about another user $k$. For $k\in I_j\setminus\{j\}$, the coefficients of $\mathbb{E}[t_k|s_k]$ and $\mathbb{E}[t_k]$ are $\gamma_4^k$ and $\gamma_6^k$ in Equation~\eqref{eq:general_eq}, and they are given by the second and third lines of Equation~\eqref{eq:expected_action_sum} in the best response. For $k\in\overline{I}_j$, the coefficient of $\mathbb{E}[t_k]$ is $\gamma_5^k$ in Equation~\eqref{eq:general_eq}, and it is given by the last line of Equation~\eqref{eq:expected_action_sum}. Equating them gives
\begin{align*}
    \gamma_4^k &= \frac{\alpha}{n-1}\bigl[\gamma_1^k+\gamma_3+(r_k-1)\gamma_4^k\bigr],\\
    \gamma_5^k &= \frac{\alpha}{n-1}\bigl[\gamma_1^k+\gamma_2^k+\gamma_3+r_k(\gamma_4^k+\gamma_6^k)+(n-2-r_k)\gamma_5^k\bigr],\\
    \gamma_6^k &= \frac{\alpha}{n-1}\bigl[\gamma_2^k+(n-1-r_k)\gamma_5^k+(r_k-1)\gamma_6^k\bigr].
\end{align*}
Each coefficient appears on both sides of its equation. Solving each equation for it gives
\begin{align}
    \gamma_4^k =& \frac{\alpha\left(\gamma_1^k+\gamma_3\right)}{(n-1) - (r_k-1)\alpha}\label{eq:gamma4}\\
    \gamma_5^k =& \frac{\alpha\left(\gamma_1^k+\gamma_2^k+\gamma_3+ r_k\left(\gamma_4^k+\gamma_6^k\right)\right)}{(n-1)-\alpha(n-2-r_k)} \label{eq:gamma5}\\
    \gamma_6^k =&\frac{\alpha\left(\gamma_2^k+(n-1-r_k)\gamma_5^k\right)}{(n-1) - (r_k-1)\alpha}  \label{eq:gamma6}
\end{align}
Each condition involves $n$, $\alpha$ and the reach $r_k$ of a single author, and it does not depend on $j$ or on the rest of the network. The system \eqref{eq:gamma1}--\eqref{eq:gamma6} therefore has the same form for every user, every $n$ and every network, and the network enters only through the reach $r_k$ of each author. We assumed that every user $i\neq j$ plays Equation~\eqref{eq:general_eq}, and we have shown that, if the coefficients satisfy Equations~\eqref{eq:gamma1}--\eqref{eq:gamma6} for every author $k$, the best response of user $j$ also takes the form of Equation~\eqref{eq:general_eq}, with the same coefficients. The user $j$ is arbitrary. Therefore, when every user plays Equation~\eqref{eq:general_eq} with these coefficients, the action of each user is a best response to the actions of the others, and the profile is a Bayes--Nash equilibrium. The priors $\mathbb{E}[t_k]$ are all equal to the same constant, so matching them author by author is a sufficient condition. Any solution of the system gives a fixed point of the best responses, and this is enough for our purpose.

\medskip
\noindent\textbf{Step 4: Solution and conclusion.} Solving the system \eqref{eq:gamma1}--\eqref{eq:gamma6} gives its unique solution
\begin{align}
    \gamma_1^k =& \frac{\alpha^2 r_k (1-\alpha)}{(n-1+\alpha)(n-1-\alpha r_k)}\label{eq:sol_gamma1}\\
    \gamma_2^k =& \frac{\alpha^2 (n-1-r_k) }{(n-1+\alpha)(n-1-\alpha r_k)}\label{eq:sol_gamma2}\\
    \gamma_3 =& 1-\alpha\label{eq:sol_gamma3}\\
    \gamma_4^k =& \frac{\alpha (n-1) (1-\alpha)}{(n-1+\alpha)(n-1-\alpha r_k)}\label{eq:sol_gamma4}\\
    \gamma_5^k =& \frac{\alpha}{n-1+\alpha}\label{eq:sol_gamma5}\\
    \gamma_6^k =&\frac{\alpha^2 (n-1-r_k)}{(n-1+\alpha)(n-1-\alpha r_k)} \label{eq:sol_gamma6}
\end{align}
Substituting these coefficients into Equation~\eqref{eq:general_eq} gives the part of Equation~\eqref{eq:optimal_action_complete} that multiplies $\beta_2$. For $k\in I_i$ with $k\neq i$, the weights $\gamma_4^k$ and $\gamma_6^k$ on $\mathbb{E}[t_k|s_k]$ and $\mathbb{E}[t_k]$ are the terms of the first sum in Equation~\eqref{eq:optimal_action_complete}. For $k=i$, the weight $\gamma_1^i$ on $\mathbb{E}[t_i|s_i]$ equals the term of the same sum at $k=i$ plus the separate term $-\alpha(1-\alpha)\mathbb{E}[t_i|s_i]/(n-1+\alpha)$, and $\gamma_2^i$ is the weight on $\mathbb{E}[t_i]$ in that term. The weight $\gamma_5^k$ gives the second sum. By Step~1, $a_i^*=\beta_1 a_i^{S}+\beta_2 a_i^{T}$, which is Equation~\eqref{eq:optimal_action_complete}.

This characterisation reduces the size of the problem substantially. The system of Equation~\eqref{eq:utility_potential_MR_BR} in Appendix~\ref{app:equilibrium_existence} has one equation for each user and each realisation of her information, that is, $\sum_{i\in N}2^{|I_i|+1}\le n2^{n+1}$ equations for a given network, and it must be solved again for each network. In contrast, the system \eqref{eq:gamma1}--\eqref{eq:gamma6} has six equations. Its form is the same for every user and every network, and the network enters only through the reach $r_k$ of each author. We therefore solve it once, in closed form, as a function of $n$, $\alpha$ and $r_k$. The six equations concern only the part for the types. The same procedure applied to the part for the posts gives three more equations, one for each of the weights $\omega_1^k$ and $\omega_2^k$ on $s_k$ and $\mathbb{E}[s_k]$ when user $i$ observes the post of $k$, and one for the weight $\omega_3^k$ on $\mathbb{E}[s_k]$ when she does not. \cite{hagenbach2010strategic} solve these three equations, and we use their solution in Step~2. The equilibrium is therefore described by nine equations and the nine coefficients of Equation~\eqref{eq:general_action_complete}, six for the types and three for the posts, each a function of $n$, $\alpha$ and a single reach. It depends on the network only through the sets $I_i$ and the reaches $r_k$.

By Steps~1--3, the profile $\mathbf{a}^*$ is an equilibrium, and by Appendix~\ref{app:equilibrium_existence} it is unique, which completes the proof. \hfill$\square$

\begin{remark}[What users need to know about the network]\label{rem:network_knowledge}
The proof takes the network $G$ as given. For instance, in Step~3 the forecast $\mathbb{E}_j[a_i^{T}]$ uses the set $I_i$ and the reach of posts that user $j$ does not observe, although user $j$ observes only $t_j$ and $(s_k,r_k)_{k\in I_j}$. Nevertheless, Equation~\eqref{eq:optimal_action_complete} uses only this information, and it remains an equilibrium when users do not know $G$.

First, the best response \eqref{eq:best_response_expected} depends on the actions of the others only through the sum $\sum_{i\neq j}\mathbb{E}_j[a_i]$. By Equation~\eqref{eq:optimal_action_complete}, all users who observe post $k$ give it the same weight, which depends only on $r_k$. For $k\in I_j$, the number of users $i\neq j$ with $k\in I_i$ is $r_k$.\footnote{If $k\neq j$, these users are the $r_k-1$ readers of post $k$ other than $j$, and the author $k$. If $k=j$, they are the $r_j$ readers of post $j$.} Hence post $k$ contributes $r_k$ times the term of a reader and $n-1-r_k$ times the term of a non-reader, and it does not matter which users read it. For $k\in\overline{I}_j$, user $j$ expects $s_k$ at $\mathbb{E}[s_k]$, and $F(\mathbb{E}[s_k],\mathbb{E}[s_k],r_k)=\mathbb{E}[s_k]$ for every $r_k$ by Equation~\eqref{eq:informational_tradeoff}. The same holds for the types, with $\mathbb{E}[t_k]$. Finally, user $j$ forecasts the type $t_i$ of another user by $\mathbb{E}[t_i\mid s_i]$ if $i\in I_j$ and by $\mathbb{E}[t_i]$ otherwise. Therefore the sum depends on $G$ only through $I_j$ and $(r_k)_{k\in I_j}$, which user $j$ observes.

Second, suppose that each user holds a belief about the network, that is, a probability distribution over the possible networks, and that, given her own type, this belief does not depend on the types of the other users. Conditional on each network, the beliefs of user $j$ about $(\mathbf{t},\mathbf{s})$ are then the same as when $G$ is known. By the first point, the sum $\sum_{i\neq j}\mathbb{E}_j[a_i]$ takes the same value for every network consistent with what user $j$ observes. The best response is affine in this sum, so averaging over networks does not change it, and Equation~\eqref{eq:optimal_action_complete} remains a best response for every user. It describes coarse expectations about the platform, in the spirit of analogy-based expectations \citep{jehiel2005analogy, jehiel2008revisiting}: users bundle the type profiles on which the platform conditions, while they remain fully Bayesian about the other users. The condition fails under rational expectations. If users know how the platform maps type profiles into networks, then $I_j$ and $(r_k)_{k\in I_j}$ are informative about the types of others, including the readers of a post, since $r_k$ counts them, and the argument fails.
We discuss this assumption in Section~\ref{sec:assumptions}.
\end{remark}

\subsection{Properties of the Optimal Action}

\begin{proof}[Proof of Proposition~\ref{prop:properties_of_F}] Let $\Delta F=F(x,y,z+1)-F(x,y,z)$. Then, 
\begin{align*}
\Delta F &= \frac{\alpha (n-1-(z+1))y+(1-\alpha)(n-1)x}{n-1-\alpha (z+1)} - \frac{\alpha (n-1-z)y+(1-\alpha)(n-1)x}{n-1-\alpha z}\\
 &=\frac{\alpha\left((n-1-z)y+(1-\alpha)(n-1)x\right)}{(n-1-\alpha z) (n-1-\alpha (z+1))} - \frac{\alpha y }{n-1-\alpha (z+1)}\\
 &=\frac{\alpha\left((n-1-z)y+(1-\alpha)(n-1)x\right) - (n-1-\alpha z) \alpha y }{(n-1-\alpha z) (n-1-\alpha (z+1))}\\
 &=\frac{\alpha(n-1)(\alpha-1)y+\alpha(1-\alpha)(n-1)x }{(n-1-\alpha z) (n-1-\alpha (z+1))}\\
 &=\frac{\alpha(n-1)(1-\alpha)}{(n-1-\alpha z) (n-1-\alpha (z+1))}(x-y)\\
\end{align*}
Since $\alpha\in (0,1)$ the result follows. 
\end{proof}

\begin{proposition}[Properties of $\Delta$]\label{prop:properties_of_Delta} 
    Let $\Delta(x,y,z)$ be defined as  
\begin{equation}\label{eq:informational_tradeoff_access}
\Delta(x,y,z)=F(x,y,z)-y
\end{equation}
Then, $\operatorname{sgn}(\Delta)=\operatorname{sgn}(x-y)$. 
\end{proposition}
\begin{proof}[Proof of Proposition~\ref{prop:properties_of_Delta}] 
Let us simplify the expression for $\Delta(x,y,z)$
\begin{align*}
\Delta(x,y,z)&=F(x,y,z)-y\\
&= \frac{\alpha (n-1-z)y+(1-\alpha)(n-1)x}{n-1-\alpha z} - y \\
&= \frac{\alpha (n-1-z)y+(1-\alpha)(n-1)x- (n-1-\alpha z) y }{n-1-\alpha z}\\
&= \frac{y \left(\alpha (n-1)-\alpha z - (n-1) + \alpha z\right)+(1-\alpha)(n-1)x}{n-1-\alpha z}\\
&= \frac{(1-\alpha)(n-1)}{n-1-\alpha z}(x-y)
\end{align*}
Since $\alpha\in (0,1)$ and $n-1>\alpha z$ then the sign of $\Delta$ depends only on the sign of $(x-y)$. 
\end{proof}

\section{Supplementary Materials for Section~\ref{sec:manipulation}}\label{a_sec:manipulation}

\begin{proof}[Proof of Proposition~\ref{prop:properties_of_expected_F}]
From the proof of Proposition~\ref{prop:properties_of_F} we know that $$\Delta F =\frac{\alpha(n-1)(1-\alpha)}{(n-1-\alpha z) (n-1-\alpha (z+1))}(x-y)= C_z (x-y)$$
Since $C_z$ only depends on the parameters of the model $\alpha$ and $n$, and the number of people who receive the signals $z$ but not the signal then the results follows from the linearity of the expected value as  
$$\mathbb{E}_P[\Delta F] = C_z\left(\mathbb{E}_P[x]-\mathbb{E}_P[y]\right)$$
\end{proof}
\begin{proposition}[Properties of ${\mathbb{E}_P[\Delta]}$]\label{prop:properties_of_expected_Delta}
Let $\Delta(x,y,z)=F(x,y,z)-y$.  Then $\operatorname{sgn}(\mathbb{E}_P[\Delta])=\operatorname{sgn}(x-y)$.
\end{proposition}
\begin{proof}[Proof of Proposition~\ref{prop:properties_of_expected_Delta}]
From the proof of Proposition~\ref{prop:properties_of_Delta} we know that $$\Delta(x,y,z)= \frac{(1-\alpha)(n-1)}{n-1-\alpha z}(x-y) = C_z (x-y) $$
Since $C_z$ only depends on the parameters of the model $\alpha$ and $n$, and the number of people who receive the signals $z$ but not the signal then the results follows from the linearity of the expected value as  
$$\mathbb{E}_P[\Delta(x,y,z)] = C_z\left(\mathbb{E}_P[x]-\mathbb{E}_P[y]\right)$$
\end{proof}

\begin{proof}[Proof of Corollary~\ref{cor:uniques_max_min}]
    We know from Proposition~\ref{prop:properties_of_expected_Delta} that revealing the information of high-type users increase the expected action. Moreover, Proposition~\ref{prop:properties_of_expected_F} tells us that once we have revealed the information of a high-type user the expected action increases when that information is revealed to everyone. The platform can do this until there are no more high-type users. Hence the only network that maximises the expected collective action is $G_{\max}$. The result is analogous for the network $G_{\min}$ that minimises the expected collective action. 
\end{proof}

\begin{proof}[Proof of Theorem~\ref{thm:aggregate_by_degree}]
Fix a network $G$ and a realisation of the posts $\mathbf{s}$, and write $\gamma=\tfrac{\alpha}{n-1+\alpha}$. Since types and posts are identically distributed across users, $\mathbb{E}[s_k]=\mathbb{E}[s]$ and $\mathbb{E}[t_k]=\mathbb{E}[t]$ for every $k$. Using Equation~\eqref{eq:informational_tradeoff}, we can therefore write the action of Equation~\eqref{eq:optimal_action_complete} for every user $m$ as a sum over authors,
\begin{equation}\label{eq:action_by_author}
\begin{split}
    a_m^* = \sum_{k\in N}\Big[\;&\mathbf{1}_{k\in I_m}\Big(\beta_1 F(s_k,\mathbb{E}[s],r_k) + \beta_2\gamma\, F(\mathbb{E}[t_k\mid s_k],\mathbb{E}[t],r_k)\Big)\\
    +\;&\mathbf{1}_{k\notin I_m}\Big(\beta_1\mathbb{E}[s] + \beta_2\gamma\,\mathbb{E}[t]\Big)\Big]
    + \beta_2\Big[(1-\alpha)t_m - \gamma(1-\alpha)\,\mathbb{E}[t_m\mid s_m]\Big].
\end{split}
\end{equation}
The term of author $k$ in the action of user $m$ depends on $m$ only through whether $m$ observes the post of $k$. If $m$ observes it, the term depends on the network only through $r_k$, and otherwise it does not depend on the network at all. The last bracket depends only on user $m$.

We now sum over $m$ and exchange the order of summation. Every user observes her own post, and $r_k$ is the number of other users who observe the post of $k$. Hence the post of $k$ is observed by $1+r_k$ users and not observed by the remaining $n-1-r_k$. It therefore contributes $1+r_k$ times the first bracket and $n-1-r_k$ times the second, while the last bracket contributes $\beta_2[(1-\alpha)t_k - \gamma(1-\alpha)\mathbb{E}[t_k\mid s_k]]$ once for each $k$. This gives, for every realisation of the posts,
\begin{equation}\label{eq:aggregate_by_degree}
\begin{split}
    \sum_{k\in N} a_k^*
    =\ &\beta_1\sum_{k\in N}\Big[(1+r_k)\,F(s_k,\mathbb{E}[s],r_k)+(n-1-r_k)\,\mathbb{E}[s]\Big]\\
    +\ &\beta_2\sum_{k\in N}\Big[\gamma\big((1+r_k)\,F(\mathbb{E}[t_k\mid s_k],\mathbb{E}[t],r_k)
        +(n-1-r_k)\,\mathbb{E}[t]\big)\\
         + \ & (1-\alpha)t_k-\gamma(1-\alpha)\,\mathbb{E}[t_k\mid s_k]\Big],
\end{split}
\end{equation}

In Equation~\eqref{eq:aggregate_by_degree}, the network enters only through the numbers of readers $r_k$, both in the counts $1+r_k$ and $n-1-r_k$ and in the third argument of $F$. Moreover, the identity holds for every realisation of the posts. Therefore two networks with the same profile $\{r_k\}_{k\in N}$ induce the same collective action for every $\mathbf{s}$, and hence also in expectation under $\mathbb{E}_P$.
\end{proof}

\begin{proof}[Proof of Corollary~\ref{cor:equivalence_classes}]
The networks in $K_i(G)$ are obtained from $G$ by adding, or by removing, one link from user $i$. They differ only in the recipient of this link, so every user has the same number of readers in all of them. The result follows from Theorem~\ref{thm:aggregate_by_degree}.
\end{proof}

\begin{lemma}[Marginal contribution of a broadcasting user]\label{lem:marginal_contribution}
Let $k\in N$ have type $t_k=c\in\{\overline{t},\underline{t}\}$. Raising $k$'s out-degree from
$r_k=0$ to $r_k=n-1$, with all other out-degrees fixed, increases the ex ante aggregate
$\sum_{m}\mathbb{E}_P[a_m^*]$ by
\begin{equation}\label{eq:marginal_contribution}
    D(c)=\beta_1(n-1+\alpha)\big(\mathbb{E}[s\mid c]-\mathbb{E}[s]\big)
    +\beta_2\,\alpha\big(\mathbb{E}[\,\mathbb{E}[t\mid s]\mid c]-\mathbb{E}[t]\big).
\end{equation}
\end{lemma}

\begin{proof}
By Theorem~\ref{thm:aggregate_by_degree}, only the $k$-indexed summand of
\eqref{eq:aggregate_by_degree} changes, all others being fixed. Take $\mathbb{E}_P[\cdot]=\mathbb{E}[\cdot\mid \mathbf{t}]$.
Since $F(x,y,n-1)=x$ and $F(x,y,0)=\alpha y+(1-\alpha)x$, the law of iterated expectations gives
\[
    \mathbb{E}_P\!\big[F(s_k,\mathbb{E}[s],r_k)\big]=
    \begin{cases}\mathbb{E}[s\mid c], & r_k=n-1,\\[2pt]
    \alpha\mathbb{E}[s]+(1-\alpha)\mathbb{E}[s\mid c], & r_k=0,\end{cases}
\]
and the analogue with $\mathbb{E}[\,\mathbb{E}[t\mid s]\mid c]$ for the type channel. The own terms
$(1-\alpha)t_k-\gamma(1-\alpha)\mathbb{E}_P[\mathbb{E}[t_k\mid s_k]]$ are independent of $r_k$ and
cancel. For the signal channel, the summand equals $n\,\mathbb{E}[s\mid c]$ at $r_k=n-1$ and
$\alpha\mathbb{E}[s]+(1-\alpha)\mathbb{E}[s\mid c]+(n-1)\mathbb{E}[s]$ at $r_k=0$, so the $r_k=n-1$
value minus the $r_k=0$ value is
\[
    n\mathbb{E}[s\mid c]-(n-1+\alpha)\mathbb{E}[s]-(1-\alpha)\mathbb{E}[s\mid c]
    =(n-1+\alpha)\big(\mathbb{E}[s\mid c]-\mathbb{E}[s]\big).
\]
The type channel carries the factor $\gamma$, giving
$\gamma(n-1+\alpha)\big(\mathbb{E}[\,\mathbb{E}[t\mid s]\mid c]-\mathbb{E}[t]\big)
=\alpha\big(\mathbb{E}[\,\mathbb{E}[t\mid s]\mid c]-\mathbb{E}[t]\big)$. Weighting by $\beta_1,\beta_2$
yields \eqref{eq:marginal_contribution}.
\end{proof}

\begin{proof}[Proof of Theorem~\ref{thm:all_high_low_equivalence}]
Write $P(\overline{s})=p\lambda+(1-p)(1-\mu)$, $P(\underline{s})=1-P(\overline{s})$,
$\mathbb{E}[s]=\overline{s}P(\overline{s})+\underline{s}P(\underline{s})$, and
$\mathbb{E}[t]=p\overline{t}+(1-p)\underline{t}$; recall
$\mathbb{E}[s\mid\overline{t}]=\lambda\overline{s}+(1-\lambda)\underline{s}$ and
$\mathbb{E}[s\mid\underline{t}]=(1-\mu)\overline{s}+\mu\underline{s}$.

The networks $\widehat{G}$ and $\widetilde{G}$ are obtained from $G$ by removing all out-edges of $i$
and adding all out-edges of $j$, respectively. Their out-degree profiles agree for every user except
$i$ and $j$, with
\[
    r_i(\widehat{G})=0,\ r_i(\widetilde{G})=n-1,\qquad
    r_j(\widehat{G})=0,\ r_j(\widetilde{G})=n-1.
\]
By Theorem~\ref{thm:aggregate_by_degree} the aggregate depends only on out-degrees, so passing from
$\widehat{G}$ to $\widetilde{G}$ raises $r_i$ and $r_j$ from $0$ to $n-1$ while leaving all other
out-degrees unchanged.\footnote{This uses only that the other users' out-degrees coincide across
$\widehat{G}$ and $\widetilde{G}$, which holds for any $G$ in which $i$ broadcasts to all and $j$ to
none; their values are arbitrary. The result is thus independent of the connectivity and type
composition of the remaining $n-2$ users.} By Lemma~\ref{lem:marginal_contribution},
\begin{equation}\label{eq:two_contributions}
    \sum_{k}\mathbb{E}_P[a_k^*(\widetilde{G})]-\sum_{k}\mathbb{E}_P[a_k^*(\widehat{G})]
    =D(\overline{t})+D(\underline{t}).
\end{equation}

Evaluate the two contributions. For the signal channel,
\[
    \mathbb{E}[s\mid\overline{t}]-\mathbb{E}[s]=(1-p)(\lambda+\mu-1)(\overline{s}-\underline{s}),
    \qquad
    \mathbb{E}[s\mid\underline{t}]-\mathbb{E}[s]=-p(\lambda+\mu-1)(\overline{s}-\underline{s}).
\]
For the type channel,
$\mathbb{E}[\,\mathbb{E}[t\mid s]\mid\overline{t}]-\mathbb{E}[t]=(1-p)(\lambda+\mu-1)\big(\mathbb{E}[t\mid\overline{s}]-\mathbb{E}[t\mid\underline{s}]\big)$
and the same with $-p$ in place of $(1-p)$ for $\underline{t}$, where
\begin{equation}\label{eq:posterior_diff}
    \mathbb{E}[t\mid\overline{s}]-\mathbb{E}[t\mid\underline{s}]
    =\frac{p(1-p)(\lambda+\mu-1)(\overline{t}-\underline{t})}{P(\overline{s})\,P(\underline{s})}.
\end{equation}\footnote{Identity \eqref{eq:posterior_diff} is the standard binary Bayesian update.
Since $\mathbb{E}[t\mid s]=\overline{t}\Pr(\overline{t}\mid s)+\underline{t}\Pr(\underline{t}\mid s)$,
one has $\mathbb{E}[t\mid\overline{s}]-\mathbb{E}[t\mid\underline{s}]
=(\overline{t}-\underline{t})\big[\Pr(\overline{t}\mid\overline{s})-\Pr(\overline{t}\mid\underline{s})\big]$.
By Bayes' rule $\Pr(\overline{t}\mid\overline{s})=p\lambda/P(\overline{s})$ and
$\Pr(\overline{t}\mid\underline{s})=p(1-\lambda)/P(\underline{s})$; over the common denominator
$P(\overline{s})P(\underline{s})$ the numerator $p\lambda P(\underline{s})-p(1-\lambda)P(\overline{s})$
simplifies to $p(1-p)(\lambda+\mu-1)$, giving \eqref{eq:posterior_diff}.}
Substituting into \eqref{eq:marginal_contribution} gives $D(\overline{t})=(1-p)(\lambda+\mu-1)Q$ and
$D(\underline{t})=-p(\lambda+\mu-1)Q$, where
\begin{equation}\label{eq:Q_def}
    Q=\beta_1(\alpha+n-1)(\overline{s}-\underline{s})
    +\beta_2\,\frac{\alpha\,p(1-p)(\lambda+\mu-1)(\overline{t}-\underline{t})}
    {P(\overline{s})\,P(\underline{s})}.
\end{equation}
Hence, by \eqref{eq:two_contributions},
\[
    \sum_{k}\mathbb{E}_P[a_k^*(\widehat{G})]-\sum_{k}\mathbb{E}_P[a_k^*(\widetilde{G})]
    =-\big(D(\overline{t})+D(\underline{t})\big)=(2p-1)(\lambda+\mu-1)\,Q.
\]

Finally, under $\alpha\in(0,1)$, $\beta_1,\beta_2>0$, $\lambda,\mu>\tfrac12$,
$\overline{s}>\underline{s}$, $\overline{t}>\underline{t}$, $n\ge 2$, both terms of $Q$ are strictly
positive for every $p\in(0,1)$, so $Q>0$; and $\lambda+\mu-1>0$. The gap therefore has the sign of
$(2p-1)$, vanishing precisely at $p=\tfrac12$. Thus $\widehat{G}$ and $\widetilde{G}$ induce the same
ex ante collective action if and only if $p=\tfrac12$.
\end{proof}

\begin{proof}[Proof of Proposition~\ref{prop:extreme_equivalence}]
We compute the aggregate action at the empty and complete networks for an arbitrary type profile, and then take the difference.

\paragraph{Empty network.} Here $r_k=0$, $I_i=\{i\}$. The $\beta_2$ bracket at $r_i=0$ contributes $\frac{\alpha}{n-1+\alpha}\big[(n-1+\alpha)\mathbb{E}[t]+(1-\alpha)\mathbb{E}[t_i\mid s_i]\big]$; adding the explicit terms $(1-\alpha)t_i-\frac{\alpha(1-\alpha)}{n-1+\alpha}\mathbb{E}[t_i\mid s_i]$, the conditional-signal terms cancel, leaving
\begin{equation*}
    a_i^*(G_{\emptyset}) = \beta_1\big[(n-1+\alpha)\mathbb{E}[s] + (1-\alpha)s_i\big]
    + \beta_2\big[\alpha\mathbb{E}[t] + (1-\alpha)t_i\big].
\end{equation*}
Summing over $i$ and applying $\mathbb{E}_P[\cdot]$, with $\mathbb{E}_P[s_i]=\mathbb{E}[s\mid t_i]$,
\begin{equation*}
\begin{split}
    \mathbb{E}_P\Big[\textstyle\sum_{i} a_i^*(G_{\emptyset})\Big]
    = \beta_1\Big[n(n-1+\alpha)\mathbb{E}[s] &+ (1-\alpha)\textstyle\sum_{k}\mathbb{E}[s\mid t_k]\Big]\\
    &+ \beta_2\Big[n\alpha\mathbb{E}[t] + (1-\alpha)\textstyle\sum_{k} t_k\Big].
\end{split}
\end{equation*}

\paragraph{Complete network.} Here $r_k=n-1$, $I_i=N$, so $F$ collapses: each $\beta_1$ term becomes
$s_k$ and each $\beta_2$ in-bracket term becomes $\mathbb{E}[t_k\mid s_k]$. Hence
\begin{equation*}
    a_i^*(G_{\text{complete}}) = \beta_1 \textstyle\sum_{k} s_k + \beta_2\Big[\tfrac{\alpha}{n-1+\alpha}\sum_{k}\mathbb{E}[t_k\mid s_k]
    + (1-\alpha)t_i - \tfrac{\alpha(1-\alpha)}{n-1+\alpha}\mathbb{E}[t_i\mid s_i]\Big].
\end{equation*}
Summing over $i$, the terms in $\mathbb{E}[t_k\mid s_k]$ carry the coefficient $\frac{n\alpha-\alpha(1-\alpha)}{n-1+\alpha}=\alpha$. Applying $\mathbb{E}_P[\cdot]$,
\begin{equation*}
\begin{split}
    \mathbb{E}_P\Big[\textstyle\sum_{i} a_i^*(G_{\text{complete}})\Big]
    = \beta_1\, n\textstyle\sum_{k}\mathbb{E}[s\mid t_k]
    &+ \beta_2\Big[\alpha\textstyle\sum_{k}\mathbb{E}\big[\mathbb{E}[t\mid s]\mid t_k\big]\\
    &+ (1-\alpha)\textstyle\sum_{k} t_k\Big].
\end{split}
\end{equation*}

\paragraph{Difference.} Subtracting the two aggregates, the terms in $\sum_k t_k$ cancel. What remains is a sum over authors, and the term of author $k$ is
\begin{equation*}
    \beta_1(n-1+\alpha)\big(\mathbb{E}[s\mid t_k]-\mathbb{E}[s]\big)
    + \beta_2\,\alpha\big(\mathbb{E}\big[\mathbb{E}[t\mid s]\mid t_k\big]-\mathbb{E}[t]\big) = D(t_k),
\end{equation*}
the marginal contribution of Lemma~\ref{lem:marginal_contribution}. It depends on author $k$ only through $t_k$. Hence
\begin{equation*}
    \mathbb{E}_P\Big[\textstyle\sum_{i} a_i^*(G_{\text{complete}})\Big]
    - \mathbb{E}_P\Big[\textstyle\sum_{i} a_i^*(G_{\emptyset})\Big]
    = \textstyle\sum_{k} D(t_k) = n_h D(\overline{t}) + n_l D(\underline{t}).
\end{equation*}
In the proof of Theorem~\ref{thm:all_high_low_equivalence} we showed that $D(\overline{t})=(1-p)(\lambda+\mu-1)Q$ and $D(\underline{t})=-p(\lambda+\mu-1)Q$, where $Q$, defined in \eqref{eq:Q_def}, is strictly positive. Substituting, and since $(1-p)n_h - p\,n_l = n_h - np$, we obtain
\begin{equation*}
    \mathbb{E}_P\Big[\textstyle\sum_{i} a_i^*(G_{\text{complete}})\Big]
    - \mathbb{E}_P\Big[\textstyle\sum_{i} a_i^*(G_{\emptyset})\Big]
    = (n_h - np)(\lambda+\mu-1)\,Q.
\end{equation*}
By \eqref{eq:posterior_diff}, $Q$ equals the bracket in \eqref{eq:extreme_gap}, which proves the formula.
Since $Q>0$ and $\lambda+\mu>1$, the difference is zero if and only if $n_h = np$.
\end{proof}

\section{Supplementary Materials for Section~\ref{sec:policy}}\label{app:policy}

Section~\ref{sec:policy} states its results in terms of the two interim utilities $U_\sigma$, the three pair moments $P_{\sigma\sigma'}$ and the criteria themselves. This appendix introduces the coefficients through which those objects are computed, proves the results of the section, and evaluates the examples.

Two facts about the timing of the game are used throughout. First, \textbf{(F1)}, conditional on $\mathbf{t}$, the signals $s_1,\dots,s_n$ are independent. Second, \textbf{(F2)}, conditional on $\mathbf{t}$, the distribution of $s_k$ depends only on $t_k$.

Write, for $\sigma \in T$,
\[
  m_\sigma = \mathbb{E}[s \mid t = \sigma], \qquad v_\sigma = \mathrm{Var}(s \mid t = \sigma),
  \qquad \nu_\sigma = m_\sigma - \mathbb{E}[s],
\]
and, for every author $k$, $\varepsilon_k = s_k - \mathbb{E}[s]$. By (F1) and (F2), the innovations $\varepsilon_k$ are conditionally independent given $\mathbf{t}$, with $\mathbb{E}_P[\varepsilon_k] = \nu_{t_k}$ and $\mathrm{Var}_P(\varepsilon_k) = v_{t_k}$. Since $\mathbb{E}[s] = p\,m_{\bar t} + (1-p)\,m_{\underline t}$,
\begin{equation}\label{eq:nu}
  \nu_{\bar t} = (1-p)\bigl(m_{\bar t} - m_{\underline t}\bigr) > 0,
  \qquad \nu_{\underline t} = -p\bigl(m_{\bar t} - m_{\underline t}\bigr) < 0,
\end{equation}
since $m_{\bar t} - m_{\underline t} = (\lambda+\mu-1)(\bar s-\underline s) > 0$.
Conditional on the type of its author, a post is thus above its unconditional mean if the author is of the high type, and below it otherwise. Write further
\begin{align*}
  \gamma &= \frac{\alpha}{n-1+\alpha},
  &\zeta &= \frac{\mathbb{E}[t \mid \bar s] - \mathbb{E}[t \mid \underline s]}{\bar s - \underline s},\\
  \kappa &= \beta_1 + \beta_2\gamma\zeta,
  &c_k &= \frac{(1-\alpha)(n-1)}{n-1-\alpha r_k}.
\end{align*}
Here $\gamma$ is the weight that the equilibrium action places on the type channel, and $c_k$ is the weight that a reader of author $k$ places on the deviation of the post from its prior mean. By Proposition~\ref{prop:properties_of_F}, $c_k$ increases in the reach $r_k$, from $1-\alpha$ at $r_k = 0$ to $1$ at $r_k = n-1$. The scalar $\zeta$ is the slope of the posterior type in the post. By \eqref{eq:posterior_diff},
\[
  \zeta = \frac{p(1-p)(\lambda+\mu-1)(\bar t-\underline t)}{P(\bar s)\,P(\underline s)\,(\bar s-\underline s)} > 0,
\]
where $P(\bar s) = 1 - P(\underline s)$ is the probability of a high post, as in the proof of Theorem~\ref{thm:all_high_low_equivalence}. Equivalently, $\zeta = \mathrm{Cov}(s,t)/\mathrm{Var}(s)$ is the regression coefficient of the type on the post. Finally, $\kappa$ collects the two informational roles of a post: its direct value as evidence on $\theta(\mathbf{s})$, weighted by $\beta_1$, and its indirect value as evidence on the type of its author, weighted by $\beta_2\gamma\zeta$.

For a block architecture $G[B]$, a type-$\sigma$ author is read by every user of a type $\sigma'$ with $B(\sigma,\sigma') = 1$ other than herself. Every type-$\sigma$ author therefore has the same reach,
\begin{equation}\label{eq:rho}
  \rho_B(\sigma) = B(\sigma,\bar t)\, n_h + B(\sigma,\underline t)\, n_l - B(\sigma,\sigma),
  \qquad
  c_B(\sigma) = \frac{(1-\alpha)(n-1)}{n-1-\alpha\rho_B(\sigma)},
\end{equation}
so that $r_k = \rho_B(t_k)$ and $c_k = c_B(t_k)$ for every $k$.

\subsection*{Proof of Lemma~\ref{lem:sw_var_identity}}\label{app:proof_identity}

For any vector $\mathbf{a}$ with mean $\bar a$,
\begin{align*}
  \sum_{i=1}^{n}\sum_{j\neq i}(a_i-a_j)^2
  &= \sum_{i=1}^{n}\sum_{j=1}^{n}(a_i-a_j)^2
  = 2n\sum_{i}a_i^{2} - 2\Bigl(\sum_{i}a_i\Bigr)^{2}\\
  &= 2n^{2}\Bigl[\tfrac1n\sum_{i}a_i^{2} - \bar a^{2}\Bigr]
  = 2n^{2}\,\mathrm{Var}(\mathbf{a}),
\end{align*}
where the first equality holds because the diagonal terms vanish. Substituting $\mathbf{a} = \mathbf{a}^*$ into \eqref{eq:SW} gives \eqref{eq:sw_var_identity}, and the statement for $\mathbb{E}_P[\cdot]$ follows by linearity. \qed

\subsection*{Equilibrium Actions and the Criteria for an Arbitrary Network}\label{app:innovation}\label{app:general}

All the computations of Section~\ref{sec:policy} rest on the following reformulation of Theorem~\ref{thm:users_equilibrium}, which holds for every network and every pair of supports.

\begin{lemma}\label{lem:innovation}
For any network $G$ and any realisation $(\mathbf{s},\mathbf{t})$,
\begin{equation}\label{eq:innovation}
  a_i^* \;=\; \beta_1 n\,\mathbb{E}[s] + \alpha\beta_2\,\mathbb{E}[t]
  \;+\; \beta_2(1-\alpha)\,t_i
  \;+\; \kappa \sum_{k \in I_i} c_k\, \varepsilon_k
  \;-\; \beta_2\gamma(1-\alpha)\zeta\, \varepsilon_i .
\end{equation}
\end{lemma}

\begin{proof}
Start from the expression \eqref{eq:action_by_author} of the equilibrium action as a sum over authors, obtained in the proof of Theorem~\ref{thm:aggregate_by_degree}. By the computation in the proof of Proposition~\ref{prop:properties_of_Delta}, $F(x,y,r_k) = y + c_k(x-y)$. Writing $\eta_k := \mathbb{E}[t_k \mid s_k] - \mathbb{E}[t]$, the terms of author $k$ in \eqref{eq:action_by_author} sum to
\[
  \beta_1\bigl(\mathbb{E}[s] + c_k\varepsilon_k\bigr) + \beta_2\gamma\bigl(\mathbb{E}[t] + c_k\eta_k\bigr)
  \ \text{ if } k \in I_i,
  \qquad
  \beta_1\mathbb{E}[s] + \beta_2\gamma\,\mathbb{E}[t]
  \ \text{ if } k \notin I_i .
\]
Because $s_k$ takes two values, $\mathbb{E}[t \mid s]$ is affine on $\{\underline s,\bar s\}$ with slope $\zeta$, and its mean is $\mathbb{E}[t]$, so
\begin{equation}\label{eq:eta_eps}
  \eta_k = \zeta\,\varepsilon_k .
\end{equation}
Summing over authors, and writing the last bracket of \eqref{eq:action_by_author} as $\beta_2(1-\alpha)t_i - \beta_2\gamma(1-\alpha)\bigl(\mathbb{E}[t] + \zeta\varepsilon_i\bigr)$, gives
\begin{align*}
  a_i^* = \beta_1 n\,\mathbb{E}[s] &+ \beta_2\gamma(n-1+\alpha)\,\mathbb{E}[t] + \beta_2(1-\alpha)t_i\\
  &+ (\beta_1 + \beta_2\gamma\zeta)\sum_{k\in I_i}c_k\varepsilon_k - \beta_2\gamma(1-\alpha)\zeta\,\varepsilon_i ,
\end{align*}
which is \eqref{eq:innovation}, since $\gamma(n-1+\alpha) = \alpha$.
\end{proof}

Equation \eqref{eq:innovation} says that every user starts from a common benchmark, adds a private adjustment proportional to her own type, and loads on the innovation of each author she reads with weight $\kappa c_k$. The last term corrects for the fact that her own post is read by others. Since $i \in I_i$, the total weight on $\varepsilon_i$ is $\kappa c_i - \beta_2\gamma(1-\alpha)\zeta$.

Every equilibrium action is therefore affine in the innovations, and all three criteria are quadratic forms in the matrix of loadings
\begin{equation}\label{eq:loadings}
  w_{ik} \;=\; \kappa\, c_k\, \mathbf{1}\{k \in I_i\} \;-\; \beta_2\gamma(1-\alpha)\zeta\,\mathbf{1}\{k = i\}.
\end{equation}
Since $\theta(\mathbf{s}) = n\mathbb{E}[s] + \sum_k \varepsilon_k$, Lemma~\ref{lem:innovation} gives the consistency residual of user $i$ and the difference between the actions of users $i$ and $j$,
\begin{align*}
  a_i^* - \beta_1\theta(\mathbf{s}) - \beta_2 t_i
  &= \alpha\beta_2\bigl(\mathbb{E}[t]-t_i\bigr) + \sum_k \bigl(w_{ik}-\beta_1\bigr)\varepsilon_k,\\
  a_i^* - a_j^* &= \beta_2(1-\alpha)(t_i - t_j) + \sum_k \bigl(w_{ik}-w_{jk}\bigr)\varepsilon_k .
\end{align*}
By (F1) and (F2), their conditional means and variances given $\mathbf{t}$ are
\begin{align}
  M_i &= \alpha\beta_2\bigl(\mathbb{E}[t]-t_i\bigr) + \sum_k (w_{ik}-\beta_1)\,\nu_{t_k},
  &V_i &= \sum_k (w_{ik}-\beta_1)^2 v_{t_k}, \label{eq:gen_consistency}\\
  M_{ij} &= \beta_2(1-\alpha)(t_i-t_j) + \sum_k (w_{ik}-w_{jk})\,\nu_{t_k},
  &V_{ij} &= \sum_k (w_{ik}-w_{jk})^2 v_{t_k}. \label{eq:gen_pair}
\end{align}
Since the conditional second moment of each quantity is its conditional variance plus its squared conditional mean, \eqref{eq:SW} and Lemma~\ref{lem:sw_var_identity} give
\begin{align}
  \mathbb{E}_P[SW(G)] &= -\sum_{i}\Bigl[(1-\alpha)\bigl(V_i + M_i^2\bigr)
  + \frac{\alpha}{n-1}\sum_{j\neq i}\bigl(V_{ij} + M_{ij}^2\bigr)\Bigr], \label{eq:gen_sw}\\
  \mathbb{E}_P\bigl[\mathrm{Var}(\mathbf{a}^*)\bigr]
  &= \frac{1}{2n^{2}}\sum_i\sum_{j\neq i}\bigl(V_{ij} + M_{ij}^2\bigr).\label{eq:gen_criteria}
\end{align}
Throughout, $M_i$ and $V_i$ indexed by a user refer to \eqref{eq:gen_consistency}, and $M_\sigma$ and $V_\sigma$ indexed by a type refer to their common value for the users of that type in a block architecture, given in \eqref{eq:Msigma}--\eqref{eq:Vsigma} below.

Two consequences of Lemma~\ref{lem:innovation} are used in Section~\ref{sec:policy}. The first shows how a feed rule moves the conditional mean of each action.

\begin{lemma}\label{lem:mean_action}
For any network $G$ and any type profile $\mathbf{t}$,
\begin{equation}\label{eq:mean_action}
  \mathbb{E}_P[a_i^*] = \beta_1 n\,\mathbb{E}[s] + \alpha\beta_2\,\mathbb{E}[t] + \beta_2(1-\alpha)\,t_i
  + \kappa \sum_{k \in I_i} c_k\,\nu_{t_k} - \beta_2\gamma(1-\alpha)\zeta\,\nu_{t_i}.
\end{equation}
In particular, the post of an author $k \neq i$ contributes $\kappa c_k \nu_{t_k}$ to the conditional mean of the action of user $i$ if $k \in I_i$, and nothing otherwise. The contribution is positive if $t_k = \bar t$ and negative if $t_k = \underline t$.
\end{lemma}

\begin{proof}
By (F2), $\mathbb{E}_P[\varepsilon_k] = \nu_{t_k}$, and taking $\mathbb{E}_P[\cdot]$ in \eqref{eq:innovation} gives \eqref{eq:mean_action}. The sign follows from $\kappa, c_k > 0$ and \eqref{eq:nu}.
\end{proof}

Lemma~\ref{lem:mean_action} is where the assumption that users do not infer types from their feeds enters the welfare analysis. A reader who does not observe the post of $k$ expects it at $\mathbb{E}[s]$, its unconditional mean, whereas conditional on $\mathbf{t}$ its mean is $m_{t_k}$. A feed rule that conditions on $t_k$ therefore moves the conditional mean of every action, and not only its conditional variance. The second consequence splits expected welfare into three components.

\begin{corollary}\label{cor:three_components}
For any network $G$ and any type profile $\mathbf{t}$,
\begin{equation}\label{eq:three_components}
  \mathbb{E}_P[SW(G)]
  = \underbrace{-(1-\alpha)\sum_{i} M_i^{2}}_{\text{level}}
  \;\underbrace{-\,(1-\alpha)\sum_{i} V_i}_{\text{precision}}
  \;\underbrace{-\,\frac{2\alpha n^{2}}{n-1}\,\mathbb{E}_P\bigl[\mathrm{Var}(\mathbf{a}^*)\bigr]}_{\text{coordination}}.
\end{equation}
\end{corollary}

\begin{proof}
By \eqref{eq:gen_consistency}, $M_i$ and $V_i$ are the conditional mean and variance of $a_i^* - \beta_1\theta(\mathbf{s}) - \beta_2 t_i$ given $\mathbf{t}$, so its conditional second moment is $V_i + M_i^{2}$. The result follows from Lemma~\ref{lem:sw_var_identity} under $\mathbb{E}_P[\cdot]$.
\end{proof}

The level component records how far actions are from ideal points on average, the precision component how noisily they track the state around that average, and the coordination component how far apart they are. Table~\ref{tab:welfare_decomposition} reports the three components for Examples~\ref{ex:welfare_fails} and \ref{ex:welfare_fails_het}.

\subsection*{Block Architectures: Proof of Proposition~\ref{prop:closed_forms}}\label{app:proof_closed}\label{app:exchangeability}

In a block architecture the loading of a reader on an author depends only on their two types, and on whether they coincide.

\begin{lemma}\label{lem:exchangeability}
Let $G[B]$ be a block architecture. For every user $i$ and every author $k$,
\begin{equation}\label{eq:weights}
  w_{ik} = w_{t_i}(t_k) \ \text{ if } k \neq i,
  \qquad w_{ii} = w^{\mathrm{own}}(t_i),
\end{equation}
where, for $\sigma,\sigma' \in T$,
\[
  w_\sigma(\sigma') = \kappa\, B(\sigma',\sigma)\, c_B(\sigma'),
  \qquad
  w^{\mathrm{own}}(\sigma) = \kappa\, c_B(\sigma) - \beta_2\gamma(1-\alpha)\zeta .
\]
Consequently, $M_i$ and $V_i$ depend on $i$ only through $t_i$, and $M_{ij}$ and $V_{ij}$ depend on the pair $(i,j)$ only through $(t_i,t_j)$.
\end{lemma}

\begin{proof}
By Definition~\ref{def:block}, $k \in I_i$ for $k \neq i$ if and only if $B(t_k,t_i) = 1$, and $c_k = c_B(t_k)$ by \eqref{eq:rho}. Substituting into \eqref{eq:loadings} gives \eqref{eq:weights}. In \eqref{eq:gen_consistency}, the sum over authors then depends on $i$ only through $t_i$ and through the number of authors of each type other than $i$, which is $n_{\sigma'} - \mathbf{1}\{\sigma' = t_i\}$. The same argument applies to \eqref{eq:gen_pair}.
\end{proof}

\begin{proof}[Proof of Proposition~\ref{prop:closed_forms}]
By Lemma~\ref{lem:exchangeability}, the interim utility $U_\sigma = \mathbb{E}_P[u_i]$ of a user $i$ of type $\sigma$ and the pair moment $P_{\sigma\sigma'} = \mathbb{E}_P[(a_i^*-a_j^*)^2] = V_{ij} + M_{ij}^{2}$ of two distinct users of types $\sigma$ and $\sigma'$ do not depend on the particular users chosen. Summing $\mathbb{E}_P[u_i]$ over $i$ and grouping users by type gives the expression for welfare. The $n(n-1)$ ordered pairs of distinct users consist of $n_h(n_h-1)$ pairs of two high types, $n_l(n_l-1)$ pairs of two low types and $2n_hn_l$ mixed pairs, and \eqref{eq:gen_criteria} gives the expression for polarisation.

For the closed forms, fix a user $i$ of type $\sigma$, and write $d_\sigma(\sigma') = w_\sigma(\sigma') - \beta_1$ and $d^{\mathrm{own}}(\sigma) = w^{\mathrm{own}}(\sigma) - \beta_1$. Among the other users, $n_{\sigma'} - \mathbf{1}\{\sigma'=\sigma\}$ are of type $\sigma'$, so \eqref{eq:gen_consistency} and \eqref{eq:weights} give
\begin{align}
  M_\sigma &= \alpha\beta_2\bigl(\mathbb{E}[t]-\sigma\bigr)
    + \sum_{\sigma'} \bigl(n_{\sigma'} - \mathbf{1}\{\sigma'=\sigma\}\bigr)\, d_\sigma(\sigma')\,\nu_{\sigma'}
    + d^{\mathrm{own}}(\sigma)\,\nu_{\sigma}, \label{eq:Msigma}\\
  V_\sigma &= \sum_{\sigma'} \bigl(n_{\sigma'} - \mathbf{1}\{\sigma'=\sigma\}\bigr)\, d_\sigma(\sigma')^{2}\, v_{\sigma'}
    + d^{\mathrm{own}}(\sigma)^{2}\, v_{\sigma}. \label{eq:Vsigma}
\end{align}
Now fix two distinct users $i$ and $j$ of types $\sigma_1$ and $\sigma_2$. By \eqref{eq:weights}, the differences of their loadings are
\[
  w_{ik}-w_{jk} =
  \begin{cases}
    \delta(t_k) := w_{\sigma_1}(t_k) - w_{\sigma_2}(t_k), & k \notin \{i,j\},\\[2pt]
    \delta_i := w^{\mathrm{own}}(\sigma_1) - w_{\sigma_2}(\sigma_1), & k = i,\\[2pt]
    \delta_j := w_{\sigma_1}(\sigma_2) - w^{\mathrm{own}}(\sigma_2), & k = j .
  \end{cases}
\]
Among the users other than $i$ and $j$, $\tilde n_{\sigma'} = n_{\sigma'} - \mathbf{1}\{\sigma'=\sigma_1\} - \mathbf{1}\{\sigma'=\sigma_2\}$ are of type $\sigma'$, so \eqref{eq:gen_pair} gives
\begin{align}
  M_{\sigma_1\sigma_2} &= \beta_2(1-\alpha)(\sigma_1-\sigma_2)
    + \sum_{\sigma'} \tilde n_{\sigma'}\,\delta(\sigma')\,\nu_{\sigma'}
    + \delta_i\,\nu_{\sigma_1} + \delta_j\,\nu_{\sigma_2},
    \label{eq:Mpair}\\
  V_{\sigma_1\sigma_2} &= \sum_{\sigma'} \tilde n_{\sigma'}\,\delta(\sigma')^{2}\, v_{\sigma'}
    + \delta_i^{2}\, v_{\sigma_1} + \delta_j^{2}\, v_{\sigma_2},
    \label{eq:Vpair}
\end{align}
and $P_{\sigma_1\sigma_2} = V_{\sigma_1\sigma_2} + M_{\sigma_1\sigma_2}^{2}$. Finally, a user of type $\sigma$ has $n_\sigma - 1$ opponents of her own type and $n_{-\sigma}$ of the other, so \eqref{eq:gen_sw} gives
\begin{equation}\label{eq:Usigma}
  U_\sigma = -(1-\alpha)\bigl(V_\sigma + M_\sigma^{2}\bigr)
  - \frac{\alpha}{n-1}\Bigl[(n_\sigma-1)\,P_{\sigma\sigma} + n_{-\sigma}\,P_{\sigma,-\sigma}\Bigr].
\end{equation}
Every coefficient in \eqref{eq:Msigma}--\eqref{eq:Usigma} is a rational function of $(\alpha,\beta_1,\beta_2,n_h,n_l)$, of $(p,\lambda,\mu)$ and of the supports, which proves the first claim of the proposition.
\end{proof}

\subsection*{Evaluation of the Examples}\label{app:examples}

All the examples use rational parameter values, so the expressions above can be evaluated in exact arithmetic and the reported comparisons are exact rather than numerical. Appendix~\ref{app:numerical} documents an independent verification of every value against enumeration over all $2^{n}$ post profiles.

\medskip
\noindent\textbf{Example~\ref{ex:equivalence_class}.} The three architectures are obtained from the network $G$ of Example~\ref{ex:political_1} by adding one out-link from user $2$. Neither $G$ nor the three architectures are block architectures, so all four rows of the table are evaluated from \eqref{eq:loadings}--\eqref{eq:gen_criteria}. In $G$ we have $r_2 = 0$. In each of the three architectures $r_2 = 1$ while every other reach is unchanged, which is why the three induce the same collective action, in accordance with Theorem~\ref{thm:aggregate_by_degree}. The common value is $\sum_i\mathbb{E}_P[a_i^*] = 26501/1280$, against $3081527/147200$ under $G$. The remaining entries of the table follow from \eqref{eq:gen_criteria} and, for heterophilic density, from \eqref{eq:HD} applied to the six edges of $G$, together with the added edge for the three architectures. The changes reported in the example are differences between the row of $G$ and the row of each architecture.

\medskip
\noindent\textbf{Example~\ref{ex:welfare_fails}.} The primitives are $\mathbb{E}[s] = 3/5$, $\mathbb{E}[t] = 1/2$, $m_{\bar t} = 9/10$, $m_{\underline t} = 3/10$, $v_{\bar t} = 9/100$, $v_{\underline t} = 21/100$, $\zeta = 5/8$, $\gamma = 7/57$ and $\kappa = 631/456$. For the complete network $\rho_B(\sigma) = 5$ and $c_B(\sigma) = 1$ for both types, while for $G_{\max}$ we have $\rho_B(\bar t) = 5$ with $c_B(\bar t) = 1$, and $\rho_B(\underline t) = 0$ with $c_B(\underline t) = 1-\alpha = 3/10$. Substituting into \eqref{eq:Msigma}--\eqref{eq:Usigma} at $(n_h,n_l) = (5,1)$ gives the following, where the three components are those of Corollary~\ref{cor:three_components}, with decimal values in Table~\ref{tab:welfare_decomposition}.
\[
\begin{array}{lcc}
\toprule
 & G_{\mathrm{complete}} & G_{\max} \\
\midrule
U_{\bar t} & -\frac{1944397}{2310400} = -0.841585 & -\frac{387885993}{577600000} = -0.671548 \\
U_{\underline t} & -\frac{2747143}{924160} = -2.972584 & -\frac{170383059}{46208000} = -3.687307 \\
\midrule
\text{level} & -\frac{1531005}{369664} & -\frac{7319270667}{2310400000} \\
\text{precision} & -\frac{295911}{1848320} & -\frac{1051294923}{2310400000} \\
\text{coordination} & -\frac{2660469}{924160} & -\frac{395315361}{115520000} \\
\midrule
\mathbb{E}_P[SW] & -\frac{6635937}{924160} = -7.180507 & -\frac{1627687281}{231040000} = -7.045045 \\
\bottomrule
\end{array}
\]
Hence
\[
  \mathbb{E}_P\bigl[SW(G_{\max})\bigr] - \mathbb{E}_P\bigl[SW(G_{\mathrm{complete}})\bigr]
  = \frac{31296969}{231040000} > 0 .
\]

\medskip
\noindent\textbf{Example~\ref{ex:welfare_fails_het}.} The primitives are $\mathbb{E}[s] = \mathbb{E}[t] = 1/2$, $m_{\bar t} = 7/10$, $m_{\underline t} = 3/10$, $v_{\bar t} = v_{\underline t} = 21/100$, $\zeta = 2/5$, $\gamma = 1/11$ and $\kappa = 3/22$. Both architectures are symmetric in the two types at $n_h = n_l = 3$. For the complete network $\rho_B(\sigma) = 5$ and $c_B(\sigma) = 1$, and for $G_{\mathrm{het}}$, $\rho_B(\sigma) = n_{-\sigma} = 3$ and $c_B(\sigma) = 5/7$. Substituting into \eqref{eq:Msigma}--\eqref{eq:Usigma} gives the following, with decimal values of the three components in Table~\ref{tab:welfare_decomposition}.
\[
\begin{array}{lcc}
\toprule
 & G_{\mathrm{complete}} & G_{\mathrm{het}} \\
\midrule
\text{level} & -\frac{233523}{1210000} & -\frac{15228027}{59290000} \\
\text{precision} & -\frac{1323}{302500} & -\frac{54567}{4235000} \\
\text{coordination} & -\frac{661599}{1512500} & -\frac{2206341}{6737500} \\
\midrule
\mathbb{E}_P[SW] & -\frac{3840471}{6050000} = -0.634789 & -\frac{16094439}{26950000} = -0.597196 \\
\bottomrule
\end{array}
\]
Hence
\[
  \mathbb{E}_P\bigl[SW(G_{\mathrm{het}})\bigr] - \mathbb{E}_P\bigl[SW(G_{\mathrm{complete}})\bigr]
  = \frac{44577}{1185800} > 0 .
\]
By \eqref{eq:Mpair}, the gap $\mathbb{E}_P[a_i^* - a_j^*]$ between a high-type user $i$ and a low-type user $j$ is $271/550 = 0.4927$ under the complete network and $1597/3850 = 0.4148$ under $G_{\mathrm{het}}$, and polarisation falls from $73511/1210000 = 0.0608$ to $245149/5390000 = 0.0455$.

\medskip
\noindent\textbf{Example~\ref{ex:echo_fails}.} The primitives are $\mathbb{E}[s] = \mathbb{E}[t] = 1/2$, $m_{\bar t} = 7/10$, $m_{\underline t} = 3/10$, $v_{\bar t} = v_{\underline t} = 21/100$, $\zeta = 2/5$, $\gamma = 9/59$ and $\kappa = 7402/1475$. Both architectures are symmetric in the two types at $n_h = n_l = 3$. For $G_{\mathrm{echo}}$ we have $\rho_B(\sigma) = n_\sigma - 1 = 2$ and $c_B(\sigma) = 5/32$, and for $G_{\mathrm{het}}$, $\rho_B(\sigma) = n_{-\sigma} = 3$ and $c_B(\sigma) = 5/23$. Substituting into \eqref{eq:Mpair}--\eqref{eq:Vpair} gives the following.
\[
\begin{array}{lccc}
\toprule
 & P_{\bar t\bar t} = P_{\underline t\,\underline t} & P_{\bar t\,\underline t} & \mathbb{E}_P[\mathrm{Var}(\mathbf{a}^*)] \\
\midrule
G_{\mathrm{echo}} & 0.000001 & 1.714768 & \frac{1193822143983}{2784800000000} = 0.428692 \\
G_{\mathrm{het}}  & 0.498183 & 1.711180 & \frac{23516479274351}{46036225000000} = 0.510826 \\
\bottomrule
\end{array}
\]
Hence
\[
  \mathbb{E}_P\bigl[\mathrm{Var}(\mathbf{a}^*;G_{\mathrm{het}})\bigr]
  - \mathbb{E}_P\bigl[\mathrm{Var}(\mathbf{a}^*;G_{\mathrm{echo}})\bigr]
  = \frac{4839816904489}{58926368000000} > 0 .
\]
The within-type moment $P_{\sigma\sigma}$ is of order $\bigl(\beta_2\gamma(1-\alpha)\zeta\bigr)^{2}$ under $G_{\mathrm{echo}}$, since by \eqref{eq:weights} two users of a common type then load identically on every author, so $\delta(\sigma') = 0$ and only $\delta_i = -\delta_j = \beta_2\gamma(1-\alpha)\zeta$ survives. Under $G_{\mathrm{het}}$ they load on each other with weight $w_\sigma(\sigma) = 0$, so $\delta_i = w^{\mathrm{own}}(\sigma)$ and the moment is of order $\bigl(\kappa c_B(\sigma)\bigr)^{2}$. This accounts for the whole of the difference in polarisation.

The welfare-maximising architecture by composition and the corresponding thresholds are reported in Appendix~\ref{app:composition}.

\section{Numerical Protocol for Section~\ref{sec:policy}}\label{app:numerical}

This appendix records the parameterisation, the verification procedure and the grids behind the numerical statements of Section~\ref{sec:policy} and Appendix~\ref{app:policy}. It also reports the welfare-maximising architecture by composition and collects further descriptive evidence on the three criteria. The exact values reported in Examples~\ref{ex:equivalence_class}--\ref{ex:echo_fails} are derived in Appendix~\ref{app:policy}. A replication script producing every number in Section~\ref{sec:policy} and in Appendices~\ref{app:policy}--\ref{app:numerical} is available with the paper.

\subsection*{Parameterisation}\label{app:param}

In all numerical illustrations we set $\underline{s} = 0$, $\bar{s} = 1$, $\underline{t} = 0$ and $\bar{t} = 1$, so that
\[
  \mathbb{E}[s] = p\lambda + (1-p)(1-\mu), \qquad \mathbb{E}[t] = p,
  \qquad m_{\bar t} = \lambda, \qquad m_{\underline t} = 1-\mu,
\]
\[
  v_{\bar t} = \lambda(1-\lambda), \qquad v_{\underline t} = \mu(1-\mu),
  \qquad
  \zeta = \frac{p\lambda}{\mathbb{E}[s]} - \frac{p(1-\lambda)}{1-\mathbb{E}[s]} .
\]
This choice fixes the scale of the two supports and is not restrictive for the comparisons reported. Adding a constant to every $t_i$ shifts all ideal points and all equilibrium actions by the same amount, leaving both $SW$ and $\mathrm{Var}(\mathbf{a}^*)$ unchanged, while rescaling $\bar t - \underline t$ is absorbed into $\beta_2$. With rational $(\alpha,\beta_1,\beta_2,p,\lambda,\mu)$ all the primitives above are rational, and so are $\gamma$, $\zeta$, $\kappa$ and every quantity in \eqref{eq:Msigma}--\eqref{eq:Usigma}. The comparisons in Examples~\ref{ex:welfare_fails} and \ref{ex:echo_fails} are therefore exact rational inequalities, computed with exact arithmetic rather than in floating point.

\subsection*{Independent Verification}\label{app:verification}

The closed forms of Proposition~\ref{prop:closed_forms} were checked against a direct implementation of Theorem~\ref{thm:users_equilibrium}. For a given architecture and type profile, the equilibrium action of every user was computed from the equilibrium formula at each of the $2^{n}$ post profiles, and $\mathbb{E}_P[SW]$, $\mathbb{E}_P[\mathrm{Var}(\mathbf{a}^*)]$, the interim utilities $U_\sigma$ and the pair moments $P_{\sigma\sigma'}$ were obtained by averaging with the conditional probabilities $\Pr(\mathbf{s}\mid\mathbf{t}) = \prod_k \Pr(s_k\mid t_k)$. Both computations use exact rational arithmetic, and they agree exactly for all sixteen block architectures at $n = 6$ for the compositions $(n_h,n_l) \in \{(5,1),(4,2),(3,3)\}$, and for five architectures at each of the compositions $(5,3)$ and $(4,4)$ at $n = 8$.

Three further checks were run on the intermediate results. The innovation form of Lemma~\ref{lem:innovation} was verified against Theorem~\ref{thm:users_equilibrium} on randomly drawn networks and post profiles, for general supports $\underline s < \bar s$ and $\underline t < \bar t$ rather than only the parameterisation above. The general expressions \eqref{eq:gen_consistency}--\eqref{eq:gen_criteria} were verified against enumeration on randomly drawn networks that are not block architectures, which is the case relevant to Example~\ref{ex:equivalence_class}. Finally, the identity of Lemma~\ref{lem:sw_var_identity} was verified on randomly drawn networks.

\subsection*{The Parameter Grid}\label{app:grid}

The frequency statement in Section~\ref{subsec:block} refers to the following grid, on which all sixteen block architectures are evaluated.
\[
\begin{array}{ll}
  \text{compositions } (n_h,n_l): & (3,3),\ (4,2),\ (5,1),\ (2,4),\ (6,2),\ (4,4);\\[2pt]
  \alpha: & 0.1,\ 0.3,\ 0.5,\ 0.7,\ 0.85,\ 0.95;\\[2pt]
  \beta_1/\beta_2 \ (\text{with } \beta_2 = 1): & 0.05,\ 0.1,\ 0.3,\ 1,\ 3,\ 10,\ 30;\\[2pt]
  \lambda,\ \mu: & 0.55,\ 0.7,\ 0.85,\ 0.95,\ 0.99;\\[2pt]
  p: & 0.3,\ 0.5,\ 0.7,
\end{array}
\]
giving $18{,}900$ configurations. Two architectures are counted as tied when their criterion values agree to within $10^{-9}$. Ties arise mechanically whenever a block is vacuous, since with $n_l = 1$, for instance, the low--low block has no effect and two quadruples differing only in that entry describe the same architecture.

\medskip
\noindent\textbf{Polarisation.} On this grid $G_{\mathrm{echo}}$ attains the maximum of $\mathbb{E}_P[\mathrm{Var}(\mathbf{a}^*)]$ among the sixteen block architectures in $96.3\%$ of configurations and fails in $3.7\%$. The failures are monotone in $\beta_1/\beta_2$ but not in $\alpha$, as the following counts show.

\begin{center}
\begin{tabular}{lccccccc}
\toprule
$\beta_1/\beta_2$ & $0.05$ & $0.1$ & $0.3$ & $1$ & $3$ & $10$ & $30$ \\
failures (of $2{,}700$) & $0$ & $4$ & $20$ & $59$ & $116$ & $223$ & $275$ \\
\midrule
$\alpha$ & $0.1$ & $0.3$ & $0.5$ & $0.7$ & $0.85$ & $0.95$ & \\
failures (of $3{,}150$) & $20$ & $1$ & $35$ & $85$ & $189$ & $367$ & \\
\bottomrule
\end{tabular}
\end{center}

In $\alpha$ the failures are concentrated at high values, with $556$ of the $697$ at $\alpha \geq 0.85$, but they also include a cluster of $20$ at $\alpha = 0.1$, all with $\beta_1/\beta_2 \geq 3$ and all with $\lambda = 0.55$ or $\mu = 0.55$, where posts are barely informative about their authors. The failure does not require a coordination motive. Repeating the exercise with $\alpha = 0.001$ on the remaining dimensions of the grid, the echo chamber fails to maximise polarisation in $33$ of the $3{,}150$ configurations, whereas the complete network maximises welfare in all of them.

\medskip
\noindent\textbf{Welfare.} On the same grid $G_{\mathrm{complete}}$ attains the maximum of $\mathbb{E}_P[SW]$ among the sixteen block architectures in $71.3\%$ of configurations and fails in $28.7\%$, that is in $5{,}418$ cases. When it fails, the maximiser is $G[1,1,1,0]$ in $1{,}967$ cases, $G[0,1,1,1]$ in $1{,}077$, $G_{\mathrm{het}}$ in $1{,}074$, $G_{\max}$ in $798$, $G[0,1,0,0]$ in $233$, $G[0,0,1,0]$ in $209$ and $G_{\min}$ in $60$. The first three delete only links between users of the same type, and the last four also delete links between users of different types, in $1{,}300$ cases in all. Both frequencies refer to the maximum over the sixteen block architectures. An architecture that maximises a criterion over all networks also maximises it over the block architectures, so the frequencies over all networks can only be lower, and the two conjectures fail at least as often as reported here.

On this grid no welfare-maximising network shows a type's posts to its own members while hiding them from the other type. The property is not general. At $n_h = 8$, $n_l = 7$, $\alpha = 0.1316$, $\beta_1 = 0.00731$, $\beta_2 = 0.14486$, $p = 0.6$, $\lambda = 0.947$ and $\mu = 0.95$, the maximiser is $G[1,0,1,0]$, in which high-type posts reach only high-type users, and it exceeds every architecture without that feature by an exact positive margin. In a further sample of $25{,}000$ configurations drawn over wider ranges than the grid, this was the only such case found. Relatedly, the complete network need not beat the maximal echo chamber: at $n_h = n_l = 3$, $\alpha = 0.97$, $p = \tfrac12$, $\lambda = \mu = 0.7$, $\beta_1 = 0.01$ and $\beta_2 = 1$ the echo chamber yields strictly higher welfare, although neither architecture is optimal there.

\medskip
\noindent\textbf{Compositions matching the prior.} Restricting the grid to the compositions with $n_h = np$, namely $(3,3)$ and $(4,4)$ at $p = \tfrac12$, the complete network fails to maximise welfare in $533$ of the $2{,}100$ configurations, all of them with $\beta_1/\beta_2 \leq 0.3$ ($290$ at $0.05$, $239$ at $0.1$ and $4$ at $0.3$). The maximiser is $G_{\mathrm{het}}$ in $410$ of them, $G[0,0,1,0]$ in $61$, $G[0,1,0,0]$ in $36$, and $G[1,1,1,0]$ and $G[0,1,1,1]$ in $13$ each. These are failures of the kind in Example~\ref{ex:welfare_fails_het}, in which the realised composition matches the prior.

\medskip
\noindent\textbf{Welfare before types are drawn.} For a block rule $B$, let
\[
  W^{\mathrm{ex}}(B) := \sum_{\mathbf{t} \in T^{n}} \Pr(\mathbf{t})\;\mathbb{E}_P\bigl[SW(G[B])\bigr]
\]
be expected welfare before Nature draws types, where the platform applies the rule $B$ at every type profile, including the profiles in which a single type is present, and $\mathbb{E}_P[SW]$ is computed from \eqref{eq:gen_sw} at each profile. Since the belief of a user about the network does not depend on the types of others, Theorem~\ref{thm:users_equilibrium} applies at every profile. At the parameters of Example~\ref{ex:welfare_fails}, $W^{\mathrm{ex}}$ equals $-9.6706$ under the complete network and $-14.7925$ under $G_{\max}$. At those of Example~\ref{ex:welfare_fails_het}, it equals $-0.5518$ under the complete network and $-0.5759$ under $G_{\mathrm{het}}$. In both cases the complete network maximises $W^{\mathrm{ex}}$ among the sixteen block architectures.

We evaluated $W^{\mathrm{ex}}(B)$ for all sixteen block rules at $n \in \{6, 8\}$, on the remaining dimensions of the grid, giving $3{,}150$ configurations for each $n$. No architecture that hides the posts of a type from every user maximises $W^{\mathrm{ex}}$ in any configuration, so the gain from muting a type in Example~\ref{ex:welfare_fails} does not survive averaging over type profiles. The complete network fails to maximise $W^{\mathrm{ex}}$ in $143$ configurations at $n = 6$ and in $166$ at $n = 8$. All of these have $\beta_1/\beta_2 \leq 0.1$ and $\alpha \leq 0.5$, and the margin never exceeds $0.0092$ at $n = 6$ and $0.0188$ at $n = 8$. In every one of them the maximiser deletes only links between users of the same type, being $G[1,1,1,0]$, $G[0,1,1,1]$ or $G_{\mathrm{het}}$. Failures of the kind in Example~\ref{ex:welfare_fails_het} therefore survive, in a small region and by small margins, when welfare is evaluated before types are drawn.

\subsection*{Composition and Thresholds}\label{app:composition}

The composition of the population, and not only the parameters, determines the welfare-maximising architecture. Table~\ref{tab:optimal_by_composition} reports it at the parameters of Example~\ref{ex:welfare_fails}, computed by evaluating all sixteen block architectures at each composition. At these parameters the minority's influence is withdrawn progressively as the minority shrinks, from deleting the low--low block at $n_l \in \{2,3\}$ to muting the minority altogether at $n_l = 1$. At $n=6$ the pattern is mirror-symmetric, with $G_{\max}$ at $(n_h,n_l)=(5,1)$ against $G_{\min}$ at $(1,5)$, and $G[1,1,1,0]$ at $(4,2)$ against $G[0,1,1,1]$ at $(2,4)$. These are features of the parameters of the table rather than properties of the class. The symmetry of the model guarantees the mirror pattern only when $p = \tfrac12$ and $\lambda = \mu$, and neither the progressive withdrawal nor the particular architectures in the table persist across the parameter space. How the welfare-maximising architecture depends on the primitives is taken up in Section~\ref{sec:conclusions}.

\begin{table}[h!]
\centering
\begin{tabular}{lccccc}
\toprule
 & $n_l=1$ & $n_l=2$ & $n_l=3$ & $n_l=4$ & $n_l=5$ \\
\midrule
$n=6$  & $G_{\max}$ & $G[1,1,1,0]$ & $G_{\mathrm{complete}}$ & $G[0,1,1,1]$ & $G_{\min}$ \\
$n=8$  & $G_{\max}$ & $G[1,1,1,0]$ & $G[1,1,1,0]$ & $G_{\mathrm{complete}}$ & $G[0,1,1,1]$ \\
$n=10$ & $G_{\max}$ & $G[1,1,1,0]$ & $G[1,1,1,0]$ & $G_{\mathrm{complete}}$ & $G_{\mathrm{complete}}$ \\
$n=12$ & $G_{\max}$ & $G[1,1,1,0]$ & $G[1,1,1,0]$ & $G_{\mathrm{complete}}$ & $G_{\mathrm{complete}}$ \\
\bottomrule
\end{tabular}
\caption{Welfare-maximising block architecture by type composition, at the parameters of Example~\ref{ex:welfare_fails}. $G[1,1,1,0]$ is the complete network with the low--low block deleted, so that low-type users no longer observe one another. At $n_l = 1$ the low--low block is vacuous, so $G_{\max}$ ties with $G[1,1,0,1]$, and likewise $G_{\min}$ with $G[1,0,1,1]$ at $(n_h,n_l) = (1,5)$.}
\label{tab:optimal_by_composition}
\end{table}

The welfare shortfall of the complete network relative to the optimum is as follows.
\[
\begin{array}{lccccc}
\toprule
 & n_l = 1 & n_l = 2 & n_l = 3 & n_l = 4 & n_l = 5\\
\midrule
n = 6  & 0.1355 & 0.5067 & 0 & 0.8080^{\dagger} & 0.3169^{\dagger}\\
n = 8  & 0.3963 & 0.7906 & 0.0784 & 0 & 0.7623^{\dagger}\\
n = 10 & 0.6380 & 0.9542 & 0.7472 & 0 & 0\\
n = 12 & 0.8703 & 1.0603 & 1.2064 & 0 & 0\\
\bottomrule
\end{array}
\]
Entries marked $\dagger$ correspond to compositions in which the low type is the majority, where the optimum is the mirror image of the corresponding minority-high architecture.

Restricting attention to the pair $\{G_{\mathrm{complete}}, G_{\max}\}$, we evaluated $\mathbb{E}_P[SW(G_{\mathrm{complete}})] - \mathbb{E}_P[SW(G_{\max})]$ from \eqref{eq:Usigma} at the parameters of Example~\ref{ex:welfare_fails}, for each $n_l \in \{1,\dots,6\}$ and each $n_h$ from $1$ to $2499$. For $n_l = 1, 2, 3, 4$ the complete network is preferred up to $n_h = 4$, $9$, $23$ and $67$ respectively, and $G_{\max}$ thereafter. For $n_l \in \{5,6\}$ the complete network is preferred at every $n_h$ in the range. The implied critical ratios $n_h/n_l$, namely $4$, $4.5$, $7.7$ and $16.8$, diverge, so no condition of the form $n_h/n_l > \bar c$ describes the region in which the complete network is dominated.

\subsection*{The Three Criteria across the Design Space}\label{app:scatter}

\begin{figure}[p]
\centering
\begin{minipage}{\textwidth}
\centering
\includegraphics[width=0.64\textwidth]{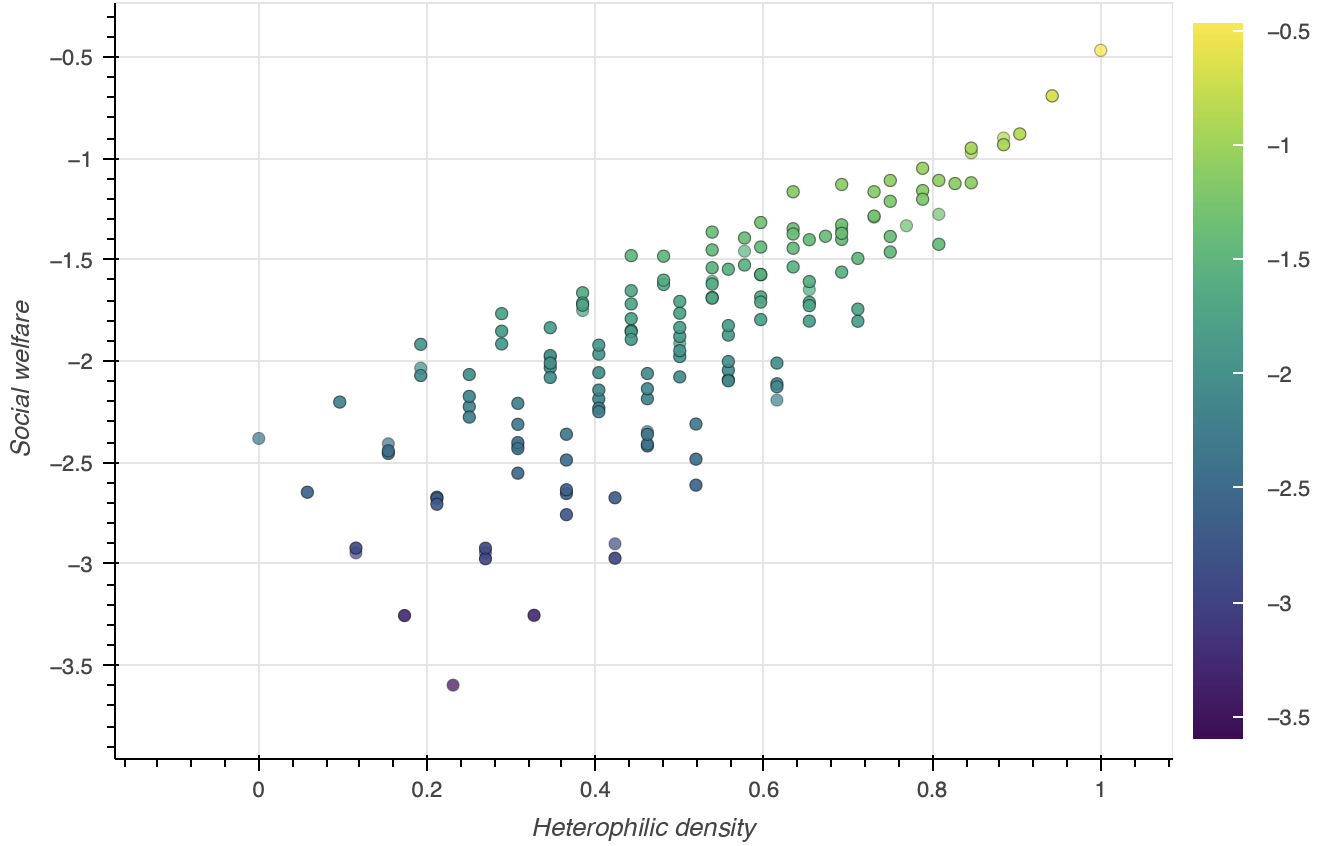}
\subcaption{$p = 0.5$, with $n_h = n_l = 2$.}
\end{minipage}
\\[4pt]
\begin{minipage}{\textwidth}
\centering
\includegraphics[width=0.64\textwidth]{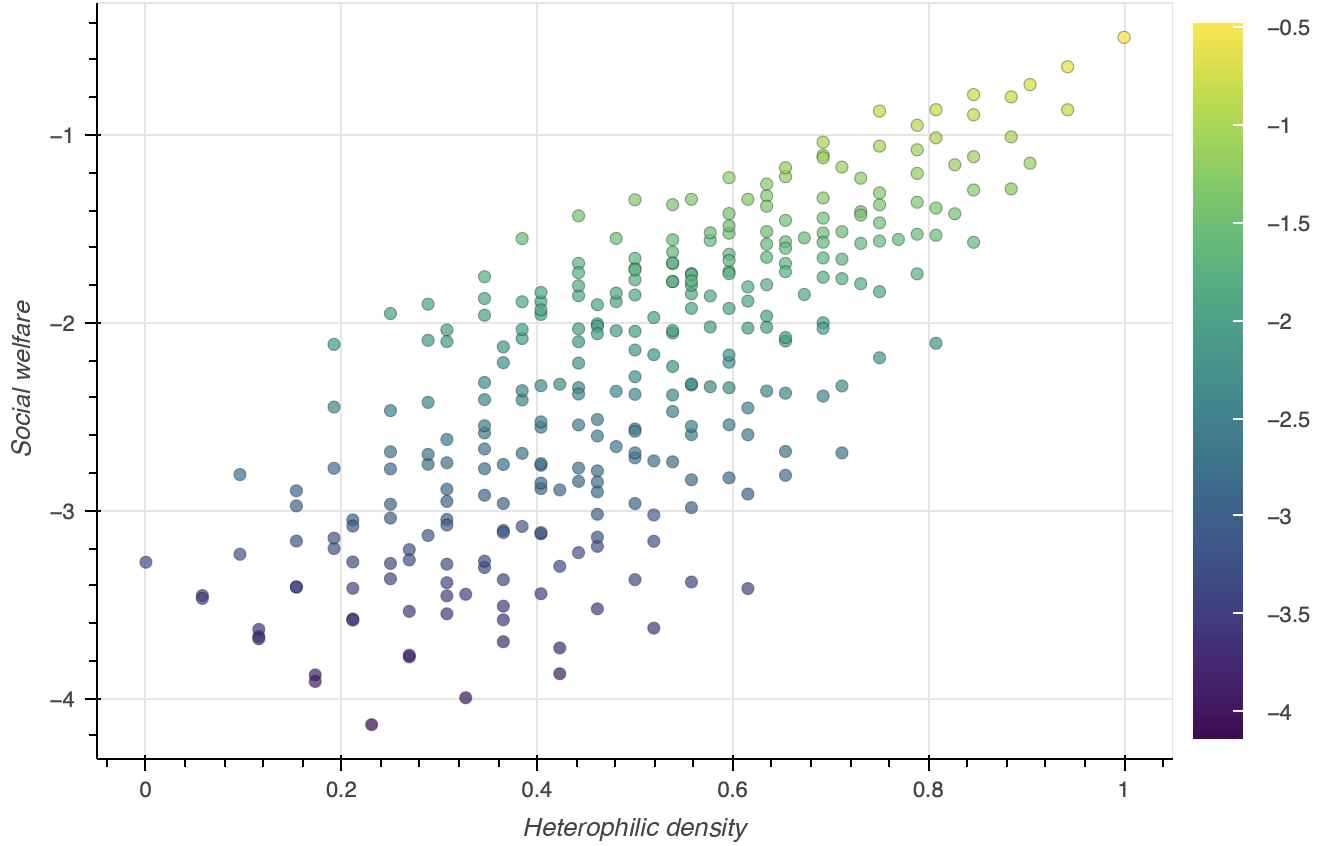}
\subcaption{$p = 0.75$, with $n_h = n_l = 2$.}
\end{minipage}
\\[4pt]
\begin{minipage}{\textwidth}
\centering
\includegraphics[width=0.64\textwidth]{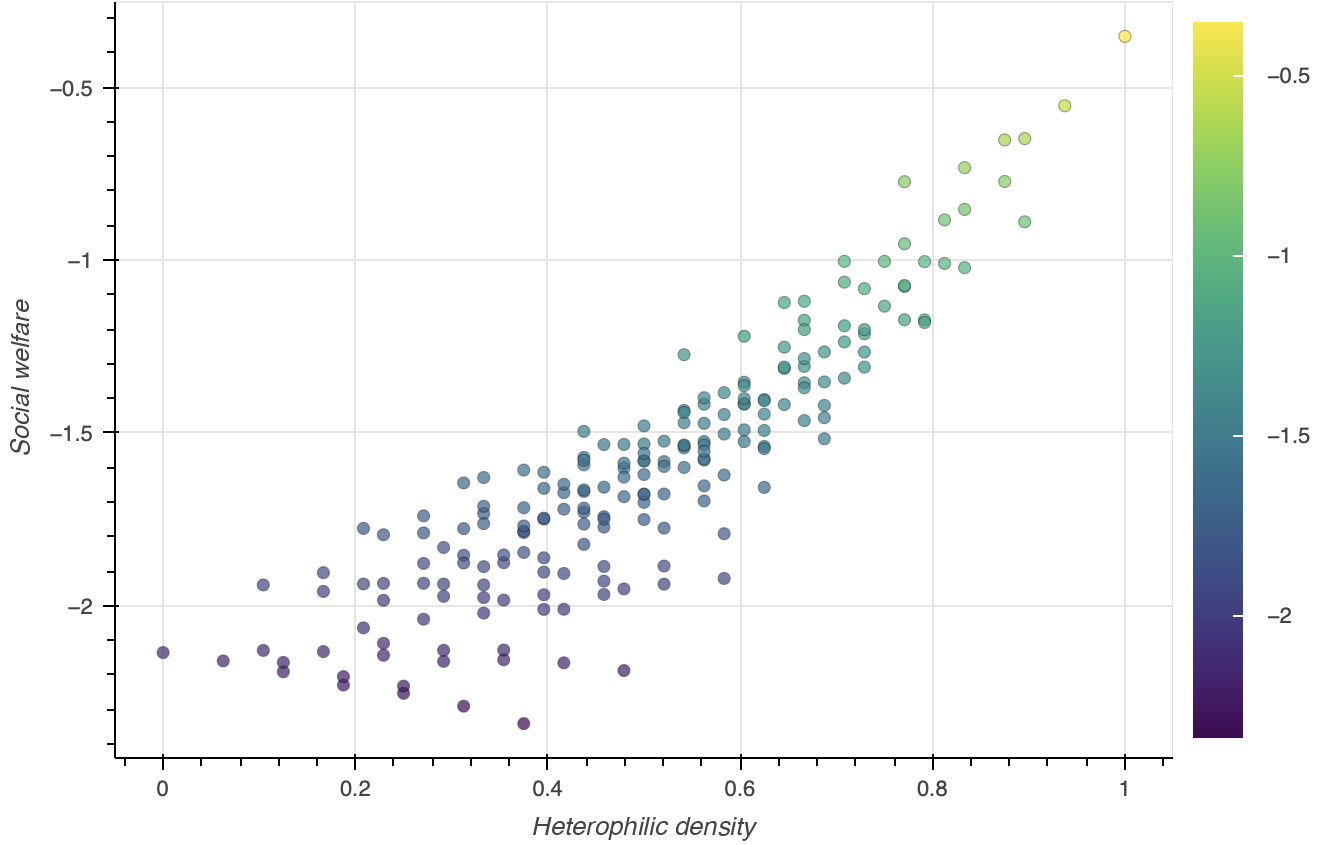}
\subcaption{$p = 0.75$, with $n_h = 3$ and $n_l = 1$.}
\end{minipage}
\caption{Social welfare $\mathbb{E}_P[SW]$ against heterophilic density for all $4096$ architectures on $n=4$ users, at $\alpha = 0.4$, $\beta_1 = \beta_2 = 1$, $\lambda = \mu = 0.8$, $\overline{t}=\overline{s}=1$, $\underline{t}=\underline{s}=0$ and $\tau = 0.6$. Colour also indicates social welfare.}
\label{fig:app_utility}
\end{figure}

\begin{figure}[p]
\centering
\begin{minipage}{\textwidth}
\centering
\includegraphics[width=0.64\textwidth]{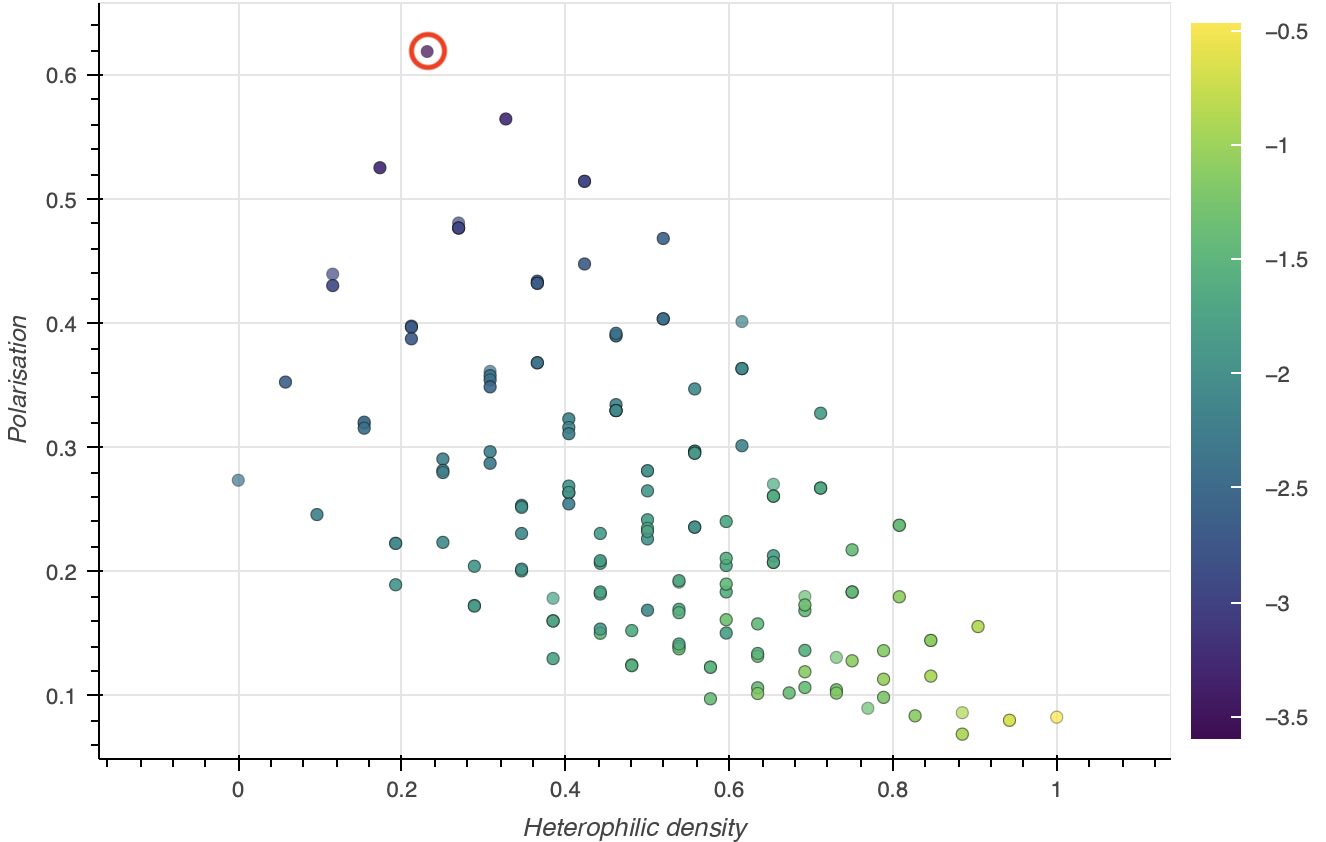}
\subcaption{$p = 0.5$, with $n_h = n_l = 2$.}
\end{minipage}
\\[4pt]
\begin{minipage}{\textwidth}
\centering
\includegraphics[width=0.64\textwidth]{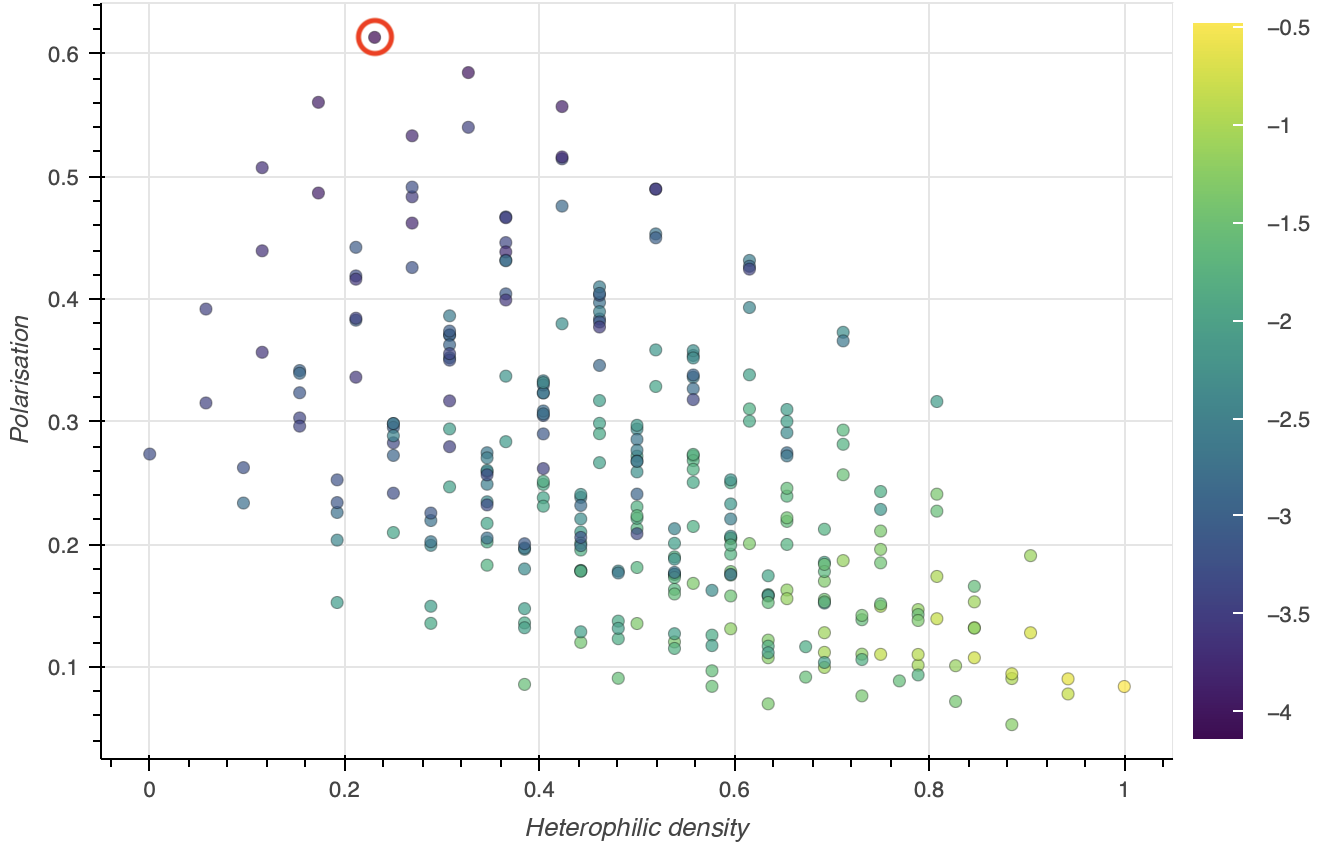}
\subcaption{$p = 0.75$, with $n_h = n_l = 2$.}
\end{minipage}
\\[4pt]
\begin{minipage}{\textwidth}
\centering
\includegraphics[width=0.64\textwidth]{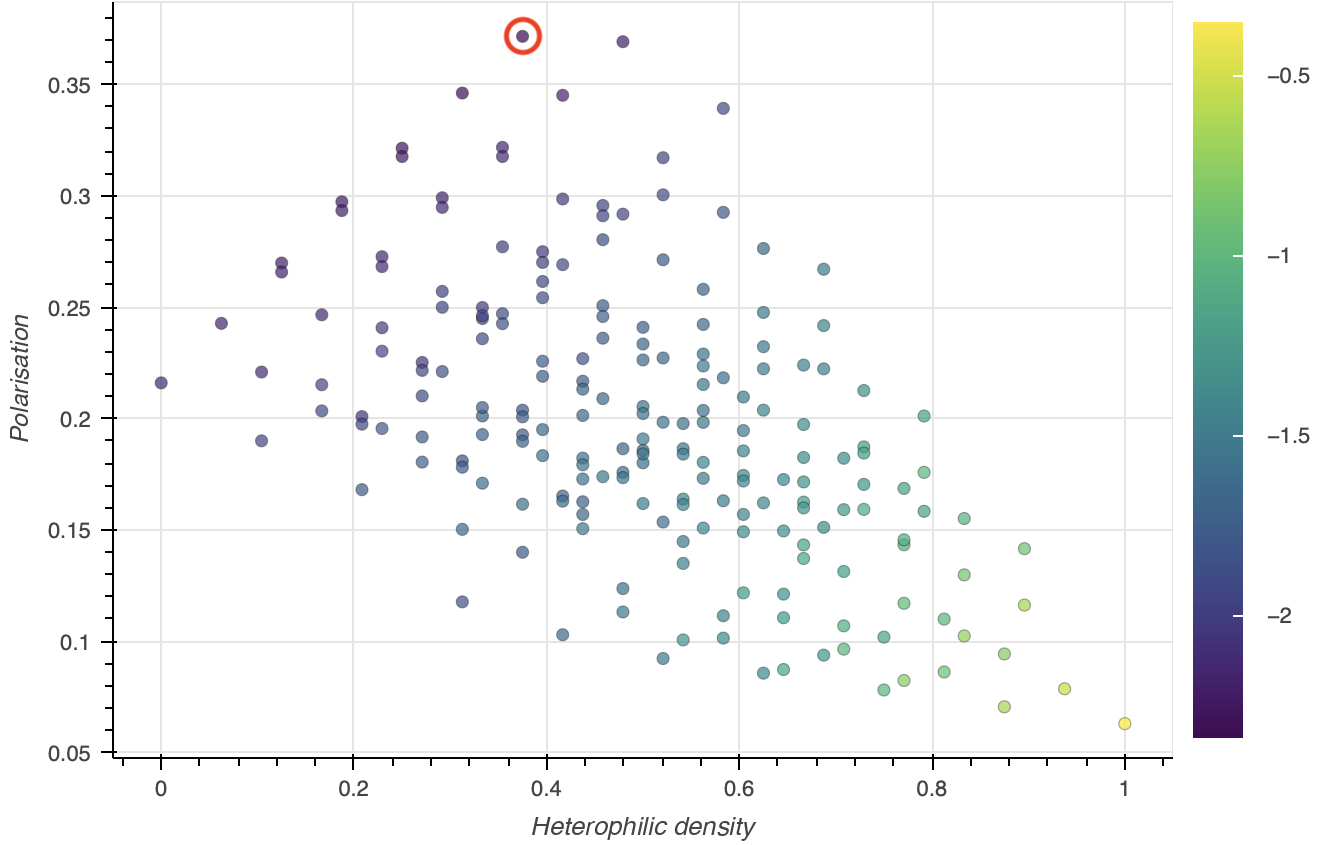}
\subcaption{$p = 0.75$, with $n_h = 3$ and $n_l = 1$.}
\end{minipage}
\caption{Polarisation against heterophilic density for all $4096$ architectures on $n=4$ users, at the parameters of Figure~\ref{fig:app_utility}. Colour indicates social welfare $\mathbb{E}_P[SW]$. The circled point marks the architecture attaining the highest polarisation, which in each panel is the maximal echo chamber.}
\label{fig:app_var}
\end{figure}

Figures~\ref{fig:app_utility} and \ref{fig:app_var} display social welfare and polarisation against heterophilic density for all $4096$ architectures on four users, at three configurations of the type distribution and composition. They are descriptive and are reported for completeness, since the results of Section~\ref{sec:policy} rest on the exact comparisons of Appendix~\ref{app:policy} rather than on these panels.

\end{document}